\documentclass[12pt]{article}
\pdfoutput=1

\usepackage{amssymb,amsmath,bm,bbm}
\usepackage{cite}
\usepackage{graphicx}
\usepackage{placeins}
\usepackage{longtable,color}

\definecolor{brown}{rgb}{0.6,0.7,0.}
\definecolor{gray}{rgb}{.38,.38,.38}

\def\simge{\mathrel{\rlap{\raise 0.511ex \hbox{$>$}}{\lower 0.511ex \hbox{$\sim$}}}}
\def\simle{\mathrel{\rlap{\raise 0.511ex \hbox{$<$}}{\lower 0.511ex \hbox{$\sim$}}}}
\def\slash#1{\setbox0=\hbox{$#1$}\dimen0=\wd0
      \setbox1=\hbox{/} \dimen1=\wd1 \ifdim\dimen0>\dimen1
      \rlap{\hbox to \dimen0{\hfil/\hfil}} #1                        \else
      \rlap{\hbox to \dimen1{\hfil$#1$\hfil}}
      /   \fi}

\newcommand{\newsection}[1]{\section{#1}\setcounter{equation}{0}}

\newcommand{\lsim}{
\mathrel{\hbox{\rlap{\hbox{\lower4pt\hbox{$\sim$}}}\hbox{$<$}}}}

\newcommand{\gsim}{
\mathrel{\hbox{\rlap{\hbox{\lower4pt\hbox{$\sim$}}}\hbox{$>$}}}}

\newcommand{\vud}{|V_{ud}|}
\newcommand{\vus}{|V_{us}|}
\newcommand{\vcs}{|V_{cs}|}
\newcommand{\vcd}{|V_{cd}|}
\newcommand{\vcb}{|V_{cb}|}
\newcommand{\vtd}{|V_{td}|}
\newcommand{\vub}{|V_{ub}|}
\newcommand{\vts}{|V_{ts}|}
\newcommand{\vtb}{|V_{tb}|}

\def\eps{\varepsilon}
\def\epe{\varepsilon'/\varepsilon}

\newcommand{\gev}{\, {\rm GeV}}
\newcommand{\mev}{\, {\rm MeV}}

\newcommand{\mw}{M_{\rm W}}

\def \cez {{c_{12}}}   
\def \ced {{c_{13}}}   
\def \cev {{c_{14}}}   
\def \czd {{c_{23}}}   
\def \czv {{c_{24}}}   
\def \cdv {{c_{34}}}   
\def \sez {{s_{12}}}   
\def \sed {{s_{13}}}   
\def \sev {{s_{14}}}   
\def \szd {{s_{23}}}   
\def \szv {{s_{24}}}   
\def \sdv {{s_{34}}}   
\def \ded {{\delta_{13}}}   
\def \dev {{\delta_{14}}}   
\def \dzv {{\delta_{24}}}

\allowdisplaybreaks[2]

\newcommand{\be}{\begin{equation}}
\newcommand{\ee}{\end{equation}}
\newcommand{\bea}{\begin{eqnarray}}
\newcommand{\eea}{\end{eqnarray}}
\newcommand{\nn}{\nonumber}
\newcommand{\bi}{\begin{itemize}}
\newcommand{\ei}{\end{itemize}}
\newcommand{\ord}{{\cal O}}

\newcommand{\RE}{{\rm Re}}
\newcommand{\IM}{{\rm Im}}

\def\kpn{K^+\rightarrow\pi^+\nu\bar\nu}

\def\klpn{K_L\rightarrow\pi^0\nu\bar\nu}

\providecommand{\update}[1]
{\ifx\undefined\updatebf #1\else {\bf \boldmath #1\unboldmath}\fi}

\begin{document}
\begin{titlepage}
\vspace*{-1.0truecm}

{\Large \today}
\begin{flushright}
TUM-HEP-750/10\\
\end{flushright}

\vspace{0.4truecm}

\begin{center}
\boldmath

{\Large\textbf{Patterns of Flavour Violation
in the Presence of \\[0.2cm] a Fourth Generation 
of Quarks and Leptons}}

\unboldmath
\end{center}

\vspace{0.4truecm}

\begin{center}
{ Andrzej J.~Buras$^{a,b}$, Bj\"orn Duling$^a$, Thorsten Feldmann$^a$,\\
Tillmann Heidsieck$^a$, Christoph Promberger$^a$, Stefan Recksiegel$^a$
}
\vspace{0.4truecm}

{\footnotesize
 $^a${\sl Physik Department, Technische Universit\"at M\"unchen,
James-Franck-Stra{\ss}e, \\D-85748 Garching, Germany}\vspace{0.2truecm}

 $^b${\sl TUM Institute for Advanced Study, Technische Universit\"at M\"unchen,  Arcisstr.~21,\\ D-80333 M\"unchen, Germany}

}

\end{center}

\vspace{0.4cm}
\begin{abstract}
\noindent \footnotesize

We calculate a number of observables related to particle-antiparticle mixing 
and the branching ratios for the most interesting rare and CP-violating
$K$ and $B$ decays in the Standard Model (SM3) extended by a fourth generation 
of quarks and leptons (SM4).
A model-independent parametrisation of these observables in
terms of gauge-independent functions 
is adopted, which is useful for studying the breaking of the
universality between $K$, $B_d$ and $B_s$ systems through non-minimal
flavour violating interactions.
We calculate first the mass differences $\Delta M_i$ in the neutral
$K$ and $B$ systems, the mixing-induced
CP asymmetries $S_{\psi K_S}$, $S_{\psi\phi}$, 
$S_{\phi K_S}$, $S_{\eta^\prime K_S}$ and 
$\varepsilon_K$. Subsequently, a detailed analysis of 
$K^+\to\pi^+\nu\bar\nu$,
$K_L\to\pi^0\nu\bar\nu$, $B_{s,d}\to \mu^+\mu^-$, $B\to
X_{s,d}\nu\bar\nu$, $K_L \to\pi^0\ell^+\ell^-$, $B\to X_s\gamma$ and $B\to
X_{s,d}\ell^+\ell^-$ is presented, and also $\epe$ is considered. For
some  of these observables the departures from SM3 predictions can still be 
spectacular.

We discuss how the new mixing parameters (3 angles, 2 CP phases)
can be determined using the flavour observables in question. 
We identify the different hierarchical
structures in the SM4 flavour mixing matrix, allowed
by phenomenological and theoretical constraints, and 
define the corresponding generalised Wolfenstein expansion.
Most importantly,
we show how the characteristic patterns of correlations among the considered 
flavour observables
allow to distinguish this New Physics scenario from supersymmetric flavour
models,
the Littlest Higgs model
with T-parity and the Randall-Sundrum model with custodial protection.
 Of particular importance are the
correlations involving $S_{\psi\phi}$, $S_{\phi K_S}$, 
 ${\rm Br}(\kpn)$, ${\rm Br}(\klpn)$ and 
${\rm Br}(B_{s,d}\rightarrow\mu^+\mu^-)$, which could in principle rule out
the SM4. Interestingly $\epe$ turns out to be suppressed below the data for
large positive values of $S_{\psi\phi}$  unless the relevant hadronic matrix 
elements differ strongly from the large $N$ results.
 The important role of $\epe$ in bounding large enhancements of rare $K$ decay branching ratios is reemphasised.
 We also show how the existing anomalies
in the unitarity triangle fits as well as in  $S_{\psi\phi}$ and $S_{\phi K_S}$
can be simultaneously explained in the SM4 scenario.

\end{abstract}

\end{titlepage}
\setcounter{page}{1}
\pagenumbering{arabic}

\newsection{Introduction} \label{sec:intro}

One of the simplest extensions of the Standard Model 
(hereafter referred to as SM3)
is the addition of a sequential fourth generation (4G)
of quarks and leptons \cite{Frampton:1999xi} (hereafter referred to as SM4).
\update{While the setup we are considdering here (e.g. perturbative Yukawa couplings) does not address any of the known hierarchy and naturalness problems,}
if present in nature, a 4G is likely to have a number of
profound implications that makes it interesting to consider. These are:
\begin{enumerate}

 \item While being consistent with electroweak precision data (EWPT) 
\cite{Maltoni:1999ta,He:2001tp,Alwall:2006bx,Kribs:2007nz,Chanowitz:2009mz,Novikov:2009kc}, 
a 4G can remove the tension between the SM3 fit and the lower bound
on the Higgs mass $m_H$ from LEP~II. Indeed, as pointed out in 
\cite{He:2001tp,Kribs:2007nz,Hashimoto:2010at}, a heavy Higgs boson does not contradict 
EWPT as soon as the 4G exists.

\item $SU(5)$ gauge coupling unification could in principle 
be achieved without 
 supersymmetry \cite{Hung:1997zj}, although the present lower
bound on the masses of 4G quarks and the appearance 
of Landau poles in Yukawa couplings well below the GUT scales practically 
excludes this possibility if one wants to stay within a perturbative
framework at short-distance (SD) scales.

\item Electroweak baryogenesis might be viable \cite{Hou:2008xd,Kikukawa:2009mu,Fok:2008yg}.

\item Most importantly, from the point of view of the present paper,
certain anomalies present in flavour-changing processes could in principle
be resolved \cite{Hou:2005yb,Hou:2006mx,Soni:2008bc}.

\item  The structure of the lepton sector in the SM4 can be interesting as shown recently in \cite{Burdman:2009ih}: a heavy (mostly Dirac)
4G neutrino in addition to very light (mostly Majorana) neutrinos
can be obtained in a setup with electroweak symmetry breaking in
warped extra dimensions.

\end{enumerate}
\update{On the other hand there are scenarios where a fourth generation might trigger dynamical breaking of electroweak symmetry \cite{Holdom:1986rn,Hill:1990ge,King:1990he,Burdman:2007sx,Hung:2009hy,Hung:2009ia,Holdom:2010za}.
However since it is in the very nature of non-perturbative Yukawa couplings to defy direct calculations there is until now no explicit model for this scenario.}

There is a rich literature on the implications and phenomenology of
the SM4. Reviews and summary statements can be found in 
\cite{Frampton:1999xi, Holdom:2009rf}.
In particular during the last years, a number of analyses were
published with the goal to investigate the impact of the existence of
a 4G on Higgs physics \cite{Kribs:2007nz}, 
electroweak precision tests 
\cite{Alwall:2006bx,Kribs:2007nz,Chanowitz:2009mz,Novikov:2009kc} and 
flavour physics \cite{Arhrib:2002md,Hou:2005yb,Hou:2006mx,Soni:2008bc,Herrera:2008yf,Bobrowski:2009ng,Eilam:2009hz}. 
In this context, Bobrowski et al.~\cite{Bobrowski:2009ng}
have studied the constraint on the mixing between 
the fourth and third generation 
by using FCNC processes, and Chanowitz \cite{Chanowitz:2009mz} by 
using global fits
to EWPT. Interesting bounds on this
mixing could be derived in this manner. We will address these bounds
in the course of our analysis.

 Detailed analyses of supersymmetry in the presence of a 4G
 have recently been performed in 
\cite{Murdock:2008rx,Godbole:2009sy}.  These analyses
show that the marriage of SUSY with the SM4 is rather
challenging because of Landau poles  in Yukawa couplings 
present at relatively low scales.
The last paper contains a good up to date collection of references to
 papers on the SM4. 

In the present paper we will take a different strategy by considering, in
addition to $\Delta F=2$ transitions and $B \to X_s\gamma$ and $B \to X_s\ell^+\ell^-$
decays as done in \cite{Soni:2008bc,Bobrowski:2009ng}, 
also rare $K$ and $B$ decays, as considered in \cite{Hou:2005yb},
on which new data will be available in the coming decade. In particular,
\renewcommand{\labelenumi}{\roman{enumi})}
\begin{enumerate}

\item
We will establish a number of correlations between various observables that
should allow us to distinguish this NP scenario from the Littlest Higgs 
model with T-parity (LHT), the Randall-Sundrum model with custodial 
protection (RSc) and supersymmetric flavour models that have been analysed in \cite{Blanke:2006sb,Blanke:2006eb,Blanke:2009am,Blanke:2008zb,Blanke:2008yr,Altmannshofer:2009ne}
recently. We will also carefully study how these correlations depend on
the size of the 4G mixing angles and phases.

\item
For the most interesting observables,
we will {investigate the} departures from models with (constrained) minimal flavour violation ((C)MFV),
taking into account all existing constraints.

\item
We will demonstrate transparently how certain anomalies observed in the  unitarity
triangle fits and in the CP asymmetries $S_{\psi\phi}$ and $S_{\phi K_S}$
 can be resolved simultaneously and how these solutions affect other 
observables.

 \item We will address the question how the additional five
parameters of the $4\times 4$ quark mixing matrix, $\theta_{14}$,
$\theta_{24}$, $\theta_{34}$, $\delta_{14}$ and $\delta_{24}$, could
--- in principle --- be determined by means of the mixing-induced CP asymmetries $S_{\psi\phi}$
and $S_{\psi K_S}$, the $B^0_{d,s}-\bar B_{d,s}$ mixing mass 
differences $\Delta M_{d,s}$, $\epsilon_K$ and the
branching ratios for the rare decays $K^+\to \pi^+\nu\bar \nu$, 
$K_L \to \pi^0 \nu\bar \nu$, $B_{s,d} \to \mu^+\mu^-$, $B\to X_s\nu\bar\nu$.

\item On the theoretical side, we will discuss how the SM4 can be
understood as a particular realisation of a next-to-minimal flavour violation
scenario. This introduces certain consistency conditions between the SM3 and
SM4 mixing angles which can be used to eliminate ``fine-tuned'' values in
parameter space. The remaining cases can be classified according to the
scaling of the 4G mixing angles with the Cabibbo angle,
and for each individual case an SM4 generalisation of the Wolfenstein parametrisation
can be constructed. 

\end{enumerate}
\renewcommand{\labelenumi}{\arabic{enumi}}

Our paper is organised as follows:
In Section~\ref{sec:V4G}, we discuss the flavour symmetries in the SM4, 
provide the $4\times 4$ quark mixing matrix and 
address the question of possible hierarchies between the 4G mixing angles. 
In Section~\ref{sec:EffHam}, we generalise the effective weak Hamiltonians to the SM4.
In Section~\ref{sec:MasterFormulae}, we compile basic formulae for FCNC processes considered
in our paper. As the SM4 goes beyond MFV,
but the operator structure of the SM3 effective Hamiltonian remains intact,
it is very efficient to introduce -- as done in the LHT model \cite{Blanke:2006sb,Blanke:2006eb,Blanke:2009am} -- generalised complex
master functions $S_i,\,X_i,\,Y_i,\,Z_i,\,D_i^\prime,\,E_i^\prime,\,E_i,\,(i=K,d,s)$,
that enter
the expressions for various observables. On the one hand, in the absence
of the 4G, these functions reduce to the SM3 flavour-universal
functions $S_0$, $X_0$, $Y_0$, $Z_0$, $D'_0$, $E'_0$ and $E_0$. 
On the other hand, using these functions,
non-MFV effects in this model can be very transparently compared to those
found in the LHT model.

In Section~\ref{sec:EWP}, we summarise the insights gained by other authors,
who studied in particular the impact of the 4G on
the electroweak precision observables.
In Section~\ref{sec:strategy}, we will summarise our strategy 
for  the  phenomenological analysis of the SM4.
In Section~\ref{sec:numerics}, we present a detailed global 
numerical analysis, using
the formulae of previous sections. In particular, we 
identify a number of correlations between various observables,
and we estimate approximate upper bounds on several CP asymmetries
and branching ratios. In Section~\ref{sec:Anatomy}, 
we study the anatomy of flavour effects in the SM4 scenario. 
In this context, we address the most important
phenomenological anomalies found in the present flavour data.

In Section~\ref{sec:DetV4G},  we outline an efficient procedure how to constrain the new
mixing angles $\theta_{14}$, $\theta_{24}$, $\theta_{34}$ and the new phases
$\delta_{14}$ and $\delta_{24}$ from
\begin{eqnarray}
&& S_{\psi\phi} \,, \quad
 S_{\psi K_S} \,, \quad
 \Delta M_{d,s} \,, \quad
 \epsilon_K \,,
\cr 
 && {\rm Br}(K^+ \to \pi^+ \nu\bar \nu) \,, \ \,
 {\rm Br}(K_L \to \pi^0 \nu\bar\nu) \,, \ \,
 {\rm Br}(B_{s,d} \to \mu^+\mu^-) \,, \ \,
 {\rm Br}(B \to X_{s,d} \nu\bar \nu) \,. 
\end{eqnarray}
For this purpose, we will provide
a classification of the 4G mixing parameters which allows us to connect 
the different patterns found in flavour observables to particular
properties of the 4G mixing angles and CP~phases.
Finally, in Section~\ref{sec:concl},
we summarise our main findings. Here we will also  compare the patterns of flavour violation in the SM4 with
those found in supersymmetric flavour models, the LHT model and the RSc model.
An appendix collects the one-loop functions used in our paper.

\boldmath\newsection{The $4\times 4 $ Mixing Matrix $V_\text{SM4}$}\label{sec:V4G}\unboldmath
\subsection{Yukawa Couplings and (Approximate) Flavour Symmetries}

The SM4 quark Yukawa sector involves two $4\times 4$ Yukawa
matrices,
\begin{equation}
 - {\cal L}_{\rm Yuk} 
 = \bar Q_L \, Y_U^{\rm SM4} \, \tilde H \, U_R
 + \bar Q_L \, Y_D^{\rm SM4} \, H \, D_R 
+ \mbox{h.c.}\,.
\end{equation}
Following \cite{D'Ambrosio:2002ex}, we identify the Yukawa couplings
as spurions, being the (only) sources for the breaking of the flavour symmetry
in the SM4,
\begin{align}
 G_F^{\rm SM4} &= SU(4)_{Q_L} \times SU(4)_{U_R} \times SU(4)_{D_R} \times U(1)_{U_R} \times U(1)_{D_R} \,,
\end{align}
where we have divided out one $U(1)$ factor corresponding to the conserved baryon number.
Under the flavour symmetry, the Yukawa spurions transform as
\begin{align}
 Y_U^{\rm SM4} \sim (4,\bar 4,1)_{1,0} \,, \qquad 
 Y_D^{\rm SM4} \sim (4,1,\bar 4)_{0,1} \,.
\end{align}
As in the SM3 case (see also \cite{Feldmann:2009dc,Santamaria:1993ah,Berger:2008zq}), 
the counting of symmetry generators (related to 18 angles + 29 phases)
 vs.\ complex Yukawa entries
(32 + 32) gives the number of physical parameters 
$$
  14 + 3 = \mbox{8 masses} + \mbox{6 angles} + \mbox{3 phases}\,,
$$
i.e.\ 2 additional masses, 3 additional angles and 2 additional phases
compared to the SM3 case.

Alternatively, we can consider the flavour symmetries from the SM3 point of view:
Taking only the large $t'$ and $b'$ Yukawa couplings into account, the effective
flavour symmetry is
\begin{align}
 G_{\rm eff}^{\rm SM3} &= SU(3)_{Q_L} \times SU(3)_{U_R} \times SU(3)_{D_R} \times U(1)_{U_R} \times U(1)_{D_R} \times U(1)_{Q'} \,,
\end{align}
which, compared to the SM3 case, involves an additional $U(1)_{Q'}$ symmetry, since --
without mixings involving the first three generations --
the 4G quark number is approximately conserved. 
The mixings
between the fourth and the first three generations can be parametrised
by an $SU(3)_{Q_L}$ triplet which is charged under the $U(1)_{Q'}$,
\begin{align}
 \chi_L \sim (3,1,1)_{0,0,1} \,.
\end{align}
Of the six independent entries of $\chi_L$, one phase is unobservable
due to the $U(1)_{Q'}$ symmetry, leaving exactly the additional
3 mixing angles and 2 CP-violating phases of the SM4. The remaining parameters
for the first three generations are contained in SM3 Yukawa matrices
\begin{align}
 Y_U^{\rm SM3} \sim (3,\bar 3,1)_{1,0,0} \,, \qquad 
 Y_D^{\rm SM3} \sim (3,1,\bar 3)_{0,1,0} \,.
\end{align}

The so-defined SM4 thus provides a particular example
of the next-to-minimal flavour violating construction discussed in
\cite{Feldmann:2006jk}, with new non-minimal flavour structures generated
by means of the $\chi_L$ spurion. Using a non-linear parametrisation 
(as it has been used for the SM3 case in \cite{Feldmann:2008ja,Kagan:2009bn}),
\begin{align}
 Y_U^{\rm SM4} &= \xi_L^\dagger \left(
\begin{array}{cc}
 Y_U^{\rm SM3} & \begin{array}{c} 0 \\ 0 \\ 0 \end{array}
\\
\begin{array}{ccc} 0 & 0 & 0 \end{array} & y_{t'}
\end{array}
\right) \,,
\qquad 
 Y_D^{\rm SM4} = \xi_L \, \left(
\begin{array}{cc}
 Y_D^{\rm SM3} & \begin{array}{c} 0 \\ 0 \\ 0 \end{array}
\\
\begin{array}{ccc} 0 & 0 & 0 \end{array} & y_{b'}
\end{array}
\right) \,,
\end{align}
we introduce the $4\times 4$ matrix
\begin{align}
\xi_L &= \exp\left[ i 
\left( \begin{array}{cc}
       \begin{array}{ccc} 0 & 0 & 0 \\ 0 & 0 & 0 \\ 0 & 0 & 0 \end{array}
& \chi_L
\\
\chi_L^\dagger & 0            
                  \end{array}
\right)\right] \,.
\end{align}
It transforms as $V_{Q_L} \xi_L V_{Q_L}^\dagger$ 
where $V_{Q_L}$ contains an $SU(3)_{Q_L}$ block matrix for
the SM3 generations, while the 4G fields remain constant.
The consistency conditions discussed in \cite{Feldmann:2006jk} translate
into
\begin{align}
& |(\chi_L)_i| |(\chi_L)_j| \ \lesssim \ \theta_{ij} \,,
\qquad && (i,j=1\ldots 3) \cr 
\Leftrightarrow \qquad & \theta_{ik} \theta_{jk} \ \lesssim \ \theta_{ij} \qquad &&(i,j,k=1\ldots 4)
\label{consist}
\end{align}
where no summation over ''k'' is understood.
Here $\theta_{12}\sim \lambda$, $\theta_{23} \sim \lambda^2$,
$\theta_{13} \sim \lambda^3$ are the SM3 mixing angles and 
$\theta_{i4}$ the 4G mixing angles (see below). Notice that for the SM3 case, 
one has one of the inequalities (\ref{consist}) saturated,
\begin{align}
& \theta_{12} \theta_{23} \sim \theta_{13} \,, \qquad \mbox{but} 
\qquad \theta_{12}\theta_{13} \ll \theta_{23}\,, \quad
      \theta_{13}\theta_{23} \ll \theta_{12} \,.
\end{align}

A particular way to realise the constraints (\ref{consist})
is to consider a simple Froggatt-Nielsen (FN)
setup \cite{Froggatt:1978nt}, where the scaling of the mixing angles is
controlled by different $U(1)$ charge factors $b_i$ for the left-handed doublets of
different generations, leading to
\begin{align}
 \theta_{ij} \sim \lambda^{|b_i - b_j|} \,.
\label{FN}
\end{align}
Here $\lambda$ is given by the VEV of some $U(1)$-breaking scalar field 
divided by a large UV-scale.
The triangle inequalities for $|b_i - b_k|+|b_k - b_j|$
then guarantee that (\ref{consist}) always holds.
Since (\ref{FN}) only involves charge \emph{differences}, 
we  may set $b_4 \equiv 0$, while
the charges $b_{1-3}$ and the related 4G mixing angles
are not completely fixed. However, we can identify certain benchmark cases
which may later be compared with the phenomenological constraints from EWPT, 
tree-level decays and rare decays of SM3 quarks:
Scenarios with some 4G mixing angles being of order ${\cal O}(1)$
are already ruled out by EWPT \cite{Alwall:2006bx,Chanowitz:2009mz}
and tree-level quark decays (see below).
Among the interesting scenarios with sufficiently small mixing angles, 
we identify:
\begin{align}
\mbox{(a) \quad 
 $b_3=1$, $b_2=3$, $b_1=4$:}
& \quad \Rightarrow \quad 
V_\text{SM4} 
\sim \left( \begin{array}{cccc}
1	& \lambda	& \lambda^3	& \lambda^4 \\
\lambda	& 1	& \lambda^2 &	\lambda^3 \\
\lambda^3&	\lambda^2 &	1	& \lambda \\
\lambda^4&	\lambda^3 &	\lambda	& 1
\end{array}
\right) \,,
\label{default1}
\end{align}
which yields a very symmetric scaling pattern for the 4G mixing matrix,
and shares the feature of the SM3 CKM matrix that 
more off-diagonal elements get smaller.
Notice that of the 9 inequalities in (\ref{consist}) involving
the 4G mixing angles, 3 are saturated, namely
\begin{align}
\mbox{(a)} & \quad
\theta_{12}\theta_{24} \sim \theta_{13}\theta_{34}  \sim \theta_{14} \,,
\quad
\theta_{24} \sim \theta_{23}\theta_{34} \,.
\end{align}
Scenarios with even smaller mixing angles 
($\theta_{i4} \to \lambda^n  \theta_{i4}$)
can simply be obtained from (\ref{default1})
by increasing $b_{1-3} \to b_{1-3}+n$.
An FN example for a scenario with (relatively) large 4G mixing angles
is given by
\begin{align}
\mbox{(b) \quad 
 $b_3=1$, $b_2=-1$, $b_1=-2$:}
& \quad \Rightarrow \quad 
V_\text{SM4} 
\sim \left( \begin{array}{cccc}
1	& \lambda	& \lambda^3	& \lambda^2 \\
\lambda	& 1	& \lambda^2             & \lambda \\
\lambda^3&	\lambda^2 &	1	& \lambda \\
\lambda^2&	\lambda &	\lambda	& 1
\end{array}
\right) \,, 
\label{default2}
\end{align}
In this case, the saturated inequalities are
\begin{align}  
\mbox{(b)} & \quad
\theta_{14}\theta_{34} \sim \theta_{13}\,,
\quad 
\theta_{24}\theta_{34} \sim \theta_{23}\,,
\quad
\theta_{12}\theta_{24} \sim \theta_{14} \,.
\end{align}
 
For later use, we further identify two  interesting non-FN scenarios.
The first one is given by
\begin{align}
\mbox{(c)} &\quad
V_\text{SM4} 
\sim \left( \begin{array}{cccc}
1	& \lambda	& \lambda^3	& \lambda^2 \\
\lambda	& 1	& \lambda^2             & \lambda^3 \\
\lambda^3&	\lambda^2 &	1	& \lambda \\
\lambda^2&	\lambda^3 &	\lambda	& 1
\end{array}
\right)  \quad \mbox{with} \ \
\theta_{14}\theta_{34} \sim \theta_{13} \,, \
\theta_{12}\theta_{14} \sim 
\theta_{23}\theta_{34} \sim \theta_{24} \,.
\label{default3}
\end{align}
In this case, the SM4 mixing matrix takes a very symmetric
form, where the mixing angle $\theta_{14}$
between the fourth and first generation is larger than $\theta_{24}$.
Finally, for
 \begin{align}
\mbox{(d)} &\quad
V_\text{SM4} 
\sim \left( \begin{array}{cccc}
1	& \lambda	& \lambda^3	& \lambda^3 \\
\lambda	& 1	& \lambda^2             & \lambda^2 \\
\lambda^3&	\lambda^2 &	1	& \lambda \\
\lambda^3&	\lambda^2 &	\lambda	& 1
\end{array}
\right)  \qquad \mbox{with} \quad
\theta_{12}\theta_{24} \sim \theta_{14} \,
\label{default4}
\end{align}
we encounter the situation that the mixing of the fourth 
and third generation with the first and second one is
similar in size, $\theta_{13} \sim \theta_{14}$, 
$\theta_{23} \sim \theta_{24}$.

Depending on which of the inequalities (\ref{consist}) are saturated,
the effects of the 4G on CP-violating observables
can be quite different. 
The different scaling of the 4G mixing angles will thus provide
a useful tool to classify different scenarios in the anatomy
for flavour observables, which will be further discussed in
 Section~\ref{Sec:anatomy_scenarios}.

\subsection{Parametrisation of \boldmath $V_\text{SM4}$ \unboldmath\label{sec:V4G-param}}

We will use a standard parametrisation of the SM4 mixing matrix
from \cite{Fritzsch:1986gv,Harari:1986xf}, which has also been used 
in \cite{Bobrowski:2009ng}. Defining 
\begin{align}
s_{ij} & = \sin\theta_{ij}\,, \qquad c_{ij}  = \cos\theta_{ij} \,,
\end{align}
we have for the generalised $4\times 4$ mixing matrix $V_\text{SM4}$:  
\begin{equation}   
\mbox{\footnotesize $\left( \begin{array}{cccc}   
\cez \ced \cev  & \ced \cev \sez & \cev \sed e^{-i\ded} & \sev e^{-i\dev} \\   
& & &\\ 
-\czd \czv \sez -\cez \czv \sed \szd e^{i\ded} &\cez \czd \czv -\czv \sez \sed \szd e^{i\ded} 
&     \ced \czv \szd & \cev \szv e^{-i\dzv} \\   
-\cez \ced \sev \szv e^{i(\dev-\dzv)} & -\ced \sez \sev \szv e^{i(\dev-\dzv)} &   
      -\sed \sev \szv e^{-i(\ded+\dzv-\dev)} &  \\   
& & & \\   
-\cez \czd \cdv\sed e^{i\ded} +\cdv \sez \szd & -\cez \cdv \szd - \czd \cdv \sez \sed e^{i\ded} &   
      \ced\czd\cdv & \cev \czv \sdv  \\   
-\cez \ced \czv \sev \sdv e^{i \dev} & -\cez \czd \szv\sdv e^{i\dzv} & -\ced\szd\szv\sdv e^{i\dzv} &   
 \\   
+\czd\sez\szv\sdv e^{i\dzv}& -\ced \czv \sez\sev\sdv e^{i\dev} & -\czv\sed\sev\sdv e^{i(\dev-\ded)} &   
      \\   
+\cez \sed \szd\szv\sdv e^{i(\ded+\dzv)} & +\sez \sed \szd \szv \sdv e^{i(\ded+\dzv)} & & \\   
& & &\\
-\cez \ced \czv \cdv \sev e^{i\dev } & -\cez \czd \cdv \szv e^{i\dzv}+\cez \szd \sdv &   
       -\ced \czd \sdv & \cev \czv \cdv\\   
+\cez \czd \sed \sdv e^{i \ded} & -\ced \czv \cdv \sez \sev e^{i \dev}&   
       -\ced \cdv \szd \szv e^{i\dzv} & \\   
+\czd \cdv \sez \szv e^{i\dzv}-\sez \szd \sdv & +\czd \sez \sed \sdv e^{i \ded} &   
       -\czv \cdv \sed \sev e^{i(\dev-\ded)} & \\   
+\cez \cdv \sed \szd \szv e^{i(\ded+\dzv)} & +\cdv \sez  \sed \szd \szv e^{i(\ded + \dzv)} & &   
\end{array} \right)\,. $}\label{eqn:v4g}
\end{equation} 
In the limiting case of vanishing mixing with the 4G quarks,
the standard parametrisation of the CKM matrix is recovered.
Note that in (\ref{eqn:v4g}), just as in the SM3 case, all angles $\theta_{ij}$ can be chosen to lie in the interval $[0,\pi/2]$. This can be shown in the following way \cite{Anselm:1985rw}: In the above parametrisation, $V_\text{SM4}$ is a product of six matrices $V_{ij}\in SU(2)\,,\, i,j=1,...,4$, consecutively mixing the different quark generations. 
The phases of each of these matrices can be factored out via
\begin{eqnarray}
V_{ij}&=&I_i(\alpha)\bar V_{ij}I_i(\beta)I_j(\gamma)=I_j(\alpha^\prime)\bar V_{ij}I_i(\beta^\prime)I_j(\gamma^\prime)\nonumber\\&=&I_i(\alpha^{\prime\prime})I_j(\beta^{\prime\prime})\bar V_{ij}I_i(\gamma^{\prime\prime})=I_i(\alpha^{\prime\prime\prime})I_j(\beta^{\prime\prime\prime})\bar V_{ij}I_j(\gamma^{\prime\prime\prime})\,,\label{eq:redundant_phases}
\end{eqnarray}
where
\begin{equation}
\left[I_i(\alpha)\right]_{jk}=\delta_{j,k}e^{i\alpha\delta_{i,j}}
\end{equation}
are phase operators and $\bar V_{ij}$ are $SO(2)$ rotation matrices with angles $\theta_{ij}$ in the interval $[0,\pi/2]$. 
Each of the four sets of angles in (\ref{eq:redundant_phases}) obeys one relation $\alpha+\beta+\gamma=0$ etc., such that each $V_{ij}$ has three parameters. Clearly, with $6\times3$ parameters $V_\text{SM4}$ must have 9 redundant phases. However, these unphysical phases will either cancel or be rotated away by quark field redefinitions. To see this, write $V_\text{SM4}$ as
\begin{equation}
V_\text{SM4}=V_{34}V_{24}V_{14}V_{23}V_{13}V_{12}\,,
\end{equation}
factor out the phases as done in (\ref{eq:redundant_phases}), and re-arrange the phase operators in that expression, repeatedly using $[I_i(\alpha),I_j(\beta)]=0$ and (\ref{eq:redundant_phases}). In doing so, all but three of those operators can be moved to the extreme left or right of $V_\text{SM4}$, where they can be absorbed into phase redefinitions of the quark fields, such that one ends up with the standard parametrisation
\begin{equation}
V_\text{SM4}=\bar V_{34}I_4(\delta_{24})\bar V_{24}I_4(-\delta_{24})I_4(\delta_{14})\bar V_{14}I_4(-\delta_{14})\bar V_{23}I_3(\delta_{13})\bar V_{13}I_3(-\delta_{13})\bar V_{12}\,,
\end{equation}
with all angles $\theta_{ij}$ in $[0,\pi/2]$ and phases $\delta_{ij}$ in $[0,2\pi]$.

While it is advisable to use this exact parametrisation in numerical
calculations, its structure turns out to be much simpler than its
appearance in (\ref{eqn:v4g}). Indeed, from the tree level measurements 
of $\vud$, $\vus$, $\vub$, $\vcd$, $\vcs$ and the unitarity of $V_\text{SM4}$,
we know already
that $c_{ij}\approx 1$ and that $s_{14}$ and $s_{24}$ are not larger than 
$s_{12}$ \cite{Kribs:2007nz}. On the other hand, from the global fits
of the precision electroweak data one finds $s_{34}\le 0.35$
\cite{Alwall:2006bx,Chanowitz:2009mz}. Our refined analysis of the implications of tree
level measurements and of $V_\text{SM4}$ unitarity in Section~\ref{sec:numerics} combined 
with the results of \cite{Chanowitz:2009mz}  leads to the upper bounds
\update{
\begin{equation}
s_{14}\le 0.04\,, \qquad s_{24}\le 0.17\,, \qquad s_{34}\le \frac{M_W}{m_{t'}} \le0.27 \,,
\label{eq:sij_bounds}
\end{equation}
}
that we will adopt in our paper.
\update{This particularly implies a decrease of the upper bound on $s_{34}$ with increasing $m_{t'}$
\cite{Chanowitz:2009mz}.}
 The \update{maximal} value of $s_{34}$ in (\ref{eq:sij_bounds}) corresponds to $m_{t^\prime}=300\gev$.
\update{In order to be consistent with EWPT, one should choose the $b^\prime$ mass to be \cite{Kribs:2007nz} 
\begin{align}
m_{b^\prime} &\approx m_{t^\prime} - 55\gev\,.
\end{align}
However, $m_{b'}$ will not enter $B$ and $K$ observables and therefore is not crucial for our analysis.}

In the limit $c_{ij}\approx 1$, as indicated above, the six mixing angles
are directly determined by the moduli of the off-diagonal elements 
in the upper right corner of $V_\text{SM4}$:
\begin{align}
s_{12} &\simeq |V_{us}|\,, & s_{13} &\simeq |V_{ub}| \,, & s_{23} &\simeq |V_{cb}| \,, \cr
s_{14} &\simeq |V_{ub^\prime}| \,, & s_{24} &\simeq |V_{cb^\prime}| \,, & s_{34} &\simeq |V_{tb^\prime}| \,. \label{eqn:relvs}
\end{align}
For a given scaling of the 4G mixing angles --- like in our
benchmark scenarios (\ref{default1})--(\ref{default4}) ---  {the subleading
terms of $V_{\rm SM4}$ can then be characterised in terms of a generalised
Wolfenstein expansion}  \cite{Wolfenstein:1983yz,Buras:1994ec}.
\begin{eqnarray}
&& \lambda \equiv s_{12}\,, \quad 
 s_{23} \equiv A \lambda^2 \,, \quad
 s_{13} e^{i\delta_{13}}  \equiv A \lambda^3 (\rho+i\eta) \equiv A \lambda^3 \, z_\rho
\end{eqnarray}
for the SM3 parameters, 
and 
\begin{eqnarray}
&&  s_{14} e^{i\delta_{14}} = \lambda^{n_1} z_\tau \,, \quad
 s_{24} e^{i\delta_{24}} = \lambda^{n_2} z_\sigma \,, \quad
 s_{34} =  \lambda^{n_3} B \,,
\label{eq:wolf}
\end{eqnarray}
where $A$, $B$, $z_i$, $z_i^*$ are coefficients of order one, 
and the exponents $n_i$ depend on the scaling of the 4G mixing angles.
\begin{itemize}

\item For instance, in the case (\ref{default1})
 we obtain the expansion of $V_\text{SM4}$ as
\begin{align}
\begin{split}
&\mbox{\footnotesize $
\left(
\begin{array}{cccc}
 1-\frac{\lambda^2}{2} 
 & \lambda  
 & \lambda^3 A z_\rho^* 
 & 0 
\\
 -\lambda  
 & 1-\frac{\lambda ^2}{2} 
 & \lambda^2 A 
 & \lambda^3 z_\sigma^* 
\\
 \lambda^3 A (1-z_\rho) 
 & -\lambda^2 A 
 & 1-\frac{\lambda^2}{2} B^2 
 & \lambda B  
\\
 0 
 & \lambda^3 \left(AB-z_\sigma \right) 
 & -\lambda B  
 & 1-\frac{\lambda^2}{2} B^2
\end{array}
\right)
$}\\
&\mbox{\footnotesize $+\lambda^4
\left(
\begin{array}{cccc}
 -\frac{1}{8} 
 & 0 
 & 0 
 &  z_\tau^* 
\\
 0 
 & -\frac{1}{8}\left(1+4 A^2\right) 
 & 0 
 & 0 
\\
 0 
 & \left(\frac{1}{2} A (1+ B^2- 2 z_\rho)- B z_\sigma \right) 
 & -\frac{1}{8} \left(4 A^2+B^4\right) 
 & 0 
\\
  \left(AB (z_\rho -1)+ z_\sigma -z_\tau \right) 
 & 0 
 & 0 
 &-\frac{B^4}{8}
\end{array}
\right)
$}+ {\mathcal O}(\lambda^5) \,.
\end{split}\label{eqn:v4gapprox431}
\end{align}
Here, we included terms up to order $\lambda^4$, where $V_{ts}$ receives
an additional imaginary part from the 4G parameter $z_\sigma$, while the other
off-diagonal elements of the effective $3\times 3$ CKM sub-matrix 
keep their SM3 expansion at this order.

\item On the other hand, in the case of (\ref{default2}), we obtain the expansion
\begin{align}
\begin{split}
&\mbox{\footnotesize $
\left(
\begin{array}{cccc}
 1-\frac{\lambda ^2}{2} 
 & \lambda  
 & \lambda^3 A z_\rho^* 
 &  \lambda^2 z_\tau^* 
\\
 -\lambda  
 & 1-\frac{\lambda ^2}{2}  \left(1+|z_\sigma|^2\right) 
 & \lambda^2 A 
 & \lambda z_\sigma^*
\\
 \lambda^3 A (1-z_\rho) 
 & - \lambda^2 \left(A+B z_\sigma \right) 
 & 1-\frac{\lambda ^2}{2} B^2 
 &  \lambda B 
 \\
  \lambda ^2 \left(-z_\tau + z_\sigma \right) 
 & - \lambda z_\sigma 
 & - \lambda B  
 & 1-\frac{ \lambda^2}{2}\left(|z_\sigma|^2+B^2\right)
\end{array}
\right)
$ }\\
&\mbox{\footnotesize $
+\lambda^3 \left(
\begin{array}{cccc}
 0 & 0 &  0 & 0 \\
 \frac{1}{2} \left(-2 z_\tau z_\sigma^*+|z_\sigma|^2\right) & 0 & 0 & 0 \\
 B (-z_\tau + z_\sigma) & 0 & 0 &-\frac{1}{2}  B |z_\sigma|^2 \\
 0 & \frac{1}{2} \left( z_\sigma(1+B^2) +2 A B - 2 z_\tau \right) & -A z_\sigma & 0
\end{array}
\right)
$}+{\mathcal O}(\lambda^4)\end{split}
\label{eqn:v4gapprox211}
\end{align}
This reveals that, in this case,
$V_{ts}$ and $V_{td}$ receive order-one CP-violating phases 
from $z_\sigma$ and $(z_\sigma-z_\tau)$, respectively.
The associated CP-violating observables will thus strongly
constrain the magnitude of $\delta_{14}$ and $\delta_{24}$ in
a correlated way, which we will verify numerically in Section~\ref{Sec:anatomy_scenarios}.

\item  Similarly, for (\ref{default3}), the generalised Wolfenstein expansion reads
\begin{align}
\begin{split}
&\mbox{\footnotesize $
\left(
\begin{array}{cccc}
 1-\frac{\lambda ^2}{2} 
 & \lambda  
 & \lambda^3 A z_\rho^* 
 &  \lambda^2 z_\tau^* 
\\
 -\lambda  
 & 1-\frac{\lambda ^2}{2}  
 & \lambda^2 A 
 & \lambda^3 z_\sigma^*
\\
 \lambda^3\left( A (1-z_\rho) -B z_\tau \right) 
 & -\lambda^2 A 
 & 1-\frac{\lambda ^2}{2} B^2 
 & \lambda B 
 \\
 - \lambda ^2 z_\tau 
 & \lambda^3\left(AB-z_\tau-z_\sigma\right) 
 & -\lambda B  
 & 1-\frac{ \lambda ^2 }{2} B^2
\end{array}
\right)
$ }
+{\mathcal O}(\lambda^4)\end{split}
\label{eqn:v4gapprox231}
\end{align}
In this case, $V_{td}$ receives a new phase from $z_\tau$ at leading order.

\item  {Finally, for (\ref{default4}), the SM4 mixing matrix is approximated as
\begin{align}
\begin{split}
&\mbox{\footnotesize $
\left(
\begin{array}{cccc}
 1-\frac{\lambda ^2}{2} 
 & \lambda  
 & \lambda^3 A z_\rho^* 
 &  \lambda^3 z_\tau^* 
\\
 -\lambda  
 & 1-\frac{\lambda ^2}{2}  
 & \lambda^2 A 
 & \lambda^2 z_\sigma^*
\\
 \lambda^3  A (1-z_\rho) 
 & -\lambda^2 \left( A + B \lambda z_\sigma \right) 
 & 1-\frac{\lambda ^2}{2} B^2 
 & \lambda B 
 \\
 \lambda ^3 (z_\sigma-z_\tau) 
 & \lambda^2 \left(\lambda AB-z_\sigma\right) 
 & -\lambda B  
 & 1-\frac{ \lambda ^2 }{2} B^2
\end{array}
\right)
$ }
+{\mathcal O}(\lambda^4)\end{split}
\label{eqn:v4gapprox321}
\end{align}
In this case, the new contributions to the phases of
$V_{td}$ and $V_{ts}$, compared to the SM3 parametrisation,
are suppressed by an additional power of the Wolfenstein parameter $\lambda$.}
\end{itemize}
While approximations like (\ref{eqn:v4gapprox431}--\ref{eqn:v4gapprox321}) 
will not be used in our actual
numerical calculations, they reveal the different potential for new CP-violating effects as discussed below.

\update{Finally, it should be remarked that other parameterisations of $V_{\rm SM4}$ can be found in the literature. 
One of them is the parameterisation in \cite{Hou:1987hm} in which the 4th row, i.e. $V_{t'd}$, $V_{t's}$ and $V_{t'b}$, 
are identified with the 3 new rotation angles and two new phases. While such a choice is certainly valid, we follow other recent studies and
use the parameterisation presented above, which has the benefit of keeping the first row simple.
}

\subsection{Unitarity}

In writing the formulae for the observables of interest,
it is useful to use the unitarity of the matrix $V_\text{SM4}$. To this
end, we define the factors ($i=u,c,t,t^\prime$)
\begin{align}
\lambda^{(K)}_i &= V^{*}_{is}V_{id}\,,& \lambda^{(d)}_{i} &= V^{*}_{ib}V_{id}\,, &\lambda^{(s)}_{i} &=V^*_{ib}V_{is}\,. \label{eqn:4gunitaritydef}
\end{align}
 The unitarity relations  are then written as
\begin{align}
\lambda^{(K)}_{u}+\lambda^{(K)}_{c}+\lambda^{(K)}_{t}+\lambda^{(K)}_{t^\prime}&=0 \,, \label{eqn:4gunitarity}
\end{align}
with analogous  expressions for the $B_d$ and $B_s$~system. Relation (\ref{eqn:4gunitarity}) allows to eliminate $\lambda^{(K)}_u$
so that only $\lambda^{(K)}_{c}$, $\lambda^{(K)}_{t}$ and $\lambda^{(K)}_{t^\prime}$ enter the final expressions.
We summarise the
scaling of $\lambda^{K,d,s}_i$ compared to the SM3 expressions 
for our four benchmark scenarios in Table~\ref{tab:phase_scale}.
As a general feature, we observe the approximate relations
between the 4G contributions  to the flavour coefficients
with $t$ and $t'$, 
\begin{align}
 {\rm Im}\left[\lambda_t^{(d)}/\lambda_t^{(d)}|_{\rm SM} \right]
 &\simeq - {\rm Im}\left[\lambda_{t'}^{(d)}/\lambda_t^{(d)}|_{\rm SM}\right] \,,
\\ 
 {\rm Im}\left[\lambda_t^{(s)}/\lambda_t^{(s)}|_{\rm SM} \right]
 & \simeq - {\rm Im}\left[\lambda_{t'}^{(s)}/\lambda_t^{(s)}|_{\rm SM} \right] \,.
\end{align}
Furthermore, the 4G effects on the light-quark coefficients $\lambda_c^{(d,s,K)}$
are always strongly suppressed.

\begin{table}[t!!!bp]
 \begin{center}
\begin{tabular}{|l||c|c|c|c|}
\hline
Scenario $n_1 n_2 n_3$ 
 & (a) 431 &  (b) 211 & (c) 231 & (d) 321
\\
\hline \hline
${\rm Im}\left[\lambda_t^{(d)}/\lambda_t^{(d)}|_{\rm SM} \right]
\simeq - {\rm Im}\left[\lambda_{t'}^{(d)}/\lambda_t^{(d)}|_{\rm SM}\right]$ 
 & ${\cal O}(\lambda^2)$ 
 & ${\cal O}(1)$
 & ${\cal O}(1)$
 & ${\cal O}(\lambda)$ 
\\
${\rm Im}\left[\lambda_t^{(s)}/\lambda_t^{(s)}|_{\rm SM} \right]
\simeq - {\rm Im}\left[\lambda_{t'}^{(s)}/\lambda_t^{(s)}|_{\rm SM} \right]$ 
 & ${\cal O}(\lambda^2)$ 
 & ${\cal O}(1)$ 
 & ${\cal O}(\lambda^2)$ 
 & ${\cal O}(\lambda)$ 
\\
${\rm Im}\left[\lambda_t^{(K)}/\lambda_t^{(K)}|_{\rm SM} \right]$
& ${\cal O}(\lambda^2)$ 
 & ${\cal O}(1)$ 
& ${\cal O}(1)$ 
& ${\cal O}(\lambda)$ 
\\
\hline 
${\rm Im}\left[\lambda_{t'}^{(K)}/\lambda_t^{(K)}|_{\rm SM} \right]$
& ${\cal O}(\lambda^2)$ 
 & ${\cal O}(\lambda^{-2})$ 
 & ${\cal O}(1)$ 
 & ${\cal O}(1)$ 
\\
\hline 
${\rm Im}\left[\lambda_c^{(d)}/\lambda_c^{(d)}|_{\rm SM}\right]
\simeq \, {\rm Im}\left[\lambda_c^{(K)}/\lambda_c^{(K)}|_{\rm SM} \right]$ 
 & ${\cal O}(\lambda^6)$ 
 & ${\cal O}(\lambda^2)$ 
 & ${\cal O}(\lambda^4)$ 
 & ${\cal O}(\lambda^4)$ 
\\
${\rm Im}\left[\lambda_c^{(s)}/\lambda_c^{(s)}|_{\rm SM} \right]$ 
 & ${\cal O}(\lambda^8)$ 
 & ${\cal O}(\lambda^4)$ 
 & ${\cal O}(\lambda^6)$ 
 & ${\cal O}(\lambda^6)$ 
\\
\hline
\end{tabular}
 \end{center}
\caption{\label{tab:phase_scale}
Scaling of new 4G-induced flavour coefficients in CP-violating observables
for different benchmark scenarios.
}
\end{table}

Depending on the respective scenario, this has a different impact on
CP-violating observables, for instance for the
mixing-induced CP asymmetries $S_{\psi K_S}$ and $S_{\psi\phi}$:
\begin{itemize}

\item[(a)] For the scenario (\ref{default1}), 
the 4G contributions in Table~\ref{tab:phase_scale} are
suppressed by at least order $\lambda^2$.
Therefore, the mixing-induced asymmetry $S_{\psi K_S}$ 
will be dominated by the argument of $\lambda_t^{(d)}$, and similarly 
$S_{\psi\phi}$ will be dominated by $\text{arg}(\lambda_t^{(s)})$, allowing
 only for small deviations compared to the SM3 analysis.

\item[(b)] The situation is completely different in the case (\ref{default2}),
where the modifications from 4G to $\lambda_{t,t'}^{(d,s)}$ and $\lambda_t^{(K)}$
are generically of ${\cal O}(1)$, and to $\lambda_{t'}^{(K)}$ even of order $\lambda^{-2}$.

For $\lambda_{t'}^{(s)}/\lambda_t^{(s)} \sim {\cal O}(1)$, 
one might be worried about
sizable corrections to $S_{\psi K_S}$ from $t^\prime$-penguin pollution, because, 
in contrast to the SM3, the interference effects from weak phases in the $b\to 
s$ transition are no longer Cabibbo-suppressed. 
In spite of the fact that the new contributions are GIM-suppressed,
without further  hadronic input, we cannot exclude corrections  as large as 10\%\ to $S_{\psi K_S}$.
In this case, the experimental bound from  $S_{\psi K_S}$ will be a bit softer.

The constraints
from $S_{\psi K_S}$ and $\epsilon_K$ then favour solutions of
\begin{equation} \label{eq:lttprimeformerfootnote}
 {\rm Im}\left[\lambda_{t,t'}^{(d)}/\lambda_t^{(d)}|_{\rm SM} \right]
   \approx \lambda^2 {\rm Im}\left[\lambda_{t'}^{(K)}/\lambda_t^{(K)}|_{\rm SM} \right] \approx 0 \,,
\end{equation}
which translates into
strong correlations for the 4G mixing parameters, implying
\begin{align}
 ({\rm b}) \quad
  \Rightarrow \quad & s_{14} \approx s_{12} s_{24} \,, \qquad \delta_{14} \approx \delta_{24} \,,
\cr 
   & \left| {\rm Im}\left[\lambda_{t,t'}^{(s)}/\lambda_t^{(s)}|_{\rm SM} \right]\right|
   \approx \left|{\rm Im}\left[\lambda_{t}^{(K)}/\lambda_t^{(K)}|_{\rm SM} \right]\right|
   \approx \frac{s_{24} s_{34} \left|\sin\delta_{14}\right|}{s_{23}} \,.
\end{align}
The last relation will lead to strong correlations between $b \to s$ and 
rare kaon decay observables.

\item[(c)] For (\ref{default3}), only the phase of $V_{td}$ is affected by leading-order
4G modifications, and therefore the $b \to s$ observables are not very much affected,
while the modifications in $b \to d$ and $s \to d$\/ CP observables are constraining the
two independent quantities 
\begin{align}
  &  {\rm Im}\left[\lambda_{t'}^{(K)}/\lambda_t^{(K)}|_{\rm SM} \right]
\quad \mbox{and} \quad
{\rm Im}\left[\lambda_{t,t'}^{(d)}/\lambda_t^{(d)}|_{\rm SM} \right]
   \approx  {\rm Im}\left[\lambda_{t}^{(K)}/\lambda_t^{(K)}|_{\rm SM} \right]
\end{align}

\item[(d)] Finally, in the case (\ref{default4}), 
the modifications from 4G to $\lambda_{t,t'}^{(d,s)}$ and $\lambda_t^{(K)}$
are of ${\cal O}(\lambda)$, and to $\lambda_{t'}^{(K)}$ of ${\cal O}(1)$.
On the one hand --- compared to (b) ---
this leads to somewhat relaxed correlations between the 4G
mixing parameters. On the other hand, it leaves room for 
moderate deviations from SM3 predictions.
\end{itemize}
These general type of observations will be verified numerically in Section~\ref{Sec:anatomy_scenarios}.

\subsection{CP-violating Invariants}

The amount of CP violation in the quark Yukawa sector -- which is relevant for
the discussion of the baryon-antibaryon asymmetry \cite{Hou:2008xd} --
can be quantified by studying appropriate basis-independent 
invariants built from the Yukawa matrices. 
The generalisation of the well-known Jarlskog determinant 
\cite{Jarlskog:1985ht,Jarlskog:1985cw}
to the SM4 case has, for instance, been discussed in  \cite{delAguila:1996pa}.
In the limit $m_{u,d,s,c}^2 \ll m_b^2 \ll m_{t,b',t'}^2$, and restricting
ourselves to scenarios that fulfil the inequalities (\ref{consist}), the invariants 
reduce to one new CP-violating quantity,
\begin{eqnarray}
I_1&=& {\rm Im} \, {\rm tr} \left[(Y_U Y_U^\dagger)^2 (Y_D Y_D^\dagger) (Y_U Y_U^\dagger)
 (Y_D Y_D^\dagger)^2\right] 
\nonumber\\
&\simeq& - m_b^2 m_t^2 m_{b'}^4 m_{t'}^2 (m_{t'}^2-m_t^2) 
\left( F_{2323} + F_{1313} + F_{2313} + F_{1323} \right)
\nonumber\\[0.2em] 
& \simeq & m_b^2 m_t^2 m_{b'}^4 m_{t'}^2 (m_{t'}^2-m_t^2) \left\{
\begin{array}{ll}
{} - s_{23} s_{24} s_{34} \sin \delta_{24} \,,
\\
 s_{34}
\left( s_{13} s_{14} \sin (\delta_{13}-\delta_{14})-s_{23} s_{24} \sin\delta_{24} \right) \,,
\end{array}
\right.
\label{invariant}
\cr &&
\end{eqnarray}
which can be related to the area of a quadrangle in the complex plane,
described by the functions
$$
F_{ijkl} = {\rm Im}\left( V_{ik}V_{jk}^* V_{il}^* V_{jl} \right) \,.
$$
The first line in (\ref{invariant}) 
refers to scenarios with $s_{24} \gtrsim s_{14}$,
like (\ref{default1},\ref{default2},\ref{default4}). The
second line is valid for $s_{24} \sim \lambda \, s_{14}$,
like in case of (\ref{default3}). It has been stressed in \cite{Hou:2008xd} that the quark mass-dependent prefactor in (\ref{invariant}) 
can lead to an enhancement of several orders of magnitude compared to the SM3 analogue.

The overall scaling with the Wolfenstein parameter $\lambda$
is given by $I_1 \sim \lambda^{2+n_2+n_3}$, which can be as large as $\lambda^4$
for scenario (\ref{default2}). For the benchmark scenario 
(\ref{default4}) one obtains $I_1 \sim \lambda^5$, whereas
(\ref{default1},\ref{default3}) would lead to $I_1 \sim \lambda^6$.
As we will illustrate in the numerical section below, the dependence of $I_1$
on the SM4 mixing parameters is directly correlated with the size of $S_{\psi\phi}$.
%


\newsection{Effective Hamiltonians in the SM4}\label{sec:EffHam}
\subsection{Effective Hamiltonians for $\Delta F = 2$ Transitions}\label{sec:EffHamDF2}

The effective Hamiltonian for $\Delta S=2$ transitions can be written as
\begin{eqnarray}\label{A1}
 {\cal H}_{\rm eff}^{\Delta S=2} &=& \frac{G_F^2}{16 \pi^2} M_W^2 
\bigg[ 
\lambda_c^{2(K)} \eta_{cc}^{(K)} S_0(x_c) + 
\lambda_t^{2(K)} \eta_{tt}^{(K)} S_0(x_t) +  
\lambda_{t^\prime}^{2(K)} \eta_{t^\prime t^\prime}^{(K)} S_0(x_{t^\prime}) 
 \nonumber \\
 && \qquad
+ 2  \lambda_t^{(K)}\lambda_c^{(K)} \eta_{ct}^{(K)}  S_0(x_t,x_c) 
+ 2 \lambda_{t^\prime}^{(K)}\lambda_c^{(K)} \eta_{ct^\prime}^{(K)} S_0(x_{t^\prime},x_c)  
\cr 
&& \qquad 
+ 2 \lambda_t^{(K)}\lambda_{t^\prime}^{(K)} \eta_{tt^\prime}^{(K)} S_0(x_{t^\prime},x_t)
\bigg] \times 
\nonumber \\
&& \quad
\times \left(\alpha_s^{(3)}(\mu) \right)^{-2/9} \left[1+\frac{\alpha_s^{(3)}(\mu)}{4 \pi} J_3 \right]\ Q(\Delta S = 2)\,,
\end{eqnarray}
where the functions $\eta_{ij}^{(K)}$ and explicit factors of $\alpha_s$ 
arise from QCD corrections.
The operator $Q(\Delta S=2)$ is defined in (\ref{eqn:ope}) below.
Absorbing the contributions of the 4G quarks into a redefinition
of the loop function $S_0(x_t)$, we can bring this Hamiltonian into a
form as in the SM3,
\begin{eqnarray}\label{A2}
{\cal H}_{\rm eff}^{\Delta S=2} &=& \frac{G_F^2}{16 \pi^2} M_W^2 \bigg[ \lambda_c^{2(K)} \eta_{cc}^{(K)} S_0(x_c) +  \lambda_t^{2(K)} \eta_{tt}^{(K)} S_K +   
2 \lambda_t^{(K)}\lambda_c^{(K)} \eta_{ct}^{(K)} S_0(x_t,x_c) \bigg] \times \nonumber \\
&& \times \left(\alpha_s(\mu) \right)^{-2/9} \left[1+\frac{\alpha_s(\mu)}{4 \pi} J_3 \right]\ Q(\Delta S = 2)\,,
\end{eqnarray}
where
\begin{eqnarray}\label{A3}
S_K = S_0(x_t) &+& \frac{\eta_{t^\prime t^\prime}^{(K)}}{\eta_{tt}^{(K)}} \left(\frac{\lambda_{t^\prime}^{(K)}}{\lambda_{t}^{(K)}} \right)^2 S_0(x_{t^\prime}) + 2 \, \frac{\eta_{tt^\prime}^{(K)}}{\eta_{tt}^{(K)}} \left(\frac{\lambda_{t^\prime}^{(K)}}{\lambda_{t}^{(K)}} \right) S_0(x_t,x_{t^\prime})\nonumber\\&+&2\,\frac{\eta_{ct^\prime}^{(K)}}{\eta_{tt}^{(K)}}\left(\frac{\lambda_c^{(K)}\lambda_{t^\prime}^{(K)}}{\lambda_t^{(K)2}}\right)S_0(x_{t^\prime},x_c)\,.
\end{eqnarray}
The standard loop functions $S_0(x)$ and $S_0(x_i,x_j)$ are given in Appendix~A.

In principle, (\ref{A2}) can be directly generalised to the $B_d,B_s$
systems. In practise, only the analogue
 of the $S_K$ term is relevant, and consequently ($q=d,s$)
\begin{equation}
{\cal H}_{\rm eff}^{(q)}=\frac{G_F^2}{16\pi^2}M_W^2 \eta_B \left(\lambda^{(q)}_t
\right)^2 S_q\left[\alpha_s^{(5)}\right]^{-6/23}\left[1
+\frac{\alpha_s^{(5)}}{4\pi} J_5\right]Q^q(\Delta B=2)\,,
\end{equation}
\begin{eqnarray}\label{Sq}
S_q=S_0(x_t)&+&\frac{\eta^{(q)}_{t^\prime t^\prime}}{\eta_{tt}^{(q)}}\left(\frac{\lambda^{(q)}_{t'}}{\lambda^{(q)}_t}\right)^2S_0(x_{t^\prime})+2\,\frac{\eta^{(q)}_{tt^\prime}}{\eta^{(q)}_{tt}}\left(\frac{\lambda^{(q)}_{t^\prime}}{\lambda^{(q)}_t} \right)S_0(x_t,x_{t^\prime})\nonumber\\&+&2\,\frac{\eta_{ct^\prime}^{(q)}}{\eta_{tt}^{(q)}}\left(\frac{\lambda_c^{(q)}\lambda_{t^\prime}^{(q)}}{\lambda_t^{(q)2}}\right)S_0(x_{t^\prime},x_c)\,.
\end{eqnarray}
In contrast to the last term in (\ref{A3}), that can be relevant for certain 
values of the parameters involved, the last term in (\ref{Sq}) turns out
to be negligibly small.

\boldmath\subsection{Effective Hamiltonians for $K_L\to\mu^+\mu^-$ and \mbox{$B_{d,s}\to\mu^+\mu^-$}}\unboldmath

As for the rare leptonic decays, we first consider $K_L\to\mu^+\mu^-$,
where we also have to address the charm contribution. 
For pedagogical reasons, we will neglect QCD corrections
for a moment (they will be discussed subsequently).
The effective Hamiltonian for  the SD part of 
$K_L\to\mu^+\mu^-$ is given in the SM3 as follows \cite{Buchalla:1995vs},
\begin{equation}
\mathcal H_\text{eff}=-\frac{G_F}{\sqrt{2}}
\, \frac{\alpha}{2\pi\sin^2\theta_W}\left(\lambda_c^{(K)}Y_0(x_c)+\lambda_t^{(K)}Y_0(x_t)\right)(\bar sd)_{V-A}(\bar\mu\mu)_{V-A}\,,\label{eq:Heff_KLmumu}
\end{equation}
with the function $Y_0(x)$ given in Appendix~A.
This Hamiltonian can be generalised in a straightforward manner by
\begin{itemize}
 \item including the effects of the heavy $t^\prime$ and 
the heavy 4G neutrino,
 with the latter exchanged only in the box diagram and 
 the former in both, the $Z^0$ penguin and box diagrams;
 \item including the mixing in the lepton sector,
  parametrised by a $4\times4$ matrix with the elements denoted by $W_{ij}$. 
  Here, the first index applies to neutrinos, the second to charged leptons.
\end{itemize}
Neglecting the masses of the lightest neutrinos and of the up-quark,
 and using the unitarity of the SM4 quark and lepton  mixing matrices,
 we find -- not unexpectedly -- 
that the final result can be written in terms of the function $Y_0$ 
and the box function
\begin{eqnarray}
F^{\mu\bar\mu}(x_i,y_4)&=&B^{\mu\bar\mu}(x_i,0)+B^{\mu\bar\mu}(0,y_4)-B^{\mu\bar\mu}(0,0)-B^{\mu\bar\mu}(x_i,y_4)\nonumber\\&=&-S_0(x_i,y_4)\,,
\end{eqnarray}
where
\begin{equation}
x_i=\frac{m_i^2}{M_W^2}\,,\quad y_4=\frac{\left(m_4^{\nu}\right)^2}{M_W^2}\,.
\end{equation}
The functions $B^{\mu\bar\mu}$ and $S_0$ are given in Appendix \ref{app:functions}.

The generalisation of (\ref{eq:Heff_KLmumu}) to include the 4G quarks
and the mixing in the 4G lepton sector is then given as follows:
\begin{eqnarray}
\lambda_c^{(K)}Y_0(x_c)&\to&\lambda_c^{(K)}\left(Y_0(x_c)+\left|W_{4\mu}\right|^2F^{\mu\bar\mu}(x_c,y_4)\right)\label{eq:lambdacY0}\,,\\
\lambda_t^{(K)}Y_0(x_t)&\to&\lambda_t^{(K)}\left(Y_0(x_t)+\left|W_{4\mu}\right|^2F^{\mu\bar\mu}(x_t,y_4)\right)\nonumber\\
&+&\lambda_{t^\prime}^{(K)}\left(Y_0(x_{t^\prime})+\left|W_{4\mu}\right|^2F^{\mu\bar\mu}(x_{t^\prime},y_4)\right)\,.\label{eq:lambdatY0}
\end{eqnarray}

Concerning QCD corrections, they have been found 
to be very small in the case of the top quark part within the SM3 \cite{Buchalla:1998ba,Misiak:1999yg}, 
and we will neglect them
for the $t$ and $t^\prime$ contributions  in the SM4.
In the case of the charm part, QCD
corrections are significant and have to be 
included \cite{Buchalla:1998ba,Gorbahn:2006bm}. 
As seen in (\ref{eq:lambdacY0}) in the absence of mixing with the heavy 
new charged lepton, the charm contribution remains intact including 
QCD corrections. If significant 
mixing is present, the charm contribution is modified and also the corresponding 
QCD corrections should be reconsidered.  Although this would be straightforward, 
we do not expect these effects to be important, 
in particular in comparison 
with the 4G effects in the top sector. Therefore we postpone their inclusion until
after the discovery of the 4G.

Clearly the terms involving $\left|W_{4\mu}\right|^2$ are only relevant if
$\left|W_{4\mu}\right|^2$ is substantial. We will comment on the size of these
terms in the next section.

The generalisation to $B_{d,s}\to\mu^+\mu^-$ is straightforward. First we can drop the charm contribution, as it is negligibly small relative to the $t$ and $t^\prime$ contributions. Second, $\lambda_t^{(K)}$ and $\lambda_{t^\prime}^{(K)}$ should be replaced by $\lambda_t^{(d,s)}$ and $\lambda_{t^\prime}^{(d,s)}$, respectively.

\boldmath\subsection{Effective Hamiltonian for $K\to\pi\nu\bar\nu$ and \mbox{$B\to X_{d,s}\nu\bar\nu$}}\unboldmath

We begin with $K^+\to\pi^+\nu\bar\nu$. The effective Hamiltonian for this
decay within the SM3, neglecting QCD corrections, is given by 
\cite{Buchalla:1995vs}
\begin{equation}
\mathcal H_\text{eff}=\frac{G_F}{\sqrt{2}}\frac{\alpha}{2\pi\sin^2\theta_W}\sum\limits_{\ell=e,\mu,\tau}\left(\lambda_c^{(K)}X_0(x_c,z_\ell)+\lambda_t^{(K)}X_0(x_t)\right)(\bar sd)_{V-A}(\bar\nu_\ell\nu_\ell)_{V-A}\,,\label{eq:Heff_Kpnunu}
\end{equation}
where
\begin{equation}
 X_0(x_c,0)=X_0(x_c)\,,\quad z_e=z_\mu\simeq0\,,\quad z_\tau=\frac{m_\tau^2}{M_W^2}\,.
\end{equation}
The $\tau$-lepton mass dependence in the top-contribution can be neglected, but not in the case of the charm contribution.
We want to generalise this Hamiltonian to include the effects of the heavy $t^\prime$ and the heavy charged lepton, as well as the effect of the mixing in the lepton sector described by the $W_{ij}$ elements of the $4\times4$ lepton-mixing matrix.

A small complication, relative to the $\mu^+\mu^-$ case,  arises from
the summation over three light neutrino species that --- being mass eigenstates -- 
will be denoted by $\nu_1$, $\nu_2$, $\nu_3$,
with $\nu_4$ denoting the heavy neutrino which is exchanged in the box diagram 
for the $K_L\to\mu^+\mu^-$ decay.
Here the 4G heavy charged lepton  that enters the box diagrams in $K^+\to\pi^+\nu\bar\nu$ will be {represented} by $z_4=m_4^{\ell2}/M_W^2$.
The final result can be written in terms of the function $X_0$ and the box function
\begin{equation}
 F^{\nu\bar\nu}(x_i,z_4)\equiv B^{\nu\bar\nu}(x_i,0)+B^{\nu\bar\nu}(0,z_4)-B^{\nu\bar\nu}(0,0)-B^{\nu\bar\nu}(x_i,z_4)\,,
\end{equation}
with  $X_0$ and $B^{\nu\bar\nu}$ given in Appendix \ref{app:functions}.

The generalisation of (\ref{eq:Heff_Kpnunu}) to include the 4G quarks and the mixing in the 4G lepton sector is given for the $\ell$-th neutrino $\nu_\ell$  in the final state as follows
\begin{eqnarray}
\lambda_c^{(K)}X_0(x_c,z_\ell)&\to&\lambda_c^{(K)}\left(X_0(x_c)+4\left|W_{\ell\tau}\right|^2F^{\nu\bar\nu}(x_c,z_\tau)+4\left|W_{\ell4}\right|^2F^{\nu\bar\nu}(x_c,z_4)\right)\,,\\\label{eq:lambdacX0}
\lambda_t^{(K)}X_0(x_t)&\to&\lambda_t^{(K)}\left(X_0(x_t)+4\left|W_{\ell4}\right|^2F^{\nu\bar\nu}(x_t,z_4)\right)\nonumber\\&+&\lambda_{t^\prime}^{(K)}\left(X_0(x_{t^\prime})+4\left|W_{\ell4}\right|^2F^{\nu\bar\nu}(x_{t^\prime},z_4)\right)\,.\label{eq:lambdatX0}
\end{eqnarray}
{Concerning QCD corrections the same comments apply as after (\ref{eq:lambdatY0}).}

The generalisation to $K_L\to\pi^0\nu\bar\nu$ and $B\to X_{d,s} \nu \bar\nu$ is straightforward: the charm contribution can be neglected and in the case of $B\to X_{d,s}\nu\bar\nu$, $\lambda_t^{(K)}$ and $\lambda_{t^\prime}^{(K)}$  should be replaced by $\lambda_t^{(d,s)}$ and $\lambda_{t^\prime}^{(d,s)}$, respectively. 
\subsection{Other Hamiltonians}
The examples of effective Hamiltonians presented above make it clear how
to find the corresponding Hamiltonians for $B\to X_s\gamma$, $B\to X_s\ell^+\ell^-$, 
$K_L\to \pi^0\ell^+\ell^-$, non-leptonic two-body decays and $\epe$. Therefore 
for these cases we will only present and/or discuss the final formulae for 
branching ratios and CP asymmetries.


\newsection{Master Formulae in the SM4}\label{sec:MasterFormulae}

\subsection{Preliminaries}
The generalisation of the known SM3 formulae for FCNC observables in 
the quark sector \cite{Buchalla:1995vs} to the SM4 is
straightforward. One just extends the summation over flavours to the 4G quarks
and uses the unitarity relation of $V_\text{SM4}$, 
like the one in (\ref{eqn:4gunitarity}), in order to eliminate 
$\lambda^{(K)}_u$, 
$\lambda^{(d)}_u$ and $\lambda^{(s)}_u$ from the final expressions. 
As discussed in the previous section,
the only complication arises from the effects of the 4G
heavy leptons in box diagrams, contributing to semi-leptonic and
leptonic decays if their mixing with SM3 leptons is non-vanishing. In the 
SM3, this effect can be neglected in view of the smallness of the SM3 lepton
masses, but in the presence of 4G heavy leptons it could, 
in principle, be non-negligible.

In fact, the numerical analysis of these effects shows that, for 
$|W_{\ell4}|$ and $|W_{4\mu}|$ of $\ord(1)$, the impact of these additional
contributions cannot be neglected. In our numerical analysis we will however
assume that $|W_{\ell4}|$ are at most $\ord(\lambda)$, and consequently these
effects can be neglected. This will avoid any detailed assumptions about 
the heavy
lepton masses and about their mixing with three SM3 lepton generations. 
However, in order to be complete, we will present the formulae for the 
branching ratios taking this mixing into account.

Next, it is useful, as in the case of the LHT model \cite{Blanke:2006eb}, 
to generalise the real
and universal Inami-Lim functions in the SM3, as introduced in \cite{Buchalla:1990qz},
to complex and non-universal functions ($i=K,d,s$)
\begin{equation}\label{eq31}
S_i\equiv|S_i|e^{i\theta_S^i},
\quad 
X^\ell_i \equiv \left|X^\ell_i\right| e^{i\theta_X^{i\ell}}, \quad 
Y_i \equiv \left|Y_i\right| e^{i\theta_Y^i}, \quad 
Z_i \equiv \left|Z_i\right| e^{i\theta_Z^i}\,,
\end{equation}
which govern particle-antiparticle mixing and rare $K$ and $B$ decays. In the
limit of three generations, $S_i,X^\ell_i,Y_i,Z_i$ reduce to the flavour universal
functions $S_0,X_0,Y_0,Z_0$ that are real. This formulation allows a
transparent comparison of our results with those
obtained for particle-antiparticle mixing and rare decays 
in the LHT model \cite{Blanke:2006sb,Blanke:2006eb,Blanke:2009am}. 
The index ''$\ell$'' in the case of $X^\ell_i$ distinguishes between different
neutrinos in the final state. This distinction is only necessary in the 
presence of $\ord(1)$ elements $|W_{\ell4}|$.

In the case of $B\to X_{s,d}\gamma$ and $B\to X_{s,d}\ell^+\ell^-$,
one has to introduce new functions representing the contributions of dipole
operators. For the magnetic photon and gluon penguin, we have respectively
($i=s,d$)
\begin{equation} \label{eq32}
D'_i \equiv \left|D_i'\right| e^{i\theta_{D'}^i}\,, \quad 
E'_i \equiv \left|E_i'\right| e^{i\theta_{E'}^i}\,.
\end{equation}
For the electromagnetic and
chromomagnetic penguins,  the corresponding
functions are
\begin{equation} \label{eq33}
D_i \equiv \left|D_i\right| e^{i\theta_D^i}\,, \quad 
E_i \equiv \left|E_i\right| e^{i\theta_E^i}\,.
\end{equation}
$D_i$ is gauge dependent and enters the gauge-independent function
$Z_i$ through
\begin{equation}\label{eq34}
Z_i = C_i + \frac{1}{4} D_i, \quad C_i \equiv \left|C_i\right| e^{i\theta_C^i} \,,
\end{equation}
with $C_i$ being the $Z$-penguin function. In the SM3, the functions
$D_i', E_i', D_i, E_i, C_i$ reduce to the known real and 
flavour-universal functions
$D'_0, E'_0, D_0, E_0, C_0$ \cite{Buchalla:1995vs}.

Thus, our next step is to find the expressions for the master
functions of (\ref{eq31})--(\ref{eq34}) in the SM4.

\subsection{Master Functions\label{sec:Master_Functions}}

As discussed, the master functions of the SM4 can be  expressed in
terms of the functions known from the SM3 and the elements of $V_\text{SM4}$. 
No new loop calculations are necessary, except for those involving 
heavy leptons, which have been performed in the previous section. 
We expect these contributions 
to be small, but for completeness we include them in the formulae 
below.

Starting with the effective Hamiltonian for $\Delta F=2$ transitions given 
in the previous section 
and proceeding as explicitly shown there, one can absorb the 
effects of $t^\prime$ into the master functions $S_i$.
We then find for the $\Delta F=2$ transitions ($q=d,s$)
\begin{equation}
S_q = S_0(x_t) + \left(\frac{\lambda_{t^\prime}^{(q)}}{\lambda_{t}^{(q)}}\right)^2 S_0(x_{t^\prime})+ 2\frac{\lambda_{t^\prime}^{(q)}}{\lambda_{t}^{(q)}}S_0(x_t,x_{t^\prime})\,,\label{eq:S_q}
\end{equation}
where $x_i=m_i^2/M_W^2$. In the $K$ system, we also have to keep the last term in
(\ref{A3}), such that
\begin{equation}
S_K = S_0(x_t) + \left(\frac{\lambda_{t^\prime}^{(K)}}{\lambda_{t}^{(K)}}\right)^2 S_0(x_{t^\prime})+ 2\frac{\lambda_{t^\prime}^{(K)}}{\lambda_{t}^{(K)}}S_0(x_t,x_{t^\prime})+2\frac{\eta_{ct}^{(K)}}{\eta_{tt}^{(K)}}\frac{\lambda_c^{(K)}\lambda_{t^\prime}^{(K)}}{\lambda_t^{(K)2}}S_0(x_c,x_{t^\prime})\,.\label{eq:S_K}
\end{equation}
In writing these expressions, we have assumed the following approximate 
relations for the QCD corrections,
\begin{equation}
\eta_{tt}^{(i)}=\eta_{tt^\prime}^{(i)}=\eta_{t^\prime t^\prime}^{(i)}, 
\qquad \eta_{ct}^{(i)}=\eta_{ct^\prime}^{(i)}\,.
\end{equation}
This approximation is justified as $m_{t^\prime}$ is only by a factor 
of 2--3 larger than $m_t$, the anomalous dimension of the involved
$(V-A)\otimes(V-A)$ operator is small, and 
the QCD corrections only very weakly depend on the actual value of $m_{t^\prime}$
(where the $t^\prime$\/--mass is defined as $m_{t^\prime}(m_{t^\prime})$).

Proceeding in an analogous manner, we obtain
\begin{eqnarray}
X^\ell_i &=& X_0(x_t) + \frac{\lambda_{t^\prime}^{(i)}}{\lambda_{t}^{(i)}} X_0(x_{t^\prime}) 
+ \Delta_X^{\ell i}\,,\label{eq:Xi} \\
Y_i &=& Y_0(x_t) + \frac{\lambda_{t^\prime}^{(i)}}{\lambda_{t}^{(i)}} Y_0(x_{t'})
+ \Delta_Y^i\,,\label{eq:Yi} 
\end{eqnarray}
as well as 
\begin{eqnarray}
Z_i &=& Z_0(x_t) + \frac{\lambda_{t^\prime}^{(i)}}{\lambda_{t}^{(i)}} Z_0(x_{t'})\,,
\qquad E_i = X_0(x_t) + \frac{\lambda_{t^\prime}^{(i)}}{\lambda_{t}^{(i)}} E_0(x_{t'})\,,\label{eq:Ei} \label{eq:Zi}\\
D'_i &=& D'_0(x_t) + \frac{\lambda_{t^\prime}^{(i)}}{\lambda_{t}^{(i)}} D'_0(x_{t'})\,,
\qquad
 E'_i = E'_0(x_t) + \frac{\lambda_{t^\prime}^{(i)}}{\lambda_{t}^{(i)}} E'_0(x_{t'})\, \label{eq:Eiprime}\,.
\end{eqnarray}
A similar expression exists for $D_i$, but we do not need it as $D_i$ is
absorbed in $Z_i$. The heavy-lepton corrections $\Delta_X^{\ell i}$ and 
$\Delta^i_Y$ can be extracted from (\ref{eq:lambdatX0}) and 
(\ref{eq:lambdatY0}), respectively.

This completes the presentation of the master functions. The SM3 functions
$S_0(x_t)$, $S_0(x_c,x_t)$, $X_0(x_t)$, $Y_0(x_t)$, $Z_0(x_t)$, $D'_0(x_t)$, $E'_0(x_t)$,
$E_0(x_t)$ and those relevant for the evaluation of $\Delta^{li}_X$ and 
$\Delta^i_Y$
are given in Appendix~A.
We should emphasise that the approximate formula for $S_0(x_c,x_t)$, valid
only for $m_c\ll m_t$ and used in the literature, cannot be used for
$S_0(x_{t^\prime},x_t)$. The exact expression \cite{Buras:1983ap} is given in 
Appendix~A.

\boldmath\subsection{$\Delta F=2$ Observables}\unboldmath

We recall that the off-diagonal element in the dispersive part of the amplitude for $K^0-\bar K^0$ mixing is given by
\begin{equation}
 2m_K\left(M_{12}^K\right)^\ast=\left\langle\bar K^0|H_\text{eff}^{\Delta S=2}|K^0\right\rangle
\,,
\end{equation}
with analogous expressions for the $B_d$ and $B_s$ systems. Our conventions 
follow \cite{Buras:1998raa}. In particular, the operators being involved are ($q=d,s$)
\be\label{eqn:ope}
Q(\Delta S=2)=(\bar s d)_{V-A}(\bar s d)_{V-A}, \qquad
Q^q(\Delta B=2)=(\bar b q)_{V-A}(\bar b q)_{V-A}.
\ee
The usual procedure \cite{Buras:1998raa} then gives
\begin{equation}
 M_{12}^K=\frac{G_F^2}{12\pi^2}F_K^2\hat B_K m_K M_W^2 \overline{M}_{12}^K\,,
\end{equation}
where
\begin{equation}\label{barM12K}
\overline{M}_{12}^K={\lambda_c^{\ast(K)}}^2 \eta_{cc} S_0(x_c)+{\lambda_t^{\ast(K)}}^2\eta_{tt}^{(K)} S_K^\ast+2\eta_{ct}^{(K)}\lambda_t^{\ast(K)}\lambda_c^{\ast(K)}S_0(x_t,x_c),
\end{equation}
and we used the fact that $S_0(x_c)$ and $S_0(x_t,x_c)$ are real-valued functions.
Then
\begin{align}
 &\Delta M_K=2\textrm{Re}\left(M_{12}^K\right)\,,\\
 &\varepsilon_K=\frac{\kappa_\varepsilon e^{i\varphi_\varepsilon}}{\sqrt{2}(\Delta M_K)_\text{exp}}\textrm{Im}\left(M_{12}^K\right)\,,
\end{align}
where the parameters $\varphi_\varepsilon=(43.51\pm0.05)^\circ$ and
$\kappa_\varepsilon=0.92\pm0.02$ \cite{Buras:2008nn} take into account that
$\varphi_\varepsilon\neq\pi/4$ and include an additional effect from
$\textrm{Im}(A_0)$, the imaginary part of the isospin-0 amplitude in
$K\to\pi\pi$. The result for $\kappa_\varepsilon$ has been recently 
confirmed in \cite{Laiho:2009eu}\footnote{The recent inclusion of additional long distance contributions modifies $\kappa_\varepsilon$ to $0.94 \pm 0.02$ \cite{Buras:2010pz} without any visible impact on our numerical results.}.
For $M_{12}^q$, describing the $B_q-\bar B_q$ mixing, 
we then have $(q=d,s)$
\begin{equation}
 M_{12}^q=\frac{G_F^2}{12\pi^2}F_{B_q}^2\hat B_{B_q}m_{B_q}M_W^2\overline{M}_{12}^q\,,
\end{equation}
where
\begin{equation}
 \overline{M}_{12}^q=\left(\lambda_t^{\ast(q)}\right)^2\eta_B S_q^\ast\,.
\end{equation}
The contributions involving charm can be neglected.

For the mass differences in the $B_{d,s}^0-\bar B_{d,s}^0$ systems we have
\begin{equation}
 \Delta M_q=2\left|M_{12}^q\right|\,.
\end{equation}
Defining 
\begin{equation}
M_{12}^q=\left|M_{12}^q\right|e^{2i\varphi_{B_q}^\text{tot}},
\end{equation}
the mixing-induced CP asymmetries $S_{\psi K_S}$ and $S_{\psi\phi}$ 
are simply given as follows,
\be\label{eq:CPNP} 
S_{\psi K_S}=\sin2\varphi_{B_d}^\text{tot}\,,\qquad
S_{\psi\phi}=-\sin2\varphi_{B_s}^\text{tot}\,.
\ee
The latter two observables are the coefficients of $\sin(\Delta M_d t)$
and $\sin(\Delta M_s t)$ in the time-dependent CP asymmetries in $B_d^0\to\psi
K_S$ and $B_s^0\to\psi\phi$, respectively. We also have
\be
2\varphi_{B_d}^\text{tot}=2\bar\beta - \theta^d_S\,, \qquad
2\varphi_{B_s}^\text{tot}=2\bar\beta_s -\theta^s_S\,,\label{eq:phiBdphiBs}
\ee
where
the phases $\bar\beta$ and $\bar\beta_s$ are defined as follows,
\begin{equation}\label{eq:bbs}
 V_{td}=\left|V_{td}\right|e^{-i\bar\beta}\,,\qquad V_{ts}=-\left|V_{ts}\right|e^{-i\bar\beta_s}\,.
\end{equation}
Similarly, the phase $2\varphi_K$ of the leading $S_K^*$ term in
(\ref{barM12K}) is given by
\be
2\varphi_K=2\bar\beta-2\bar\beta_s -\theta^K_S\,.\label{eq:phiK}
\ee

It should be emphasised that $\bar\beta$ and $\bar\beta_s$ differ from 
$\beta\approx 21^\circ$ and $\beta_s\approx -1^\circ$, familiar from 
the SM3 and unitarity triangle (UT) analyses. Indeed, the expressions for $V_{td}$ and 
$V_{ts}$ as seen in (\ref{eqn:v4g}) differ from the ones in the CKM matrix.
In the SM3 $\bar\beta$ and $\bar\beta_s$ reduce to 
$\beta$ and $\beta_s$, respectively. Moreover, the contributions of the
$t^\prime$ quark are absent, and the two asymmetries in question 
are given as follows,
\be\label{eqn:CPSM}
(S_{\psi K_S})_{\rm SM}=\sin2\beta\,,\qquad
(S_{\psi\phi})_{\rm SM}=-\sin2\beta_s\,.
\ee
In the presence of  contributions from the 4G quarks,
these two asymmetries do not measure $\beta$ and $\beta_s$, not even 
$\bar\beta$ and $\bar\beta_s$, but determine 
$\varphi_{B_q}^\text{tot}$ that includes also the contributions from the
$t^\prime$ quark. As we will see, $\varphi_{B_d}^\text{tot}$ will 
only slightly differ from $\beta$ because of the experimental 
constraint\footnote{
Concerning the suppression of $t^\prime$-penguin pollution in the
determination of $\beta$,
the comment  
before Eq.~(\ref{eq:lttprimeformerfootnote}) applies.}
on $S_{\psi K_S}$. On the other hand, $\varphi_{B_s}^\text{tot}$
must be very different from $\beta_s$, in order to reproduce the 
data on $S_{\psi\phi}$ from CDF \cite{Aaltonen:2007he} and D0 \cite{Abazov:2008fj}.
HFAG \cite{Barberio:2008fa} gives the following values for $\varphi_{B_s}^\text{tot}$
\begin{equation}
\varphi_{B_s}^\text{tot} = -(0.39^{+0.18}_{-0.14})\,\,\left[-(1.18^{+0.14}_{-0.18})\right]\,.
\end{equation}
As we will see in Section~\ref{sec:numerics}, the present tensions in the SM3 UT fits 
favour $\theta^d_S > 0$ and $\theta^K_S <0$, while the enhanced value
of $S_{\psi\phi}$ at Tevatron favours $\theta^s_S > 0$. In the case of the $\varepsilon_K$ anomaly, also an enhanced value of $|S_K|$ would help.
As we discussed already in Section~\ref{sec:V4G-param}, and as we
will numerically confirm later in Section~\ref{sec:DetV4G},
a big $S_{\psi\phi}$ will strongly correlate $\delta_{14}$ and 
$\delta_{24}$ as well as $s_{14}$ and $s_{24}$.

Finally, it should also be emphasised that in the SM4 
the ratio $\Delta M_d/\Delta M_s$
does not determine directly the side $R_t$ in the UT, as now the UT
does not close, and there are non-MFV contributions.
\update{Therefore, the determination of the ratio $\vtd/\vts$ by means of 
\be\label{eq:DMdos}
\frac{\Delta M_d}{\Delta
M_s} = \frac{m_{B_d}}{m_{B_s}} \frac{\hat  B_{B_d} F_{B_d}^2}{\hat
B_{B_s} F_{B_s}^2}
\left|\frac{V_{td}}{V_{ts}}\right|^2\frac{|S_d|}{|S_s|}\,,
\ee
will clearly be affected since the last ratio is now generally different from unity.
}

\boldmath\subsection{$B_{s,d}\to\mu^+\mu^-$}\unboldmath

One of the main targets of flavour physics in the coming years will be the
measurement of the branching ratio for the highly suppressed decay 
$B_s\to\mu^+\mu^-$. Hopefully,  the even more suppressed decay 
$B_d\to \mu^+\mu^-$ will be discovered as well. These two decays are helicity
suppressed in the SM3 and CMFV models. Their branching ratios are 
proportional to the
squares of the corresponding weak decay constants that still suffer from
sizable uncertainties. However, using
simultaneously the SM3 expressions for the very well measured mass 
differences $\Delta M_{s,d}$, this uncertainty can be 
eliminated \cite{Buras:2003td},
leaving the uncertainties in the hadronic parameters
 $\hat B_{B_s}$  and $\hat B_{B_d}$
as the only theoretical uncertainty 
in ${\rm Br}(B_{s,d}\to\mu^+\mu^-)$. As seen in Table~\ref{tab:parameters},
these parameters are already known from
lattice calculations with $5-10\%$ precision,
and they enter the branching ratios linearly. 

Generalising this idea to the SM4, we find
\be\label{eq:R2}
{\rm Br}(B_{q}\to\mu^+\mu^-)
=C\frac{\tau(B_{q})}{\hat B_{B_{q}}}
\frac{|Y_q|^2}{|S_q|~} 
\Delta M_{q}\,, \qquad (q=s,d)\,,
\ee
where $\Delta M_q$ is supposed to be taken from experiment,
and the prefactor $C$ is defined as 
\be
C={6\pi}\frac{\eta_Y^2}{\eta_B}
\left(\frac{\alpha}{4\pi\sin^2\theta_{W}}\right)^2\frac{m_\mu^2}{\mw^2}
=4.39\cdot 10^{-10}\,.
\ee
Consequently, the golden relation between ${\rm Br}(B_{d,s}\to\mu^+\mu^-)$ and 
$\Delta M_d/\Delta M_s$, valid in CMFV models \cite{Buras:2003td}, 
gets modified as
follows:
 \be\label{eq:r}
\frac{{\rm Br}(B_s\to\mu^+\mu^-)}{{\rm Br}(B_d\to\mu^+\mu^-)}= \frac{\hat
B_{B_d}}{\hat B_{B_s}} \frac{\tau(B_s)}{\tau(B_d)} \frac{\Delta
M_s}{\Delta M_d}\,r\,,\quad r= \left|\frac{Y_s}{Y_d}\right|^2
\frac{|S_d|}{|S_s|}\,,
\ee
with $r$ being generally different from unity.

Using these expressions, one finds in the SM3 the rather precise predictions
\be\label{TH}
{\rm Br}(B_s\to \mu^+\mu^-)= (3.2\pm0.2)\cdot 10^{-9}\,, \qquad
{\rm Br}(B_d\to \mu^+\mu^-)= (1.0\pm0.1)\cdot 10^{-10}\,,
\ee
where the updated value of ${\rm Br}(B_s\to \mu^+\mu^-)$ has already been 
reported in \cite{Shigemitsu:2009jy}.

These predictions should be compared to the $95\%$ C.L.\ upper limits from 
CDF  \cite{Aaltonen:2007kv} and D0 \cite{Abazov:2007iy} (in parentheses)
\be\label{CDFD0}
{\rm Br}(B_s\to \mu^+\mu^-)\le 3.3~(5.3)\cdot 10^{-8}\,, \qquad
{\rm Br}(B_d\to \mu^+\mu^-)\le 1 \cdot 10^{-8}\,.
\ee
The numbers given above are updates presented at the EPS-HEP09 conference.
More information is given by Punzi \cite{Punzi:2010nv}.
It is clear from (\ref{TH}) and (\ref{CDFD0}) 
that a lot of room is still left for NP contributions.

\boldmath\subsection{$K_L\to\mu^+\mu^-$}\unboldmath

The discussion of the NP contributions to this decay is analogous
to $B_{d,s}\to \mu^+\mu^-$. Again, only the SM3 operator $(V-A)\otimes (V-A)$
contributes, and the real function $Y_0(x_t)$ is replaced by the complex
function $Y_K$ defined in (\ref{eq:Yi}).

In contrast to $B_{s,d}\to\mu^+\mu^-$, the SD
contribution calculated here is only one part of a dispersive contribution
to $K_L\to\mu^+\mu^-$ that is by far dominated by the absorptive contribution
with two internal photon exchanges. Consequently, the SD contribution
 constitutes only a small fraction of the 
branching ratio. Moreover, because of long-distance (LD) contributions to the
dispersive part of $K_L\to\mu^+\mu^-$, the extraction of the SD
part from the data is subject to considerable uncertainties. The most recent estimate gives \cite{Isidori:2003ts}
\be\label{eq:KLmm-bound}
{\rm Br}(K_L\to\mu^+\mu^-)_{\rm SD} \le 2.5 \cdot 10^{-9}\,,
\ee
to be compared with $(0.8\pm0.1)\cdot 10^{-9}$ in the SM3 
\cite{Gorbahn:2006bm}.
In the SM4, we
have 
\be\label{eq:BrKLmumu}
 {\rm Br}(K_L\to\mu^+\mu^-)_{\rm SD} = 2.08\cdot 10^{-9}
\left(\frac{\RE \lambda_c^{(K)}}{\vus}P_c(Y_K)+
\frac{\RE(\lambda_t^{(K)}Y_K)}{\vus^5}\right)^2\,,
\ee
where 
$P_c\left(Y_K\right)=0.113\pm 0.017$
\cite{Gorbahn:2006bm}. 
The numerical results are discussed in Section~\ref{sec:numerics}.

\boldmath\subsection{$B\to X_{s,d}\nu\bar\nu$}\unboldmath

Also  $B$ decays with $\nu\bar\nu$ in the final state provide a very
good test of modified $Z$-penguin contributions 
\cite{Colangelo:1996ay,Buchalla:2000sk}, 
but their measurements appear to be
even harder than those of the rare $K$ decays discussed subsequently. 
Recent analyses
of these decays within the SM3 and several NP scenarios can be found in 
\cite{Altmannshofer:2009ma,Bartsch:2009qp}. The experimental prospects
for these decays at future Super-B machines are summarised in 
\cite{Bona:2007qt}.

Here we will concentrate on 
the theoretical clean decays ${B\to  X_{s,d}\nu\bar\nu}$. The recently 
improved SM3 prediction for $B\to X_s\nu\bar\nu$ reads 
\cite{Altmannshofer:2009ma}  
\be
{\rm Br}(B\to X_s\nu\bar\nu)_{\rm SM}=(2.7\pm 0.2)\cdot 10^{-5}\,.
\ee
As the refinements related to this result apply also to the SM4,
we will only consider the ratios
\bea
\frac{{\rm Br}(B\to X_s\nu\bar\nu)}{{\rm Br}(B\to X_s\nu\bar\nu)_\text{SM}}&=&
\frac{1}{3}\frac{\sum_{\ell=1,2,3}|X^\ell_s|^2}{|X_0(x_t)|^2} r_s\,,
\qquad r_s=\frac{\vts^2}{\vts^2_{\rm SM}} \,,
\label{eq:437}
\\
\frac{{\rm Br}(B\to X_d\nu\bar\nu)}{{\rm Br}(B\to X_d\nu\bar\nu)_\text{SM}}&=&
\frac{1}{3}\frac{\sum_{\ell=1,2,3}|X^\ell_d|^2}{|X_0(x_t)|^2} r_d\,,
 \qquad r_d=\frac{\vtd^2}{\vtd^2_{\rm SM}}\,, \\
\frac{{\rm Br}(B\to X_d\nu\bar\nu)}{{\rm Br}(B\to X_s\nu\bar\nu)}&=&
\frac{\sum_{\ell=1,2,3}|X^\ell_d|^2}{\sum_{\ell=1,2,3}|X^\ell_s|^2}
\left|\frac{V_{td}}{V_{ts}}\right|^2 \,.
\label{eq:BrXdXs}
\eea
As seen explicitly, the branching ratios are defined to include all three
light neutrinos in the final state.
The index SM3 reminds us that the extracted $\vtd$ and $\vts$ in the
presence of 4G quarks can differ from those in the SM. We note also
that for 
$X^\ell_d\ne X^\ell_s$, the relation of the last ratio to $|V_{td}/V_{ts}|$ is
modified with respect to MFV models.
In the absence of right-handed current contributions in the SM4, the formula 
for the ratio in (\ref{eq:437})  applies also to the branching ratios for 
$B\to K^*\nu\bar\nu$ and $B\to K\nu\bar\nu$ as well as to various distributions 
discussed in \cite{Buchalla:2000sk,Altmannshofer:2009ma}. 
Therefore we will not present them here.

\boldmath\subsection{$\kpn$ and $\klpn$}\unboldmath

The SM3 expressions for the specific light neutrino mass eigenstate 
can be easily generalised to the SM4 with
the result
\be\label{eq:BrK+}
{\rm Br}(K^+\to\pi^+\nu_\ell\bar\nu_\ell) =
 \frac{\kappa_+}{3}\left[\left(\frac{\IM(\lambda_t^{(K)}X^\ell_K)}{\vus^5}\right)^2
+\left(\frac{\RE \lambda_c^{(K)}}{\vus}P^\ell_c(X)+
\frac{\RE(\lambda_t^{(K)}X^\ell_K)}{\vus^5}\right)^2\right],
\ee
\be\label{eq:BrKL} 
{\rm Br}(K_{L}\to\pi^0\nu_\ell\bar\nu_\ell) = \frac{\kappa_L}{3} 
\left(\frac{\IM(\lambda_t^{(K)}X^\ell_K)}{\vus^5}\right)^2.
\ee

Summing over three light neutrinos in the final state we simply have
\be
{\rm Br}(K^+\to\pi^+\nu\bar\nu)=\sum_{\ell=1,2,3}{\rm Br}(K^+\to\pi^+\nu_\ell\bar\nu_\ell)\,,
\ee
\be
{\rm Br}(K_{L}\to\pi^0\nu\bar\nu)=\sum_{\ell=1,2,3}{\rm Br}(K_L\to\pi^0\nu_\ell\bar\nu_\ell).
\ee
In the case of small mixing of light leptons with heavy leptons, there is 
no $\ell$-dependence and the factors ''3'' in the denominator in (\ref{eq:BrK+})
and (\ref{eq:BrKL}) drop out.

In the presence of substantial mixing with the 4G leptons, also the charm 
contribution will be affected by this mixing. The inclusion of this effect
would require the reconsideration of QCD corrections  and electroweak 
corrections in the charm sector. As the charm contribution is subleading, 
we do not think that this is required before the discovery of a 4G. Therefore
we will simply set
\be
P_c^\ell(X)=P_c(X)=0.42\pm0.03\,,
\ee
with $P_c(X)$ calculated in the SM3 and including
 the NNLO QCD corrections \cite{Buras:2006gb}, electroweak
 corrections \cite{Brod:2008ss} and 
LD contributions \cite{Isidori:2005xm}.
In reducing the parametric uncertainties in $P_c(X)$, the improved 
value of the charm quark mass
$m_c(m_c)$ 
\cite{Chetyrkin:2009fv,Laiho:2009eu,Allison:2008xk} played an important role.

Next, the determination of the 
relevant hadronic matrix 
elements from tree-level leading $K$ decays,
for $\lambda=0.226$, gives \cite{Mescia:2007kn}
\be\label{eq:kappas}
 \kappa_+=(5.36\pm0.026)\cdot 10^{-11}\,,\qquad   
 \kappa_L= (2.31\pm0.01)\cdot 10^{-10}\,.
\ee
The most recent predictions in the SM3 read
\cite{Buras:2006gb,Brod:2008ss}
\be
{\rm Br}(\kpn)_{\rm SM}=(8.5\pm0.7)\cdot 10^{-11}\,,\qquad
{\rm Br}(\klpn)_{\rm SM}=(2.8\pm0.6)\cdot 10^{-11}\,,
\ee
where the errors are dominated by parametrical uncertainties, in particular
by the CKM parameters. The corresponding experimental data read
\cite{Artamonov:2008qb,Ahn:2007cd}
\be\label{kpnexp}
{\rm Br}(\kpn)=(17.3^{+11.5}_{-10.5})\cdot 10^{-11}\,,\qquad 
{\rm Br}(\klpn)\le 6.7\cdot 10^{-8}\,.
\ee
The experimental upper bound on ${\rm Br}(\klpn)$ is still by more than three 
orders of magnitude above the SM3 value, but the present upper bound
from E391a at KEK
should be significantly improved in the coming decade.
We will see in Section~\ref{sec:numerics} that in the SM4 both branching ratios 
can be spectacularly enhanced over the SM3 expectations.

\boldmath\subsection{The $B\to X_s\gamma$ Decay}\unboldmath\label{sec:Bsgamma}

One of the most popular decays, used to constrain NP
contributions, is the $B \rightarrow X_s \gamma$ decay, for which
the measured branching ratio~\cite{Barberio:2008fa} 
\begin{equation}
{\rm Br}(B \rightarrow
X_s \gamma)_\text{exp} = (3.52 \pm 0.30) \cdot 10^{-4}
\label{eq:bsgexp} 
\end{equation} 
agrees well with the NNLO
prediction in the SM3~\cite{Misiak:2006zs},  
\begin{equation}
{\rm Br}(B \rightarrow X_s
\gamma)_\text{SM} = (3.15 \pm 0.23) \cdot 10^{-4}\,.
\label{eq:bsgSM} 
\end{equation}

The effective Hamiltonian, relevant for this decay within the SM3, is
given as follows,
\begin{equation}
\mathcal H_\text{eff}^\text{SM}(\bar b \rightarrow \bar s
\gamma) = - \frac{G_F}{\sqrt{2}} \lambda_t^{(s)} \left [ \sum_{i=1}^6 C_i(\mu_b)Q_i
+ C_{7\gamma}(\mu_b)Q_{7\gamma} + C_{8G}(\mu_b)Q_{8G}  \right ]\,,
\label{eq:Heffbsg}
\end{equation}
where $Q_i$ are four-quark operators,
$Q_{7\gamma}$ is the magnetic photon penguin operator and $Q_{8G}$
the magnetic gluon penguin operator, contributing to $\bar b\to \bar s\gamma$ 
transitions. The explicit expression for
the branching ratio ${\rm Br}(B \rightarrow X_s \gamma)$, resulting
from~(\ref{eq:Heffbsg}), is very complicated and we will not be presented
here (see \cite{Misiak:2006zs} and references therein).

For our purposes, it is sufficient to know that in the LO
approximation, the Wilson coefficients $C_{7 \gamma}$ and $C_{8G}$ 
at the renormalisation scale $\mu_W=\mathcal{O}(M_W)$  are given as follows,
\be
C_{7\gamma}^{(0)}(\mu_W) = - \frac{1}{2} D_s^\prime\,, \qquad C_{8G}^{(0)}(\mu_W) = -
\frac{1}{2} E_s^\prime\,,
\label{eq:C7g0C8G0}
\ee
with the explicit expressions for $D_s^\prime$ and $E_s^\prime$ given in
Subsection~\ref{sec:Master_Functions}.

In view of the importance of QCD corrections in this decay, we will make
sure that in the limit of neglecting 4G contributions, we reproduce the 
NNLO result in the SM3 given in (\ref{eq:bsgSM}). To this end, we will use
the known LO expressions for the relevant Wilson coefficients evaluated at an 
appropriately chosen renormalisation scale, $\mu_{\rm eff}$,
which turns out to equal $3.22\gev$. The fact that this scale is 
somewhat lower than the 
bottom-quark mass, expresses the important role of QCD corrections, leading to 
an enhancement of the branching ratio. The 4G effects will then be included
through the modification of the SM3 Wilson coefficients at $\mu=M_W$ without
the inclusion of additional QCD corrections.
As the dominant QCD corrections to ${\rm Br}(B \rightarrow X_s
\gamma)$ come anyway from the renormalisation group evolution from $M_W$ down to $\mu_{\rm eff}$, and from the matrix elements of the operators $Q_2$ and 
$Q_{7\gamma}$ at $\mu_{\rm eff}$,
these dominant corrections are common to the SM3 and the SM4. While not
exact, this treatment of QCD corrections in the SM4
should be sufficient for our purposes.

 Thus, in this approximate treatment, the SM4 formulae for 
$B \rightarrow X_s \gamma $ are obtained by making the following 
replacements in the master functions,
\be
\lambda_t^{(s)} D_0'(x_t)\rightarrow \lambda_t^{(s)} D_s'\,, \qquad
\lambda_t^{(s)} E_0'(x_t)\rightarrow  \lambda_t^{(s)} E_s'\,
\label{eq:TDTE}
\ee
and choosing $\mu_{\rm eff}=3.22\gev$.
The Wilson coefficients $C_{7\gamma}(\mu_{\rm eff})$ and 
$C_{8G}(\mu_{\rm eff})$ are given in the SM4 as follows:
\be\label{eq:C78}
C_{7\gamma}(\mu_{\rm eff})=-(0.208+0.305 D_s'+0.052 E_s')\,,\qquad
C_{8G}(\mu_{\rm eff})=-(0.095+0.325 E_s')\,.
\ee  
 Consequently,  the ratio of SM4 to SM3 branching ratios is given within 
our approximations by
\be
\frac{{\rm Br}(B \rightarrow X_s\gamma)}{{\rm Br}(B \rightarrow X_s
\gamma)_\text{SM}}=r_{bs\gamma}^2
\left|\frac{0.208+0.305 D_s'+0.052 E_s'}{0.208+0.305 D_0^\prime(x_t)+0.052
  E_0^\prime(x_t)}\right|^2\,,
\ee
with $r_{bs\gamma}$ defined as
\begin{align}
r_{bs\gamma} & = \left(\frac{|V_{ts}^* V_{tb}|}{|V_{cb}|}\right)\Big/\left(\frac{|V_{ts}^* V_{tb}|}{|V_{cb}|}\right)_{\rm SM}\,.
\end{align}

\boldmath\subsection{Direct CP Violation in $B\to X_s\gamma$ and 
$S_{\phi K_S}$}\unboldmath

\subsubsection{Preliminaries}

Of particular interest is the
direct CP asymmetry $A_{\rm CP}^{bs\gamma}$  
\cite{Soares:1991te,Kagan:1998bh,Kagan:1998ym}
that, similar to $S_{\psi\phi}$, is predicted to be tiny ($-0.5\%$) in the 
SM3 but
could be much larger in some of its extensions, as analysed 
recently in detail in the context of the flavour-blind MSSM (FB\-MSSM)
\cite{Altmannshofer:2008hc} and supersymmetric flavour models 
\cite{Altmannshofer:2009ne}.
In particular in the supersymmetric flavour models with exclusively 
left-handed currents, and also in the FB\-MSSM where these currents dominate,
this asymmetry can be by one order of magnitude larger than in the SM3.
This is in contrast to models with right-handed  currents,
where the $A_{\rm CP}^{bs\gamma}$ asymmetry remains SM3-like 
\cite{Altmannshofer:2009ne}.

As pointed out in \cite{Altmannshofer:2008hc,Altmannshofer:2009ne}, it 
is interesting to consider the direct CP asymmetry in question together
with the theoretically clean
asymmetry $S_{\phi K_S}$ that is found to be significantly smaller than
the expected value of approximately $0.70$ 
\cite{Beneke:2005pu,Buchalla:2008jp,Barberio:2007cr}: 
\be\label{spK}
S_{\phi K_S}=0.44\pm 0.17, \qquad S_{\eta'K_S}= 0.59\pm 0.07\,,
\ee
while  $S_{\eta'K_S}$ is fully consistent with the expectations although
on the low side.

In supersymmetric models with exclusively left-handed currents 
and in the FB\-MSSM,
the desire to explain the $S_{\phi K_S}$ anomaly automatically implies  
that $A_{\rm CP}^{bs\gamma}$ is much
larger in magnitude than its SM3 value \cite{Altmannshofer:2008hc,Altmannshofer:2009ne} and has opposite sign.
Therefore, it is of interest to investigate whether in the SM4  
the two CP asymmetries are large and correlated.

\boldmath
\subsubsection{$A_{\rm CP}^{bs\gamma}$ in the SM4}\unboldmath

If NP effects dominate over the tiny SM3 contribution 
$A^\text{SM}_{\rm CP}(b\to s\gamma)\simeq - 0.5\%$, the following 
expression for $A_{\rm CP}(b\to s\gamma)$ holds~\cite{Kagan:1998bh,Kagan:1998ym},
\begin{multline} \label{eq:acp_bsg}
A_\text{CP}(b\to s\gamma) \equiv \frac{\Gamma(B \to X_{\bar{s}} \gamma) - \Gamma(\overline{B} \to X_s \gamma)}{\Gamma(B \to X_{\bar{s}}\gamma) + \Gamma(\overline{B} \to X_s\gamma)}\simeq \\
\simeq - \frac{1}{|C_{7\gamma}|^2} \left( 1.23~{\rm Im}[C_2 C_{7\gamma}] -
9.52~{\rm Im}[C_{8G}^* C_{7\gamma}] + 0.10 ~{\rm Im}[C_2 C_{8G}] \right)-0.5 
~~~({\rm in}~\%)~,
\end{multline}
where we assumed a cut for the photon energy at $E_{\gamma} \simeq 1.8\gev$
(see~\cite{Kagan:1998bh,Kagan:1998ym} for details). In~(\ref{eq:acp_bsg}), the
Wilson coefficients $C_i$ are evaluated at the scale $\mu_{\rm eff}$ as given in
(\ref{eq:C78}). The Wilson coefficient $C_2$ is to a very good approximation independent of the 4G parameters and given by $C_2\approx 1.14$.
This treatment of QCD corrections is certainly an approximation and a full
NNLO analysis would be much more involved. Yet, at present a NNLO analysis
would clearly be premature.

 We would like to note that our Wilson coefficients $C_i$ 
correspond to the $\bar b \to \bar s$ transition and not to $b\to s$ 
used in
\cite{Kagan:1998bh,Kagan:1998ym,Altmannshofer:2008hc,Altmannshofer:2009ne}. 
Consequently they are complex conjugates of the ones used in the latter papers. This
explains the different placing of ``*'' in (\ref{eq:acp_bsg}) relative to
these papers.

\boldmath
\subsubsection{Time-Dependent CP Asymmetries in $B_d\to\phi(\eta^{\prime}) K_S$}\unboldmath

The time-dependent CP asymmetries in the decays of neutral $B$ mesons into final CP eigenstates $f$ 
can be written as
\begin{equation}
{\cal A}_f(t)=S_f\sin(\Delta M t)-C_f\cos(\Delta M t) \,.
\end{equation}
Our presentation follows closely \cite{Buchalla:2005us}.
Within the SM, it is predicted with good accuracy that the $|S_f|$ 
and $C_f$ parameters are universal for
all the transitions $\bar b\to\bar q^\prime q^\prime\bar s$
($q^\prime=c,s,d,u$). In particular, the SM3 predicts that $-\eta_f
S_f\simeq\sin2\beta$ and 
$C_f\simeq0$ where $\eta_f=\pm1$ is the CP eigenvalue of
the final state $f$. NP effects can contribute to\footnote{We assume that the asymmetry in the tree-level transition $\bar b\to\bar cc\bar s$ is not significantly affected by NP.}
\begin{itemize}
\item[(i)] the $B_d$ mixing amplitude \cite{Amsler:2008zzb},
\item[(ii)] the decay amplitudes $\bar b\to\bar qq\bar s$ ($q=s,d,u$)~\cite{Amsler:2008zzb,London:1997zk,Grossman:1996ke}.
\end{itemize}
In case (i), the NP contribution shifts all $S_f$  asymmetries 
 away from $\sin2\beta$ in a universal way, while the $C_f$  asymmetries
will still vanish. 
In case (ii), the various $S_f$ and also the $C_f$  asymmetries are, in general, 
 different from their values in the SM3.

The CP asymmetries $S_f$ and $C_f$ in $B_d\to f$ decays are calculated as follows. 
One defines a complex quantity $\lambda_f$,
\begin{equation}
\lambda_f=e^{-2i\varphi_{B_d}^{\rm tot}}(\overline{A}_f/A_f)\,,
\end{equation}
where $\varphi^{\rm tot}_{B_d}$ is the  phase of the $B_d$-mixing amplitude, $M_{12}^d$,
and $A_f$ ($\overline{A}_f$)
is the decay amplitude for $B_d(\overline{B_d})\to f$. 
$A_f$ and $\overline{A}_f$ can be calculated from
the effective Hamiltonian relevant for $\Delta B=1$ decays~\cite{Buchalla:1995vs} in the following way
\begin{equation}
A_f=\langle f|{\cal H}_{\rm eff}|B_d\rangle ~,~~
\overline{A}_f=\langle f|{\cal H}_{\rm eff}|\overline{B_d}\rangle~,
\end{equation}
where the Wilson coefficients of the effective Hamiltonian depend on the electroweak theory while the
matrix elements $\langle f|O_i|B_d (\overline{B_d}) \rangle$ can be estimated, for instance, by means
of QCD factorisation~\cite{Buchalla:2005us}. We then have
\begin{equation}
S_f=\frac{2{\rm Im}(\lambda_f)}{1+|\lambda_f|^2} ~,~~
C_f=\frac{1-|\lambda_f|^2}{1+|\lambda_f|^2}~.
\end{equation}
The SM3 contribution to the decay amplitudes, related to $\bar b\to\bar q^\prime q^\prime\bar s$ transitions, can always be written as a sum of two terms, $A_f^{\rm SM}=A_f^c+A_f^u$, with $A_f^c\propto V_{cb}^*V_{cs}$
and $A_f^u\propto V_{ub}^*V_{us}$. Defining the ratio $a_f^u\equiv e^{-i\gamma}(A_f^u/A_f^c)$, we have
\begin{equation} \label{eq:def_a_fu}
A_f^{\rm SM}=A_f^c\left(1+a_f^u e^{i\gamma}\right)~,
\end{equation}
where the $a_f^u$ parameters have been evaluated in the QCD factorisation 
approach\footnote{A
critical discussion of the importance of power corrections and the potential size 
of long-distance final-state interactions
in (\ref{eq:def_a_fu}) can be found in \cite{Cheng:2005bg}. 
As long as a model-independent
prediction for these effects is lacking, we have to assign an irreducible
theoretical error to the predictions for the $a_f^u$.} 
\cite{Beneke:2002jn,Buchalla:2005us,Beneke:2005pu}. 
Within the SM3, it turns out that 
$S_{\phi K_S}\simeq S_{\eta^{\prime}K_S} \simeq S_{\psi K_S} \simeq \sin2\beta$,
with precise predictions
given in Tables~1 and 6 of \cite{Altmannshofer:2009ne}.
 The term $a_f^u$ provides only a negligible contribution to $B_d\to\psi K_S$, 
 thus $\lambda_{\psi K_S}^{\rm SM3}=-e^{-2i\beta}$. 
Also for charmless modes, the effects induced by $a_f^u$ are small (at the percent level), being proportional to $|(V_{ub}V_{us}^{*})/ (V_{cb}V_{cs}^{*})|$.

In the SM4, for simplicity,
we follow the analysis in \cite{Buchalla:2005us}
which only takes into account the leading-order terms in $\alpha_s$ and neglects 
$\Lambda/m_b$ corrections (except for so-called chirally enhanced terms).
The modification of $A_f$ in (\ref{eq:def_a_fu}) 
due to 4G contributions can then be written as 
\begin{equation} \label{eq:def_b_fu}
A_f=A_f^c\left[1+a_f^ue^{i\gamma}+\sum_i
\left(b_{fi}^c+b_{fi}^ue^{i\gamma}\right)C_i^{\rm NP}(M_W)\right]~,
\end{equation}
where $C_i^{\rm NP}(M_W)$ are the NP contributions to the Wilson coefficients
evaluated at the scale $M_W$. Defining the NP contributions to the master functions $F_i$ given in (\ref{eq:Xi})-(\ref{eq:Ei}) by $\Delta F_i$ with
\begin{equation}
 F_i=F_0^\text{SM}+\Delta F_i\,,
\end{equation}
the non-vanishing $C_i^\text{NP}(M_W)$ relevant for 
$\bar b\to \bar s$ transitions are given as follows:
\begin{eqnarray}
C_3^\text{NP}(M_W)&=&\frac{\alpha}{6\pi}\frac{1}{\sin^2\theta_W}\left(2\Delta Y_s-\Delta X_s\right)\,,\\
C_7^\text{NP}(M_W)&=&\frac{\alpha}{6\pi}\,4\Delta Z_s\,,\\
C_9^\text{NP}(M_W)&=&\frac{\alpha}{6\pi}\left[4\Delta Z_s-\frac{2}{\sin^2\theta_W}\left(\Delta X_s+\Delta Y_s\right)\right]\,,\\
C_{7\gamma}^\text{NP}(M_W)&=&-\frac{1}{2}\Delta D_s^\prime\,,\\
C_{8G}^\text{NP}(M_W)&=&-\frac{1}{2}\Delta E_s^\prime\,.
\end{eqnarray}
Here $\alpha=\alpha(M_W)=1/127.9$ is the QED coupling constant and $\sin^2\theta_W=0.231$. 
The parameters $b_{fi}^u$ and $b_{fi}^c$
calculated in \cite{Buchalla:2005us} are collected for the $\phi K_S$ and
$\eta^\prime K_S$ channels in Table~\ref{tab:hadparamas}. 
As the effects in $B \to X_s\gamma$ are small, we follow \cite{Buchalla:2005us}
and neglect  $C_{7\gamma}$.

\begin{table}[t!!!pb] 
\begin{center}
\begin{tabular}{|c||c|c|}
\hline
 	$f$	&  $\phi K_s$	& $\eta^\prime K_s$ \\ \hline\hline
	$b^c_{f3}$ & $-46$  & $-26$ \\ \hline
	$b^c_{f7}$ & $22$   & $3.8$ \\ \hline
	$b^c_{f9}$ & $23$   & $3.5$ \\ \hline
	$b^c_{f8G}$& $1.4$  & $0.86$ \\ \hline
\end{tabular}
\caption{Hadronic parameters at $\mu = m_b$ taken from \cite{Buchalla:2005us}. The parameters $b^u_{fi}$ can be obtained via $b^u_{fi}=(|V_{ub}V_{us}^\ast|/|V_{cb}V_{cs}^\ast|)b^c_{fi}$.} \label{tab:hadparamas}
\end{center}
\end{table}
The absence of complex conjugation on Wilson coefficients $C_i$ in
(\ref {eq:def_b_fu}) as opposed to 
\cite{Altmannshofer:2008hc,Altmannshofer:2009ne} reflects the fact that
our coefficients
correspond to the $\bar b \to \bar s$ transition and not to $b\to s$ 
used in the latter papers.

We note that within a very good approximation
\begin{equation} \label{eq:approx}
A_f\approx A_f^c\left[1+\sum_i
b_{fi}^c C_i^{\rm NP}(M_W)\right]=
 A_f^c\left[1+r_f\frac{\lambda^{(s)}_{t'}}{\lambda^{(s)}_{t}}\right]\,,
\end{equation}
where 
\begin{eqnarray}\label{rphiKs}
r_{\phi K_S}&=&-0.248~ Y_0(x_{t'})+0.004~ X_0(x_{t'})+0.075~ Z_0(x_{t'})-0.7~ E'_0(x_{t'})\,,
\\
\label{retaKs}
r_{\eta' K_S}&=&-0.106~ Y_0(x_{t'})+0.034~ X_0(x_{t'})+0.012~ Z_0(x_{t'})-0.43~ E'_0(x_{t'})\,.
\end{eqnarray}
Thus, the departure of $S_{\phi K_S}$ and $S_{\eta' K_S}$ from 
$S_{\psi K_S}$ is governed by the common phase of
${\lambda^{(s)}_{t'}}/{\lambda^{(s)}_{t}}$ with the effect being larger in 
the case of $S_{\phi K_S}$. Denoting the final phase of $A_f$ by $\varphi_f$,
we find 
\be\label{Sf}
S_f=-\eta_f\sin(2(\varphi_{B_d}^{\rm tot}+\varphi_f)),
\ee
where $\eta_f$ is the CP parity of the final state: $\eta_f=-1$ for both channels considered here.
For $\varphi_f \neq 0$ the departure from $S_{\psi K_S}$ in (\ref{eq:CPNP}) can be obtained.

\update{It is known that going beyond leading order in the calculations of the $b_f$ parameters, one would introduce a potentially sizable strong phase. We comment on this issue in Section \ref{sec73}.
}

\boldmath\subsection{$B\to X_s\ell^+\ell^-$}\unboldmath\label{sec:brbxsll}
No spectacular 4G effects are present in this decay, 
therefore we will only make sure that our numerical
analysis is in accordance with the existing data on these decays. 
Basically, what one has to do is to replace in the known SM3 expressions for the branching 
ratios and the forward-backward asymmetry the SM3 master 
functions $F_0$ by the SM4 master functions  of Section 4.2.
In this context Section 5.5 of \cite{Buras:2004ub} illustrates 
explicitly this replacement, but the formulae are complicated and 
will not be repeated here. Alternatively, all the formulae can be found in \cite{Buras:1994dj}.
The NNLO treatment can be found, for instance, in
 \cite{Asatrian:2002va,Ghinculov:2003qd,Bobeth:2003at,Huber:2007vv}.
The functions that enter this analysis are $Y_s$, $Z_s$, $E_s$, 
$E^\prime$ and $D^\prime$. Of particular interest is 
the forward-backward asymmetry in $b\to s \mu^+\mu^-$, 
in the inclusive and exclusive measurement,
see also \cite{Ali:1991is,Burdman:1998mk,Beneke:2001at,Buras:2004ub}.

The HFAG \cite{Barberio:2008fa} group gives for $M(\ell^+\ell^-)>0.2\gev$
\begin{align}
{\rm Br}(B\to X_s \ell^+\ell^- ) &= 3.66^{+0.76}_{-0.77} \cdot 10^{-6}\,.
\end{align}
Due to the presence of resonances there is no rigorous theoretical prediction for the whole $q^2$ range. Instead, theory and experiment are compared for a high $q^2$ cut, $q^2>14.4\,{\rm GeV}^2$, and a low $q^2$ range, $1\,{\rm GeV}^2<q^2<6\,{\rm GeV^2}$.
The BaBar \cite{Aubert:2004it} and Bell \cite{Iwasaki:2005sy} collaboration published the following results for both ranges:
\begin{align}
{\rm Br}(B\rightarrow ~ X_s \ell^+\ell^- )_{\rm low} &= \left\{
\begin{array}{lr}
\left(1.493\,\pm\, 0.504^{+0.411}_{-0.321}\right)\cdot 10^{-6} \quad & \quad \text{Belle}\\
(1.8\,\pm\, 0.7\,\pm\, 0.5)\cdot 10^{-6} \quad & \quad \text{BaBar}\\
(1.6\,\pm\, 0.5)\cdot 10^{-6} \quad & \quad\text{Average}
\end{array}
\right.\\
{\rm Br}(B\to X_s \ell^+\ell^-)_{q^2>14.4{\rm GeV}^2} &=\left\{
\begin{array}{lr}
\left( 0.418\,\pm\, 0.117^{+0.061}_{-0.068}\right)\cdot 10^{-6} \quad & \quad \text{Belle}\\
\left( 0.5\,\pm\, 0.25^{+0.08}_{-0.07}\right)\cdot 10^{-6} \quad & \quad \text{BaBar}\\
(0.44\,\pm\, 0.12)\cdot 10^{-6} \quad & \quad\text{Average}
\end{array}
\right.
\end{align}
 We will use the averaged measurement for our numerical analysis.
The NNLO prediction for the zero $\hat s_0$ of the forward-backward 
asymmetry $A_{FB}$ in the SM3 is \cite{Ghinculov:2003qd}
\begin{align}
\hat s_0 &= 0.162 \pm 0.008\,.
\end{align}
Note that according to \cite{Ghinculov:2003qd}, the NNLO contributions to $B\to X_s\ell^+\ell^-$ are sizable and \textit{negative}. 
To accommodate the NNLO effects we matched our NLO result to the NNLO result given in \cite{Ghinculov:2003qd} for the low and high $q^2$ ranges
independently.

\boldmath\subsection{$K_L\to\pi^0\ell^+\ell^-$}\unboldmath

The rare decays $K_L\to\pi^0e^+e^-$ and $K_L\to\pi^0\mu^+\mu^-$ are
dominated by CP-violating contributions. In the SM3, the main
contribution comes from the indirect (mixing-induced) CP violation and
its interference with the direct CP-violating contribution
\cite{D'Ambrosio:1998yj,Buchalla:2003sj,Isidori:2004rb,Friot:2004yr}. 
The direct
CP-violating  contribution to the branching ratio is in the ballpark of 
$4\cdot 10^{-12}$, while the CP-conserving contribution is at
most $3\cdot 10^{-12}$. Among the rare $K$ meson decays, the decays in
question belong to the theoretically cleanest, but certainly cannot
compete with  $K\to\pi\nu\bar\nu$ decays. Moreover,
the dominant indirect CP-violating contributions are practically
determined by the measured decays $K_S\to\pi^0\ell^+\ell^-$ and the
parameter $\eps_K$. Consequently, the decays $K_L\to\pi^0\ell^+\ell^-$
 are not as sensitive as the 
$K_L\to\pi^0\nu\bar\nu$ decay to new
physics contributions, present only in the subleading direct
CP violation. However, as pointed out 
in \cite{Buras:2004ub}, in
the presence of large new CP-violating phases, the direct CP-violating
contribution can become the dominant  one, and the branching
ratios for $K_L\to\pi^0\ell^+\ell^-$ can be significantly 
enhanced, with a stronger effect in the case of
$K_L\to\pi^0\mu^+\mu^-$ \cite{Isidori:2004rb,Friot:2004yr}. Most recent 
discussions can be found in \cite{Mescia:2006jd,Prades:2007ud}.

Adapting the formulae in \cite{Buchalla:2003sj,Isidori:2004rb,Friot:2004yr,Mescia:2006jd}
with the help of \cite{Buras:2004ub} to the SM4, we find 
\be\label{eq:BrKpiLL}
{\rm Br}(K_L\to\pi^0\ell^+\ell^-)=\left(C_\text{dir}^\ell\pm
  C_\text{int}^\ell\left|a_s\right| +
  C_\text{mix}^\ell\left|a_s\right|^2+C_\text{CPC}^\ell\right)\cdot
10^{-12}\,,
\ee
where
\begin{align}
&C_\text{dir}^e = (4.62\pm0.24)(\omega_{7V}^2+\omega_{7A}^2)\,,&\qquad&
C_\text{dir}^\mu =(1.09\pm0.05)(\omega_{7V}^2+2.32\omega_{7A}^2)\,,\\
&C_\text{int}^e = (11.3\pm0.3)\omega_{7V}\,,&\qquad&
C_\text{int}^\mu = (2.63\pm0.06)\omega_{7V}\,,\\
&C_\text{mix}^e = 14.5\pm0.05\,,&\qquad&
C_\text{mix}^\mu = 3.36\pm0.20\,,\\
&C_\text{CPC}^e \simeq 0\,,&\qquad&
C_\text{CPC}^\mu = 5.2\pm1.6\,,\\
&&&\hspace{-1.5cm}\left|a_s\right|=1.2\pm0.2
\end{align}
with
\bea
\omega_{7V} &=& \frac{1}{2\pi}\left[P_0+\frac{|Y_K|}{\sin^2\theta_W}
  \frac{\sin\beta^K_Y}{\sin(\bar\beta-\bar\beta_s)}-4|Z_K|
  \frac{\sin\beta^K_Z}{\sin(\bar\beta-\bar\beta_s)}\right]\left[\frac{\IM
  \,\lambda_t^{(K)}}{1.4\cdot10^{-4}}\right]\,,\\
\omega_{7A} &=& -\frac{1}{2\pi}\frac{|Y_K|}{\sin^2\theta_W}
  \frac{\sin\beta^K_Y}{\sin(\bar\beta-\bar\beta_s)}\left[\frac{\IM
  \,\lambda_t^{(K)}}{1.4\cdot10^{-4}}\right]\,.
\eea
Here $P_0=2.88\pm 0.06$ \cite{Buras:1994qa} includes NLO QCD corrections and
\be
\beta^K_Y=\bar\beta-\bar\beta_s-\theta^K_Y\,,\qquad 
\beta^K_Z=\bar\beta-\bar\beta_s-\theta^K_Z
\ee
with $Y_K$ and $Z_K$ defined in (\ref{eq:Yi}) and (\ref{eq:Zi}).  
The phases $\bar\beta$ and $\bar\beta_s$ are defined in (\ref{eq:bbs}).

The effect of the new physics contributions is mainly felt in
$\omega_{7A}$, as the corresponding contributions in $\omega_{7V}$
cancel each other to a large extent.

The present experimental bounds \cite{AlaviHarati:2003mr,AlaviHarati:2000hs},
\be
{\rm Br}(K_L\to\pi^0e^+e^-)<28\cdot10^{-11}\,,\qquad
{\rm Br}(K_L\to\pi^0\mu^+\mu^-)<38\cdot10^{-11}\,,
\ee
are still by one order of magnitude larger than the SM3 predictions,
\cite{Mescia:2006jd}
\begin{gather}
{\rm Br}(K_L\to\pi^0e^+e^-)_\text{SM}=
3.54^{+0.98}_{-0.85}\left(1.56^{+0.62}_{-0.49}\right)\cdot 10^{-11}\,,\label{eq:KLpee}\\
{\rm Br}(K_L\to\pi^0\mu^+\mu^-)_\text{SM}= 1.41^{+0.28}_{-0.26}\left(0.95^{+0.22}_{-0.21}\right)\cdot 10^{-11}\,,\label{eq:KLpmm}
\end{gather}
with the values in parentheses corresponding to the ``$-$'' sign in
\eqref{eq:BrKpiLL}.

\boldmath\subsection{$\epe$}\unboldmath

An important observable is the ratio
$\epe$
 that measures the size of the direct CP
violation in $K_L\to\pi\pi$ 
relative to the indirect CP violation described by $\varepsilon_K$. 
In the SM3 $\varepsilon^\prime$ is governed by QCD penguins but 
receives also an important destructively interfering
 contribution from electroweak
penguins that is generally much more sensitive to NP than the QCD
penguin contribution.

{
Now the electroweak penguin and box diagrams that enter the evaluation of the rare decay branching ratios, as ${\rm Br}(\klpn)$ and ${\rm Br}(\kpn)$, have also a considerable impact on the electroweak component of $\epe$ so that in a given model specific correlations between $\epe$ and the branching ratios for rare $K$ decays exist \cite{Buras:1998ed,Buras:1999da}.
Quite generally, the enhancement of rare decay branching ratios implies the suppression of $\epe$ although this correlation investigated first in \cite{Buras:1998ed,Buras:1999da} contains some model dependence. Yet, as pointed out in \cite{Buras:1998ed} and analysed in more detail within the MSSM in \cite{Buras:1999da}, the enhancements of rare $K$ decay branching ratios could be bounded in principle be $\epe$. In fact, in \cite{Buras:1998ed} approximate bounds of ${\rm Br}(\klpn) \lesssim 2 \cdot 10^{-10}$ and ${\rm Br}(\kpn) \lesssim 2 \cdot 10^{-10}$ have been derived, where in the latter case the bound on ${\rm Br}(K_L \to \mu^+\mu^-)$ also played a role.

Unfortunately, the low precision on the relevant hadronic parameters $B_6$ and $B_8$ in the calculation of $\epe$ and the strong cancellation of QCD penguin and electroweak penguin contributions to this ratio, introduce significant uncertainties in the correlation in question.
}

It is hoped that lattice 
calculations will provide these elements in this decade 
\cite{Christ:2009ev}.

Yet, the present experimental world average  from 
NA48 \cite{Batley:2002gn}   and 
KTeV \cite{AlaviHarati:2002ye,Worcester:2009qt}, 
\be
\epe=(16.8\pm 1.4)\cdot 10^{-4}~,
\ee
could have an important impact on several extensions of the SM3 
if $B_6$ and $B_8$ were known.
An analysis of $\epe$ in the LHT model demonstrates this problem
in explicit terms \cite{Blanke:2007wr}. If one uses $B_6=B_8=1$ as obtained 
in the large $N$ approach \cite{Buras:1987qa}, $(\epe)_{\rm SM}$ is in the ballpark of the 
experimental data, and sizable departures of ${\rm Br}(\klpn)$ from its SM3 
value are not allowed. $\kpn$, being CP conserving and consequently 
not as strongly correlated with $\epe$ as $\klpn$, could still be 
enhanced by $50\%$. On the other hand, if $B_6$ and $B_8$ are different 
from unity and $(\epe)_{\rm SM}$ disagrees with experiment, much more
room for enhancements of rare $K$ decay branching ratios through
NP contributions is available. \update{An analysis of $\epe$ in the context of the SM4 can be found in \cite{Hou:2005yb}.}

In the present paper, we will use the formulae given in Section~3 
of \cite{Blanke:2007wr} that are based on \cite{Buras:2003zz}.
In order to use these formulae, one has to replace the complex 
master functions of the LHT model by the master functions of the
SM4, relevant for the $K$ system, that we have collected in 
Section 4.2. It is a good approximation to neglect the 4G effects on
the flavour-conserving side of box diagrams contributing to $X_K$ and 
$Y_K$, so that these functions together with $Z_K$ are the same as in
rare $K$ decays considered above. $E_K$ plays a subleading role in $\epe$, but
we include it in our numerical analysis which is presented in 
Section~\ref{sec:numerics}.

\newsection{Constraints from Electroweak Precision Data}\label{sec:EWP}

The electroweak precision observables (EPO), even if flavour conserving, have an impact on the allowed parameter space of the SM4.
As the $b^\prime$-quark does not enter our analysis of FCNC processes, we are first of all interested in the range of $m_{t^\prime}$. From direct searches at the Tevatron, it follows that
$m_{t'} >256\gev$. Taking into account the EPO study in \cite{Kribs:2007nz}, 
 together with the upper limit from  perturbativity, 
a plausible range for $m_{t'}$ is given by
\begin{equation}\label{mtrange}
 300\gev \leq m_{t'} \leq 600\gev\,.
\end{equation}
Equally important are electroweak constraints on $s_{34}$ \cite{Alwall:2006bx,Chanowitz:2009mz}, 
as this mixing angle plays an important role in FCNC analyses \cite{Bobrowski:2009ng,Herrera:2008yf,Hou:2005yb,Hou:2006mx,Soni:2008bc}.

As emphasised recently by Chanowitz \cite{Chanowitz:2009mz}, in the presence of $s_{34} \neq 0$, there are two non-decoupling radiative corrections to the EPO with quadratic sensitivity to $m_{t'}$: 
the $T$-parameter and $Zb\bar b$ vertex corrections. They are both proportional to $|V_{t'b}|^2 m_{t'}^2$, and in the case of the $T$-parameter there are also large
 corrections proportional to $|V_{tb'}|^2 m_{b'}^2$ if $m_{b'}^2\gg m_t^2$.
As seen in (\ref{eqn:v4gapprox431}--\ref{eqn:v4gapprox321}),
for all relevant cases we have  $|V_{t'b}| \approx |V_{tb'}|\approx s_{34}$.
The results of a detailed analysis of Chanowitz can be summarised by an approximate upper bound on $|s_{34}|$ as a function of $m_{t'}$:
\begin{equation} \label{s34bound}
 |s_{34}| \le \frac{M_W}{m_{t^\prime}}\,.
\end{equation}
Together with the lower limit on $m_{t'}$, 
this leads to a more stringent upper bound on $s_{34}$
\begin{align}
s_{34}&\leq 0.27\,.
\end{align}
This bound eliminates examples of large $s_{34}$ studied in \cite{Bobrowski:2009ng,Herrera:2008yf} and also has some impact on the 
FCNC analyses in \cite{Hou:2005yb,Hou:2006mx,Soni:2008bc}. 
Still, even for $m_{t'}\simeq 400\gev$, $s_{34}$ can be as large as $s_{12}$,
and in the full range of $m_{t'}$ considered by us it can be 
larger than $|V_{cb}| \simeq 0.04$, although smaller values are certainly allowed. 
We will incorporate (\ref{mtrange}) and (\ref{s34bound}) 
into our numerical analysis in Section \ref{sec:numerics}.

\newsection{Strategy for the Phenomenological Analysis\label{sec:strategy}}

\subsection{Part I: Global Analysis}

In the first part of our numerical analysis, we will use the  values 
for input parameters collected in 
Table~\ref{tab:parameters}. The values of non-perturbative parameters used in our 
analysis are taken from the unquenched lattice calculations 
summarised recently in \cite{Laiho:2009eu}. They are compatible within the errors with the collection of Lubicz and Tarantino \cite{Lubicz:2008am}. The references connected with other parameters are given in this table.
\update{The ranges used for the parameters $m_{t'}$ and $s_{34}$ can be found in (\ref{mtrange}) and (\ref{s34bound}), respectively.}

This analysis will give us the information on the presently allowed ranges 
for the NP parameters and the corresponding maximal departures from SM3 
expectations for various observables that are consistent with the data 
on FCNC processes, EWPT and the unitarity of 
$V_\text{SM4}$. This will also give us a grand view of the patterns of flavour
violation in the SM4, in particular about correlations between 
various observables that are less sensitive to the particular values 
of the parameters involved.

\subsection{Part II: Anatomy}

 The goal of this part will be a detailed analysis of certain features 
of the SM4  that cannot be easily seen in a global analysis 
at present. This analysis consists of the following steps that we will 
outline in what follows.

\begin{description}
  \item[Step 1:] We will investigate how the SM4  addresses 
the present tensions in the unitarity triangle, in particular the 
tension between the values of $S_{\psi K_S}$ and $|\varepsilon_K|$ within
the SM3. To this end, we will use the input parameters of 
Table~\ref{tab:parameters}, except that 
for  $\vub$ and $\delta_{13}$, that are still not known with an
accuracy better than $10\%$ from tree level decays, we will choose three 
scenarios in which $\vub$ and $\delta_{13}$ will take specific values with
errors of $2-3\%$. In Table~\ref{tab:scenarios}, we give three scenarios for 
 $\vub$ and $\delta_{13}$,  whose origin will be explained in 
Section~\ref{sec:Anatomy}.

  \item[Step 2:] We will investigate how the future measurements of various 
branching ratios and CP asymmetries will have an impact on our analysis.
The branching ratios for $B_s\to\mu^+\mu^-$, $\kpn$ and $\klpn$ and 
the CP asymmetry $S_{\psi\phi}$ will play important roles in this study,
and various scenarios for future measurements of these observables 
will be analysed in Section~\ref{sec:Anatomy}.
\end{description}

\boldmath\subsection{Part III: Determination of the $V_\text{SM4}$
  matrix}\unboldmath

In this part, we will outline and illustrate with examples how the 
new parameters of the SM4 can in principle be determined by using
future data on FCNC processes.
In this context, we pay particular attention to the different possible
scenarios for the scaling of the 4G mixing angles. For each case, we
will identify more or less pronounced correlations between the new
CP-violating phases in the SM4 that allow us to define certain equivalence
classes, which can clearly be distinguished by their predictions for
rare decay flavour observables.

\begin{table}[t!!!pb] 
\begin{center}
\begin{tabular}{|l|l||l|l|}
\hline
parameter & value & parameter & value \\
\hline\hline
$\hat B_K$ & $0.725 \pm 0.026$ 	\cite{Laiho:2009eu}			& $|V_{ud}|$&$ 0.97418\pm 0.00027$\cite{Amsler:2008zzb}\\
$F_{B_d}$ & $(192.8 \pm 9.9) \mev$ \cite{Laiho:2009eu} 			& $|V_{us}|$&$ 0.2255 \pm 0.0019 $\cite{Amsler:2008zzb}\\
$F_{B_s}$ & $(238.8 \pm 9.5) \mev$\cite{Laiho:2009eu}			& $|V_{ub}|$&$ (3.93\pm 0.36)\times 10^{-3}$\cite{Amsler:2008zzb}\\ 
$F_K$ & $(155.8\pm 1.7) \mev$	\cite{Laiho:2009eu}			& $|V_{tb}|$&$ 0.91\pm 0.11 \pm 0.07$\cite{Aaltonen:2009jj}\\
$\hat B_{B_d}$ & $1.26\pm 0.11$ \cite{Laiho:2009eu}			& $|V_{tb}|$ & $> 0.71$ at $95\%$CL \cite{Aaltonen:2009jj}\\
$\hat B_{B_s}$ & $1.33\pm 0.06$	\cite{Laiho:2009eu}			& $|V_{cd}|$&$ 0.230\pm 0.011 $\cite{Amsler:2008zzb}\\
$\sqrt{\hat B_{B_d}} F_{B_d}$ & $(216\pm 15) \mev$\cite{Laiho:2009eu}	& $|V_{cs}|$&$ 1.04 \pm 0.06$ \cite{Amsler:2008zzb}\\
$\sqrt{\hat B_{B_s}} F_{B_s}$ & $(275\pm 13) \mev$\cite{Laiho:2009eu}	& $|V_{cb}|$&$ (41.2 \pm 1.1)\times 10^{-3}$\cite{Amsler:2008zzb}\\
$\xi$ & $1.243 \pm 0.028$\cite{Laiho:2009eu}				& $\kappa_\varepsilon$& $0.92\pm 0.02$ \cite{Buras:2008nn} \\
$\eta_{cc}$ & $1.51\pm 0.24$\cite{Herrlich:1993yv}			& $|\varepsilon_K|$& $(2.229\pm0.012)\cdot 10^{-3}$\\
$\eta_{tt}$ & $0.5765\pm 0.0065$\cite{Buras:1990fn}			& $S_{\psi K_S}$& $0.672\pm0.024$\\
$\eta_{ct}$ & $0.47\pm 0.04$\cite{Herrlich:1995hh}			& $\tau(B_d)$ & $(1.525\pm 0.009)$ps\\
$\eta_{B}$  & $0.551\pm 0.007$\cite{Buras:1990fn,Buchalla:1996ys}	& $\tau(B_s)$ & $(1.425\pm 0.041)$ps \\
$m_c(m_c)$ & $(1.268\pm 0.009) \gev$\cite{Laiho:2009eu,Allison:2008xk}	& $\Delta M_d$ & $(0.507\pm 0.005)~ {\rm ps}^{-1}$ \cite{Barberio:2007cr}\\
$m_t(m_t)$ & $(163.5\pm 1.7) \gev$\cite{:2009ec}			& $\Delta M_s$ & $(17.77\pm 0.12)~{\rm ps}^{-1}$ \cite{Barberio:2007cr}\\
\hline
\end{tabular}
\caption{Values of the input parameters used in our analysis.} \label{tab:parameters}
\end{center}
\end{table}

\newsection{Global Numerical Analysis}\label{sec:numerics}

\subsection{Preliminaries}

For our numerical analysis, we construct a large number of random
points in parameter space that are evenly distributed in all
the mixing angles and phases. We keep only those points that satisfy
all tree level CKM constraints at $2\sigma$ (it is not
possible to bring \update{$|V_{cs}|=1.04 \pm 0.06$} within $1\sigma$ of its central value)
and the experimental constraints on the $\Delta F=2$ observables,
\begin{align}
\varepsilon_K\,, && \Delta M_K\,, && \Delta M_q\,, && \Delta M_d/\Delta M_s\,,
\end{align}
as well as the CP asymmetry $S_{\psi K_s}$ within $1\sigma$ and $\cos 
\varphi_{B_d}^{\rm tot}>0$.

In order to compare our results to the experimental values, 
we calculate the propagated error from the corresponding
hadronic parameters using Gaussian error propagation. 
We accept a point if it is at $1\sigma$
within its theoretical uncertainty compatible with the experimental value 
and its respective uncertainty.

One exception is the mass difference $\Delta M_K$:
$\Delta M_K$ is very precisely measured,  but it is
subject to LD contributions that -- in spite of many efforts -- are
not well understood at present. Several analyses indicate that these 
contributions are positive and in the ballpark of $30\%$ of the measured value
\cite{Buras:1985yx,Bijnens:1990mz,Gerard:2005yk}. 
This is supported by the fact that, in the SM3, the value
of the SD box-diagram contributions \update{$\Delta M_K^{\rm SD}$} to $\Delta M_K$ amounts to only 
$(70\pm 10)\%$ of the measured value. Still, there is significant
room in $\Delta M_K$ for NP contributions, and we demand
\update{$0.7 \leq \Delta M_K^{\rm SD}/\Delta M_K^{\rm exp} \leq 1.3$}. 
 Note that the contributions from
the 4G quarks to the SD part of $\Delta M_K$ can exceed $+30\%$.
 In this context we note that the SM3 value of $|\varepsilon_K|$ also appears to be 
lower than the data \cite{Buras:2008nn}. However, in this case, LD 
contributions are
estimated to be small \cite{Buras:2008nn,Buras:2010pz}, and the cure of this anomaly can only come either from
NP, or from significant changes in the input parameters like $\vcb$ and $\vub$.

In addition to the $\Delta F=2$ observables, we impose the following 
constraints from $\Delta F=1$ observables:
\begin{align}
{\rm Br}(B\to X_s\ell^+\ell^-)_{\rm low.-int.}\,, &&{\rm Br}(B\to X_s\gamma)\,, 
&& {\rm Br}(K^+\to \pi^+\nu\bar\nu)\,,
\end{align}
each at $4\sigma$, not taking into account the theoretical errors.
We note that a constraint comes also from the upper bound on ${\rm Br}(K_L\to\mu^+\mu^-)_{\rm SD}$ in (\ref{eq:KLmm-bound}).

Since $\Delta M_K$ poses no stringent bound on $S_K$, and $S_K$ is correlated
with $X_K$, $Y_K$ etc., we chose to impose the experimental exclusion limits on 
various Kaon decays. For consistency
reasons, as well as to eliminate highly tuned points, we also constrain
$B_{s,d}\to\mu^+\mu^-$.

\update{Obviously, the chance of a point to fulfil all constraints
is much larger for small mixing angles than for larger ones.
We therefore have not plotted all the  points that we found
for small mixing angles,
but rather tried to give a comparable number of
parameter points over the full allowed range of mixing angles.
We would like to stress that (even without this procedure)
the point density in our plots must not be understood as a
probability density. 
The main information of the various correlation plots is contained in the \emph{enveloping curve} for these points, without any preference for different points within this region.
}

In presenting the results of the global analysis, it will be useful to
use  a special colour coding,  in order to emphasise
some aspects of the anatomy presented
in the next section and to stress certain points that we found in the process
of our numerical analysis:
\begin{itemize}
\item
 The large black point represents the SM3.
\item
Light blue and dark blue  points stand for the results of our global analysis of the SM4 
with
the following distinction: light blue stands for  ${\rm Br}(K_L\rightarrow \pi^0\nu\bar\nu) > 2\cdot 10^{-10}$
and dark blue for ${\rm Br}(K_L\rightarrow \pi^0\nu\bar\nu) \leq 2\cdot 10^{-10}$. 
Note that the regions with light and dark blue points are
not always exclusive but that the dark blue points are plotted above the
light blue ones.
\item
The yellow, green and red colours represent
 the three scenarios  for $S_{\psi\phi}$ and ${\rm Br}(B_s\to\mu^+\mu^-)$ 
that are
shown in 
Table~\ref{tab:Bscenarios} and related to Step 2 of the anatomy. 
\end{itemize}

\begin{table}[t!!!pb]
\begin{center}
\begin{tabular}{|c||c|c|c|}
\hline
 		&BS1 (yellow) &BS2 (green)	& BS3 (red) 	\\ \hline\hline
$S_{\psi\phi}$	& $0.04\pm 0.01$& $0.04\pm 0.01$ & $ \geq 0.4$ 	\\ \hline
${\rm Br}(B_s\to\mu^+\mu^-)$	& $(2\pm 0.2)\cdot 10^{-9}$ & $(3.2\pm 0.2)\cdot 10^{-9} $   &   $\geq 6\cdot 10^{-9}$ 	\\ \hline
\end{tabular}
\caption{Three scenarios for  $S_{\psi\phi}$ and  ${\rm Br}(B_s\to\mu^+\mu^-)$.} \label{tab:Bscenarios}
\end{center}
\end{table}

\subsection{Violation of Universality}
 
Imposing the existing constraints from tree level determinations of the
 CKM matrix, electroweak precision observables and the existing data on the 
 FCNC and CP-violating observables, it is possible to significantly constrain 
 the allowed ranges for the magnitudes and the phases of the master functions
 $F_i$ introduced in Section 4. 
We recall that in the SM3, the master functions are real and independent 
of the meson system considered.

\begin{figure}[t!!!pb] 
\includegraphics[width=.3\textwidth]{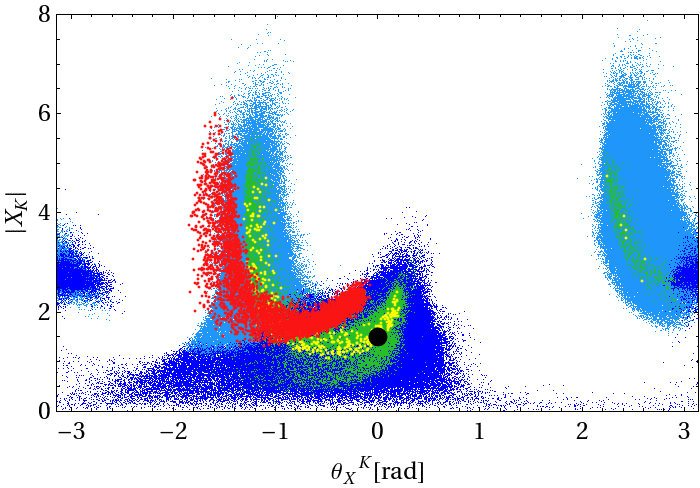}\hspace{.044\textwidth}
\includegraphics[width=.3\textwidth]{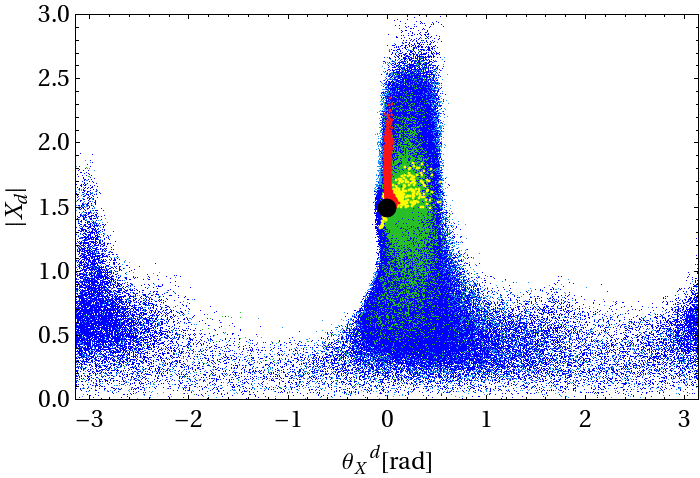}\hspace{.044\textwidth}
\includegraphics[width=.3\textwidth]{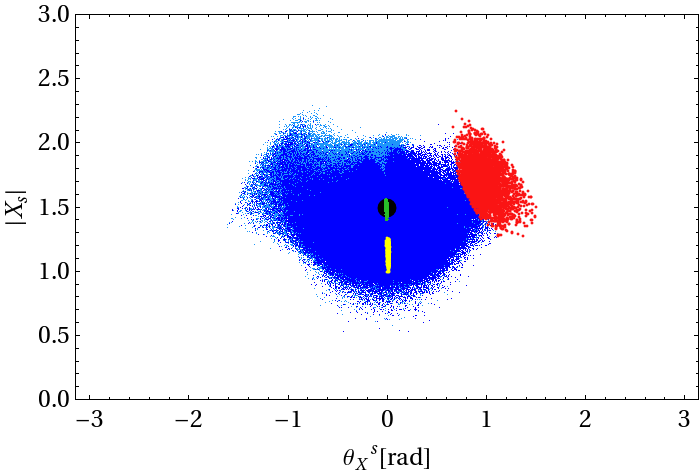}
\caption{The arguments $\theta_X^{i}$ of the functions $X_i$ plotted against the absolute values $\left|X_{i}\right|$ for $i=K$ (left panel), 
$i=d$ (centre panel) and $i=s$ (right panel).\label{fig:X_i}}
\end{figure}

 In Fig.~\ref{fig:S_i} we show the allowed ranges in the planes $(\theta^i_S,|S_i|)$ for
 $i=K,d,s$. The flavour-universal SM3 value of $|S_i|=S_0(x_t)$ is indicated 
 by a  black dot. In Fig.~\ref{fig:X_i}, a similar analysis is done for the functions $X_i$.
 We observe that the flavour 
 universality in question is significantly violated in a hierarchical manner:

\begin{figure}[t!!!pb] 
\includegraphics[width=.3\textwidth]{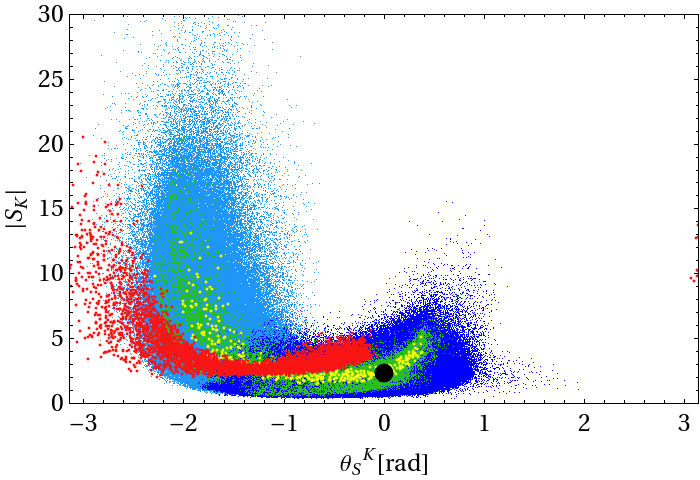}\hspace{.044\textwidth}
\includegraphics[width=.3\textwidth]{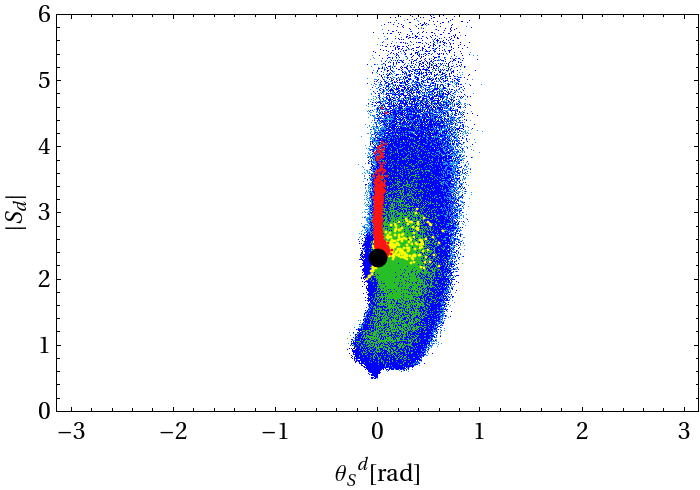}\hspace{.044\textwidth}
\includegraphics[width=.3\textwidth]{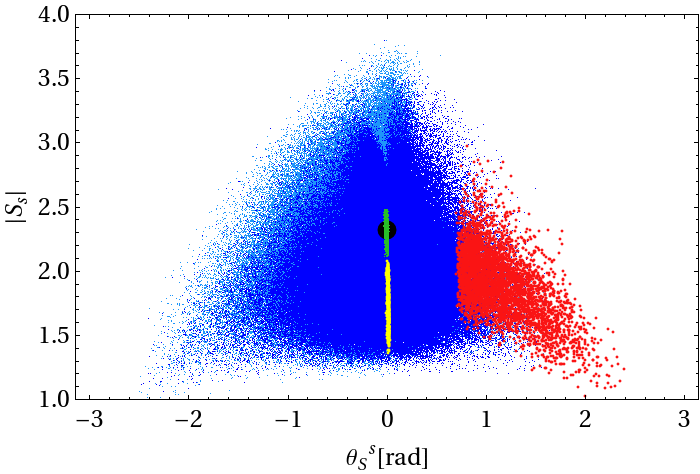}
\caption{The arguments $\theta_S^{i}$ of the functions $S_i$ plotted against the absolute values $\left|S_{i}\right|$ for $i=K$ (left panel), 
$i=d$ (centre panel) and $i=s$ (right panel).\label{fig:S_i}}
\end{figure}

\begin{itemize}

 \item Concerning $\left|S_i\right|$, the largest effects are found for $\left|S_K\right|$, followed by $\left|S_d\right|$, and with the smallest effects found 
 for $\left|S_s\right|$. This hierarchy is familiar from the LHT model and reflects the factor $1/\lambda_t^{(i)}$ in the definition of 
 $S_i$ with $|\lambda_t^{(K)}|\ll|\lambda_t^{(d)}|\leqslant|\lambda_t^{(s)}|$,
 as well as the fact that $S_K$ is not as directly constrained through $\varepsilon_K$ as
 $S_d$ is through $S_{\psi K_s}$ and $\Delta M_d$. In addition, as mentioned before, $\Delta M_K$ only poses a mild constraint.

 \item The departures of $\theta_S^i$ from zero are again largest in the $K$ system. The strong preference for $\theta_S^K<0$ is related --- as seen 
 through (\ref{eq:phiK}) --- to the $\varepsilon_K$-anomaly in the SM3 \cite{Buras:2008nn},
 whose solution favours $\varphi_K>\bar\beta-\bar\beta_s$. 

 \item The phase $\theta_S^d$ is already rather constrained through $S_{\psi K_S}$,
 but a preference for $\theta_S^d>0$ is clearly visible. This 
 reflects the fact that for central values of $\left|V_{ub}\right|$, that are dominated by inclusive decays, the phase $\varphi_{B_d}^\text{tot}$ is required to be smaller than $\bar\beta$, 
 in order to fit $S_{\psi K_S}$ (see (\ref{eq:phiBdphiBs})). 
 
\item $\theta_S^s$ is much less constrained than $\theta_S^d$,
 as the CP violation in the $B_s$ system is experimentally basically unknown. 
 The $S_{\psi\phi}$ anomaly at Tevatron, corresponding to the red points in Fig.~\ref{fig:S_i}, requires $\theta_S^s>0$, as explicitly 
 seen in (\ref{eq:phiBdphiBs}) and (\ref{eq:CPNP})\cite{Lunghi:2008aa,Buras:2008nn}.

 \item Even for no effects in $S_{\psi\phi}$ and $Br(B_s\to \mu^+\mu^-)$ (green points) the SM4 still allows for large effects in the kaon system.
\end{itemize}
Similar hierarchies in the violation of universality are observed in the case of
the functions $X_i$, with the effects in $X_s$ being smallest, not only in 
the magnitude, but also in its phase. 

 \boldmath\subsection{$S_{\psi\phi}$ vs.~$S_{\phi K_S}$ and 
$A_{\rm CP}^{bs\gamma}$} 
\begin{figure}[t!!!pb] 
\includegraphics[width=.48\textwidth]{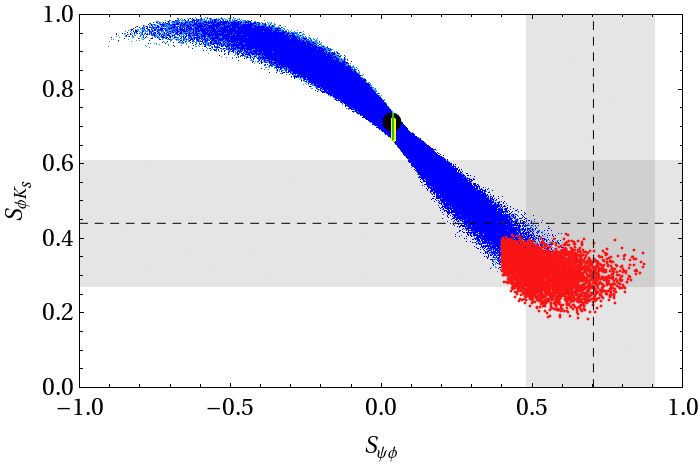}\hspace{.03\textwidth}
\includegraphics[width=.48\textwidth]{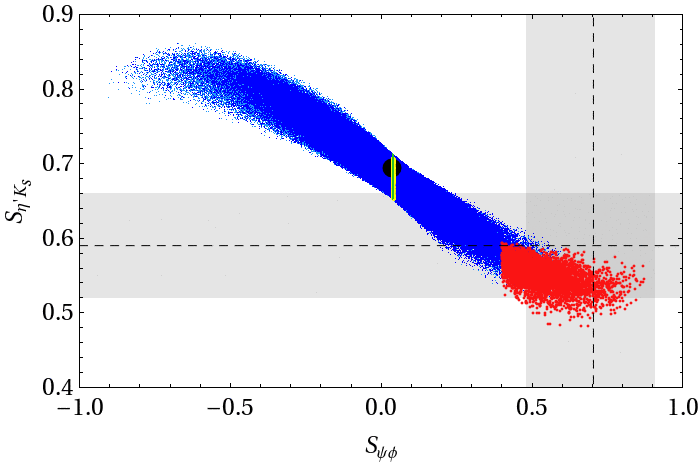}\\
\includegraphics[width=.48\textwidth]{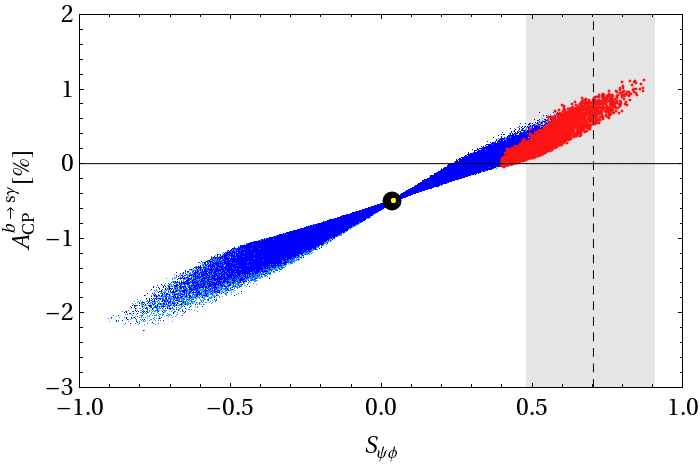}\hspace{.03\textwidth}
\includegraphics[width=.48\textwidth]{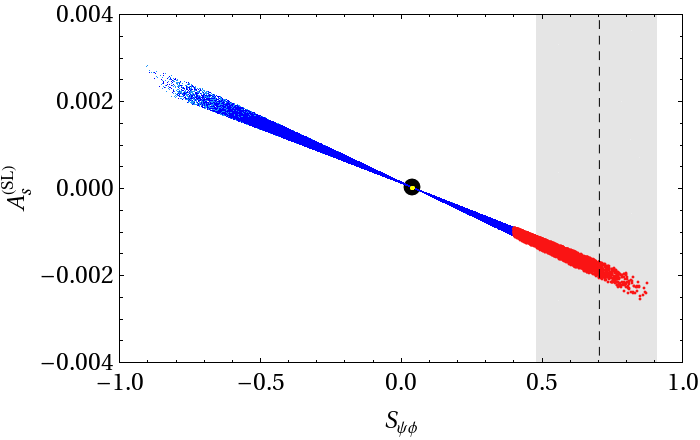}
\caption{The CP asymmetries $S_{\phi K_S}$ (upper-left panel), $S_{\eta^\prime K_S}$ (upper-right panel), $A_\text{CP}^{bs\gamma}$ (lower-left panel), $A_{SL}^s$ (lower-right panel) shown as functions of $S_{\psi\phi}$.\label{fig:CP-asymmetries}}
\end{figure}

\unboldmath
\label{sec73}
As pointed out in \cite{Soni:2008bc}, there is a strong correlation between the
CP asymmetries $S_{\psi\phi}$ and $S_{\phi K_S}$ within the SM4.
We show this correlation in the upper-left panel of Fig.~\ref{fig:CP-asymmetries}.
First of all, we observe that $S_{\psi\phi}$ can be as large as $0.8$ although even larger values are possible.
For $S_{\psi\phi}\approx 0.4$,
the asymmetry $S_{\phi K_S}$ is strongly suppressed relative to $S_{\psi K_S}$
and in the ballpark of $0.4$, close to its experimental central value represented by the horizontal dashed line.
The analogous plot for $S_{\eta'K_S}$ is shown in the upper-right panel of Fig.~\ref{fig:CP-asymmetries}. The suppression 
is now significantly weaker, and for  $S_{\psi\phi}\approx 0.4$ one finds
$S_{\eta'K_S}\approx 0.55$, in accordance with the data. 
The big black points represent the SM3 values of the asymmetries in question that 
are slightly above $S_{\psi K_S}$ \cite{Beneke:2005pu,Buchalla:2008jp,Barberio:2007cr}.

 Interestingly, for $S_{\psi\phi}\geq 0.6$ the values of $S_{\phi K_s}$ and $S_{\eta' K_s}$ predicted 
by the SM4 are below their central values indicated by data.

 The correlation seen in the upper panels in Fig.~\ref{fig:CP-asymmetries} can easily be understood by noting 
that the ratio ${\lambda^{(s)}_{t'}}/{\lambda^{(s)}_{t}}$ and, in particular,
its phase is responsible for departures of both,  $S_{\psi\phi}$ and $S_{\phi K_S}$,
from the SM3 predictions. A {\it positive} complex phase of this ratio implies 
the desired enhancement of $S_{\psi\phi}$ and, through (\ref{eq:approx}) and 
(\ref{rphiKs}),  a {\it negative}
phase $\varphi_{\phi K_S}$ of the decay amplitude $A_{\phi K_S}$. In turn, 
as seen in (\ref{Sf}), $S_{\phi K_S}$ is suppressed relative to $S_{\psi K_S}$. 

At this point some comments are in order
\begin{itemize}
	\item The theoretical errors on (the real part of) the $b_f$ parameters in (\ref{eq:def_b_fu}) are of minor importance. We checked
	numerically that varying the $b_f$ parameters by $10\%$ yields only small effects on the studied correlations.
	\item On the other hand, sizable strong phases from non-factorisable final-state interactions
     may alter the \emph{slope} for the predicted correlation between $S_{\phi K_S}$ and $S_{\psi\phi}$.
     This effect explains the difference between our result and the analysis by Soni et al.~\cite{Soni:2008bc,Soni:2010xh},
     where the non-perturbative parameters are taken from \cite{Beneke:2003zv}, which is based on a phenomenological
     optimisation of theory input in the context of the SM3.
\end{itemize}

The weaker suppression of $S_{\eta'K_S}$ originates in smaller values of 
non-perturbative parameters $b_i$ as seen in Table~\ref{tab:hadparamas}.
On the other hand, as pointed out in
\cite{Altmannshofer:2008hc,Altmannshofer:2009ne} 
in the supersymmetric flavour models with exclusively left-handed currents 
and in the FBMSSM, 
the desire to explain the $S_{\phi K_S}$  anomaly implies automatically that 
$A_{\rm CP}^{bs\gamma}$ is much
larger in magnitude than its SM3 value and has opposite sign. A qualitatively 
similar behaviour is found in the SM4, but as $S_{\phi K_S}$ is strongly correlated
with $S_{\psi\phi}$ and the latter asymmetry is theoretically cleaner, we
prefer to show the correlation between $A_{\rm CP}^{bs\gamma}$ and 
$S_{\psi\phi}$. As seen in the lower-left panel of Fig.~\ref{fig:CP-asymmetries}, for $S_{\psi\phi}\approx 0.5$ 
the asymmetry $A_{\rm CP}^{bs\gamma}$ reverses the sign but its
magnitude is SM3-like. Larger effects are found for larger 
$|S_{\psi\phi}|$, in particular negative values of
$S_{\psi\phi}$, which are however disfavoured by Tevatron data. We conclude
that $A_{\rm CP}^{bs\gamma}$ remains small also in the SM4, but the 
sign flip for large \textit{positive} $S_{\psi\phi}$ could help to distinguish the SM4 from 
the SM3.

Finally, in the lower right panel of Fig.~\ref{fig:CP-asymmetries}, we show 
the familiar correlation between $A_{SL}^s$ and $S_{\psi\phi}$ \cite{Ligeti:2006pm}. The size 
of $A_{SL}^s$ can be by an order of magnitude larger in the SM4 than in the
SM3.

\boldmath\subsection{$B_{s,d} \to \mu^+\mu^-$}\unboldmath

\begin{figure}[t!!!pb] 
\begin{center}
\includegraphics[width=.48\textwidth]{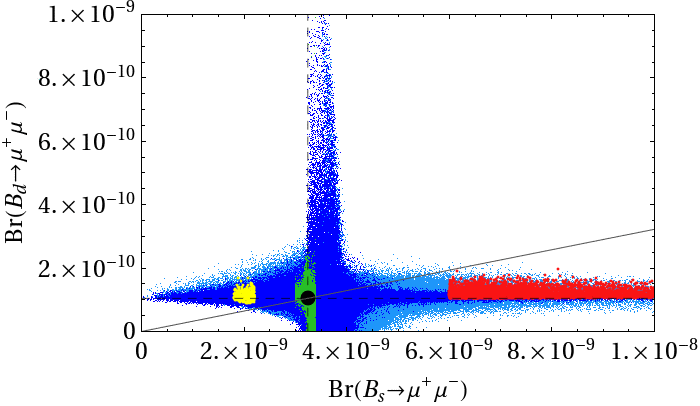}
\end{center}
\vspace{-.5cm}
\caption{${\rm Br}(B_d\to\mu^+\mu^-)$ as a function of ${\rm Br}(B_s\to\mu^+\mu^-)$.\label{fig:Bmumu}}
\end{figure}

 In Fig.~\ref{fig:Bmumu},  we show ${\rm Br}(B_d\to \mu^+\mu^-)$ as a function of 
 ${\rm Br}(B_s\to \mu^+\mu^-)$. The straight line in this plot represents the
 ``Golden Relation''  of CMFV models given in (\ref{eq:r}) with $r=1$.
We observe very strong departures from CMFV. 
We also observe that ${\rm Br}(B_d\to \mu^+\mu^-)$ can be as large as 
$8\cdot 10^{-10}$ and ${\rm Br}(B_s\to \mu^+\mu^-)$ as large as 
$1\cdot 10^{-8}$. The striking message from Fig.~\ref{fig:Bmumu},
 reflecting the non-CMFV character of NP 
contributions, is that large enhancement of ${\rm Br}(B_s\to \mu^+\mu^-)$ implies
 SM3-like values of ${\rm Br}(B_d\to \mu^+\mu^-)$ and vice-versa.

\begin{figure}[t!!!pb] 
\begin{center}
\includegraphics[width=.48\textwidth]{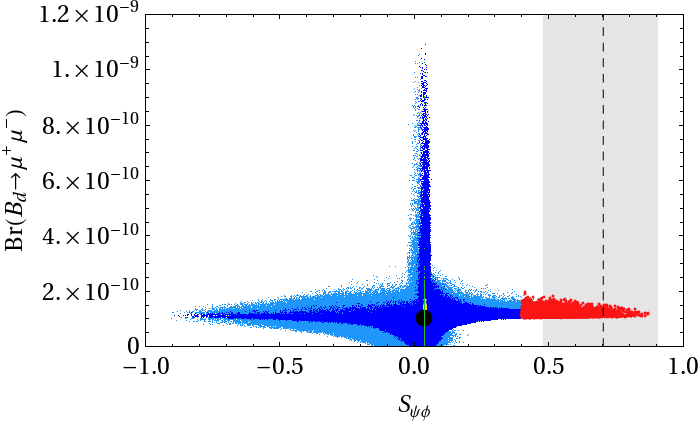}\hspace{.03\textwidth}
\includegraphics[width=.48\textwidth]{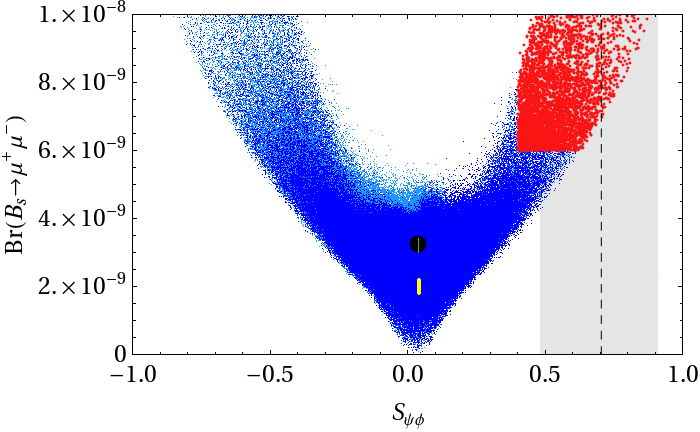}
\end{center}
\vspace{-.5cm}
\caption{${\rm Br}(B_d\to\mu^+\mu^-)$ (left panel) and ${\rm Br}(B_s\to\mu^+\mu^-)$ (right panel) each as a function of $S_{\psi\phi}$. \label{fig:Spsiphi-Bmumu}}
\end{figure}

In Fig.~\ref{fig:Spsiphi-Bmumu}, we show ${\rm Br}(B_d\to \mu^+\mu^-)$ and ${\rm Br}(B_s\to \mu^+\mu^-)$ as functions of 
 $S_{\psi\phi}$. The disparity between these plots shows the non-CMFV character of the NP contributions in the SM4. 
We observe a definite correlation between ${\rm Br}(B_s\to \mu^+\mu^-)$ and $S_{\psi\phi}$,
thus, for a given value of $S_{\psi\phi}$, only a certain range for ${\rm Br}(B_s\to\mu^+\mu^-)$ is predicted. Moreover, 
with increasing $S_{\psi\phi}$ also ${\rm Br}(B_s\to\mu^+\mu^-)$ generally increases. 
In particular, for $S_{\psi\phi}> 0.4$, an enhancement  of ${\rm Br}(B_s\to\mu^+\mu^-)$ is found. For $S_{\psi\phi}\approx 0.4$,
we find that ${\rm Br}(B_s\to\mu^+\mu^-)$ can reach values as high as
$7\cdot 10^{-9}$. For larger values of $S_{\psi\phi}$, even higher values of ${\rm Br}(B_s\to \mu^+\mu^-)$ are possible. 
 Interestingly, for SM3-like values of $S_{\psi\phi}$, the  branching ratio ${\rm Br}(B_s\to\mu^+\mu^-)$ is more likely suppressed than 
enhanced.  
 We conclude that a future measurement of $S_{\psi\phi}$ above $0.4$ accompanied by ${\rm Br}(B_s\to\mu^+\mu^-)$ 
close to or below its SM3 value would put the SM4 into difficulties.

From the right panel in Fig.~\ref{fig:Spsiphi-Bmumu} we can infer the following \textit{soft} bounds on ${\rm Br}(B_s\to\mu^+\mu^-)$ as a function of $S_{\psi\phi}$
\begin{align}
{\rm Br}(B_s\to\mu^+\mu^-) & \leq 
\left( 6 + \left( 4 \cdot (S_{\psi\phi}-\left(S_{\psi\phi}\right)_{\rm SM})\right)^4\right) \cdot 10^{-9}\,,\\
{\rm Br}(B_s\to\mu^+\mu^-) & \geq \left\{ 
\begin{array}{lr} 
\left(-0.08 + S_{\psi\phi} \right) \cdot 10^{-8}\quad  &\quad \text{for}\ S_{\psi\phi} >0\\ 
\left(-0.04 - S_{\psi\phi} \right) \cdot 10^{-8}\quad  &\quad \text{for}\ S_{\psi\phi} <0 
\end{array} \right.\,.
\end{align}
In addition we find the global \textit{soft} upper bound
\begin{align}
{\rm Br}(B_s\to\mu^+\mu^-) & \leq 1.3\cdot 10^{-8}\,.
\end{align}

\boldmath \subsection{$K_L\to\mu^+\mu^-$}\unboldmath\label{sec:klmumu}

\begin{figure}[t!!!pb] 
\begin{center}
\includegraphics[width=.48\textwidth]{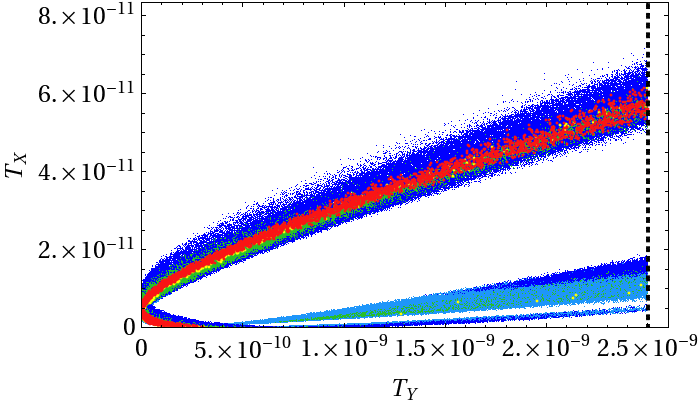}
\end{center}
\vspace{-.5cm}
\caption{$T_X$ as a function of $T_Y$, as defined in (\ref{TYX}). The vertical dashed red line represents the bound (\ref{eq:KLmm-bound}). \label{fig:txvsty}}
\end{figure}

We begin our analysis of rare $K$ decays with  the SD contribution to
 $K_L\to\mu^+\mu^-$, on which
 the bound is given in (\ref{eq:KLmm-bound}). It turns out that in 
the SM4 this bound can be
strongly violated, and imposing it has an impact on the size of possible
enhancements in other rare $K$ decays.
In order to see this transparently, let us define
\update{\be\label{TYX}
T_Y\equiv {\rm Br}(K_L\to\mu^+\mu^-)_{\rm SD},
      \qquad 
T_X\equiv {\rm Br}(K^+\to \pi^+\nu\bar\nu) - \frac{\kappa^+}{\kappa_L}{\rm Br}(K_L\to \pi^0\nu\bar\nu)\,,
\ee
}
with 
$T_X$ entering the branching ratio ${\rm Br}(\kpn)$ in (\ref{eq:BrK+}). 
In Fig.~\ref{fig:txvsty}, we show
$T_X$ as a function of $T_Y$ and find a strong correlation
between these two quantities which could be anticipated on the basis of the 
analytic expressions for $T_X$ and $T_Y$. 
As $T_Y$ is bounded from above directly through 
(\ref{eq:KLmm-bound}),
we obtain also an upper bound on $T_X$.
Throughout our numerical analysis, we impose
the bound (\ref{eq:KLmm-bound}).

\boldmath\subsection{$K^+\to \pi^+ \nu \bar\nu$ and $K_L\to \pi^0 \nu
  \bar\nu$}
\unboldmath

\begin{figure}[t!!!pb] 
\begin{center}
\includegraphics[width=.48\textwidth]{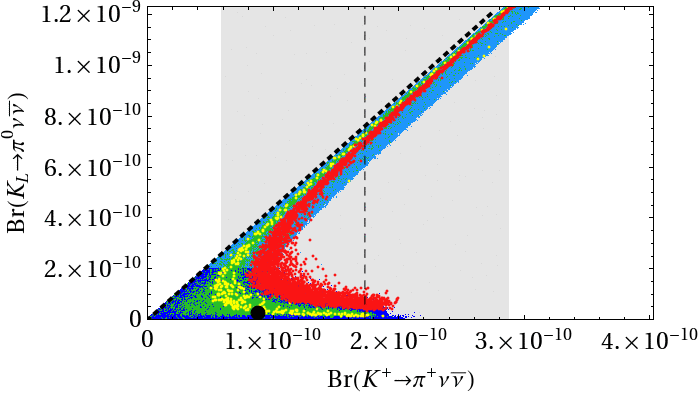}
\end{center}
\vspace{-.5cm}
\caption{${\rm Br}(K_L\to\pi^0\nu\bar\nu)$ as a function of ${\rm Br}(K^+\to\pi^+\nu\bar\nu)$. The dotted line corresponds to the model-independent GN bound.\label{fig:Kpinunu}}
\end{figure}

 In Fig.~\ref{fig:Kpinunu}, we show ${\rm Br}(\klpn)$ as a function of ${\rm Br}(\kpn)$. The black point shows the central SM3 values of the
branching ratios in question, and the shaded region corresponds to the experimental 
$1\sigma$ range for ${\rm Br}(\kpn)$, with its central value given by the light vertical dashed
line.  Unless indicated otherwise, the meaning of dashed lines and shaded areas will be the same as described here throughout our analysis. 

We observe that both branching ratios can be enhanced 
relative to the SM3 values in a spectacular manner. This is in particular
the case for ${\rm Br}(\klpn)$, which can reach values as high as 
$10^{-9}$, that is by a factor of 40 larger than found in the SM3. ${\rm Br}(K^+\to\pi^+\nu\bar\nu)$ can be by a factor of $4$ larger than in the SM3. 
We note that the Grossman-Nir (GN) \cite{Grossman:1997sk}  bound on ${\rm Br}(\klpn)$ 
can be saturated for all values of ${\rm Br}(\kpn)$ shown in the plot.
We also observe that large enhancements of ${\rm Br}(\klpn)$ imply 
necessarily enhancements of ${\rm Br}(\kpn)$. The converse is obviously 
not true, but for ${\rm Br}(\kpn)> 1.7\cdot 10^{-10}$ the ${\rm Br}(\klpn)$ is
either  below $10^{-10}$ or close to the GN bound and larger than
 $ 6\cdot 10^{-10}$. 
\update{For an earlier analysis of $K \to \pi \nu \bar \nu$, where large effects of 4G quarks can be found, see \cite{Hou:2005yb}.}

 The pattern seen in Fig.~\ref{fig:Kpinunu} can be understood as follows. The horizontal lower
 branch, on which ${\rm Br}(\klpn)$ does not depart by much from the SM3 values
 but ${\rm Br}(\kpn)$ can be strongly enhanced, corresponds to the range of
 parameters for which the term $T_X$ in (\ref{eq:BrK+}) dominates ${\rm Br}(\kpn)$.
 But as we have seen above, $T_X$ is efficiently constrained by the bound 
on ${\rm Br}(K_L\to\mu^+\mu^-)_{\rm SD}$, and consequently a stringent upper bound  is put
on ${\rm Br}(\kpn)$ on this branch
\footnote{The fact that the bound on ${\rm Br}(K_L \to \mu^+\mu^-)_{\rm SD}$ can have sizable impact on ${\rm Br}(\kpn)$ has already been pointed out in \cite{Buras:1998ed}. See also \cite{Buras:1999da}.}. 
The second branch, on which both
$K\to\pi\nu\bar\nu$ branching ratios can be strongly enhanced, corresponds
to the region of parameters for which $T_X$ is  subdominant  and
the first term in (\ref{eq:BrK+}) dominates ${\rm Br}(\kpn)$. Comparing (\ref{eq:BrK+}) and 
(\ref{eq:BrKL}),
we observe in this case a very strong correlation between ${\rm Br}(\klpn)_\text{SD}$ and 
${\rm Br}(\kpn)$: their ratio is simply given by $\kappa_L/\kappa_+\approx 4.3$
which is precisely the  GN bound.

 It is evident from this discussion that
the upper branch, in contrast to the lower branch, is not affected by 
the bound on ${\rm Br}(K_L\to\mu^+\mu^-)$. In order to see how the latter
bound affects other observables, we divide the points in 
 Fig.~\ref{fig:Kpinunu} in two groups, with the ones corresponding 
to ${\rm Br}(\klpn)> 2\cdot 10^{-10}$ represented by {\it light blue} points.

\begin{figure}[t!!!pb] 
\includegraphics[width=.48\textwidth]{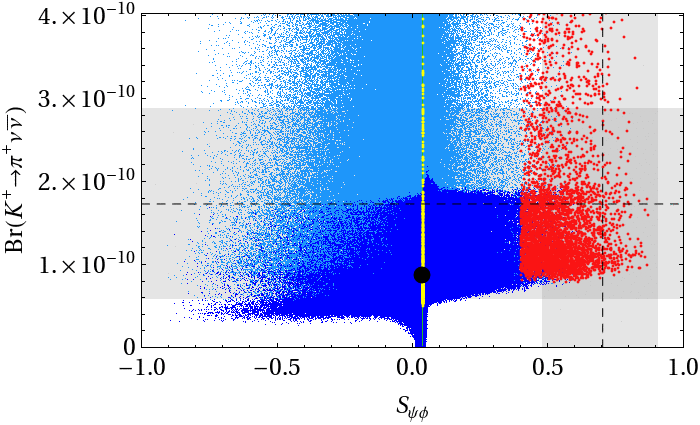}\hspace{.03\textwidth}
\includegraphics[width=.48\textwidth]{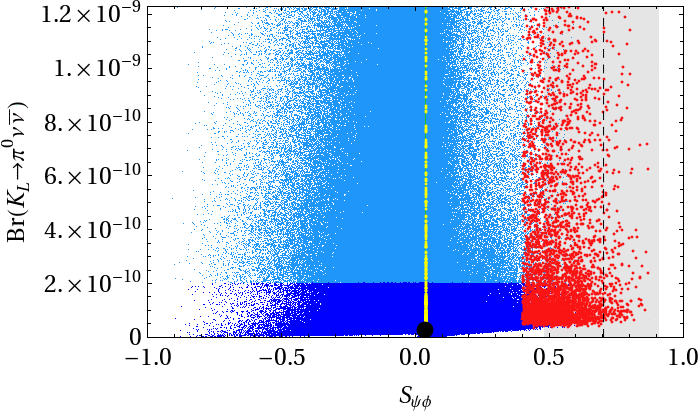}
\caption{${\rm Br}(K^+\to\pi^+\nu\bar\nu)$ (left panel) and ${\rm Br}(K_L\to\pi^0\nu\bar\nu)$ (right panel) as functions of the CP asymmetry $S_{\psi\phi}$.\label{fig:Kpinunu-Spsiphi}}
\end{figure}

In spite of the interesting pattern of deviations from the SM3 seen in 
 Fig.~\ref{fig:Kpinunu},
we conclude  that on the basis of the present constraints 
the predictive power of the SM4 is limited, except that spectacular
deviations from the SM3 are definitely possible, but suppressions cannot
be excluded. We also note that even for large values of $S_{\psi\phi}$ and ${\rm Br}(B_s\to \mu^+\mu^-)$ that are
represented by red points, large NP effects in both branching ratios are possible.
In Fig.~\ref{fig:Kpinunu-Spsiphi}, we look closer at the latter feature. 
A large enhancement of both branching ratios is clearly 
possible as already seen in Fig.~\ref{fig:Kpinunu}.
Moreover, these plots look very different from those found in the LHT and RSc models
\cite{Blanke:2006eb,Blanke:2006sb,Blanke:2009am,Blanke:2008zb,Blanke:2008yr}.\par
In \cite{Blanke:2009pq} the impact of $\eps_K$ on the correlation of ${\rm Br}(K_L\to\pi^0\nu\bar\nu)$ and ${\rm Br}(K^+\to\pi^+\nu\bar\nu)$ was studied.
It was argued that in the case of correlated  phases in the NP contributions to $\Delta F=2$ and $\Delta F=1$ processes one naturally gets the observed two branched structure
of the correlation. However even if this assumption is relaxed, the presence of additive NP contributions to $\eps_K$ implies
\begin{align}
\theta_K^X & \neq \bar\beta - \bar\beta_s \pm \frac{\pi}{2}\,,
\end{align}
and consequently the upper branch never reaches the GN bound. In the SM4 also ${\rm Im}\lambda_c^{(K)}$ is affected and $|{\rm Im}\lambda_c^{(K)}|\gg \left(|{\rm Im}\lambda_c^{(K)}|\right)_{\rm SM}$ can compensate
for large effects introduced through changes in $\lambda_t^{(K)}$ and through $\lambda_{t^\prime}^{(K)}$. Effectively the SM4 is able to maximally violate the
assumption of correlated new phases in $\eps_K$ and $K\to \pi\nu\bar\nu$ and the GN bound can be reached. The effects of $|{\rm Im}\lambda_c^{(K)}|\gg \left(|{\rm Im}\lambda_c^{(K)}|\right)_{\rm SM}$  on ${\rm Br}(K_L\to\pi^0\nu\bar\nu)$
can be neglected, which is evident from the structure of (\ref{eq:BrK+}) and (\ref{eq:BrKL}).

\boldmath\subsection{$K_L\to\mu^+\mu^-$ and $\kpn$}
\begin{figure}[t!!!pb] 
\begin{center}
\includegraphics[width=.48\textwidth]{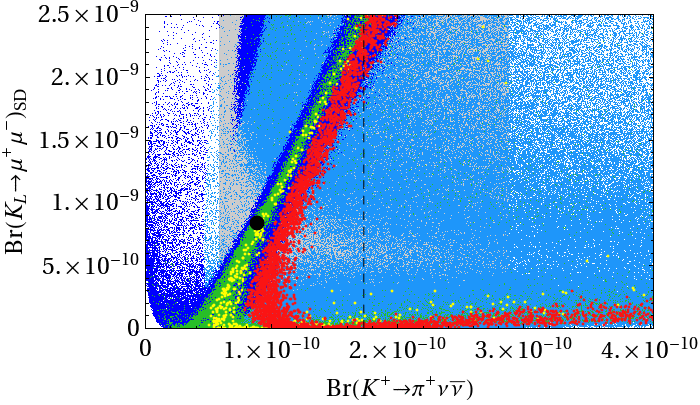}
\end{center}
\vspace{-.5cm}
\caption{The SD contribution to ${\rm Br}(K_L\to\mu^+\mu^-)$ as a function of ${\rm Br}(K^+\to\pi^+\nu\bar\nu)$.\label{fig:Kpinunu-KLmumu}}
\end{figure}

\unboldmath
In order to understand still better the structure of NP effects in 
 Fig.~\ref{fig:Kpinunu}, we show
 in Fig.~\ref{fig:Kpinunu-KLmumu} the correlation between ${\rm Br}(K_L\to\mu^+\mu^-)_{\rm SD}$ and 
${\rm Br}(\kpn)$. We observe that most points cluster around two branches
corresponding to the two branches in Fig.~\ref{fig:Kpinunu}.
On one of them ${\rm Br}(K_L\to\mu^+\mu^-)_{\rm SD}$ is suppressed relative to the SM3 
value while ${\rm Br}(\kpn)$ can be large. On the second branch ${\rm Br}(K_L\to\mu^+\mu^-)_{\rm SD}$
can reach the upper limit at which point ${\rm Br}(\kpn)$ is most likely 
in the ballpark of the central experimental value. Still, as seen in 
Fig.~\ref{fig:Kpinunu-KLmumu}, other combinations of the values of both branching ratios cannot be excluded at present.

\boldmath
\subsection{$B_{s} \to \mu^+\mu^-$ vs.~$K^+\to \pi^+ \nu \bar \nu$}
\unboldmath

\begin{figure}[t!!!pb] 
\begin{center}
\includegraphics[width=.48\textwidth]{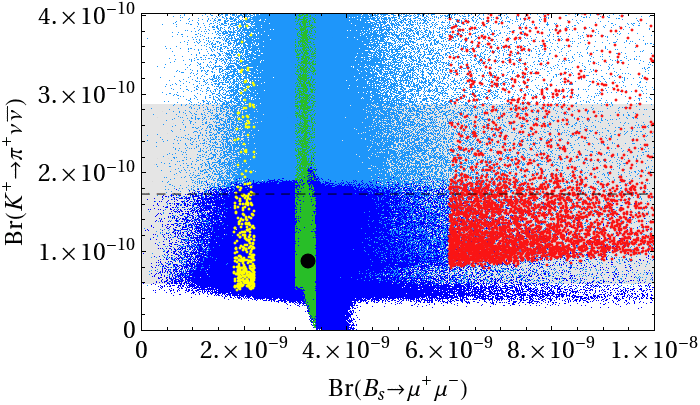}
\end{center}
\vspace{-.5cm}
\caption{${\rm Br}(K^+\to\pi^+\nu\bar\nu)$ as a function of ${\rm Br}(B_s\to\mu^+\mu^-)$.\label{fig:Bsmumu-Kpinunu}}
\end{figure}

In Fig.~\ref{fig:Bsmumu-Kpinunu}, we show ${\rm Br}(\kpn)$ as a function of 
${\rm Br}(B_{s} \to \mu^+\mu^-)$. The striking feature in this plot is the
disparity of possible enhancements of both branching ratios relative to
the SM3 values. While ${\rm Br}(\kpn)$ can be strongly enhanced, as already 
seen in Figs.~\ref{fig:Kpinunu} and \ref{fig:Kpinunu-Spsiphi}, the possible enhancement of ${\rm Br}(B_{s} \to \mu^+\mu^-)$ is
more modest. 
 We also conclude that there is
no strong correlation between both branching ratios, so that they 
can be enhanced significantly at the same time, but this is not necessarily the case.

\boldmath\subsection{$B\to X_s \nu\bar\nu$, $\klpn$ and $B_s\to\mu^+\mu^-$}\unboldmath

\begin{figure}[t!!!pb] 
\begin{center}
\includegraphics[width=.48\textwidth]{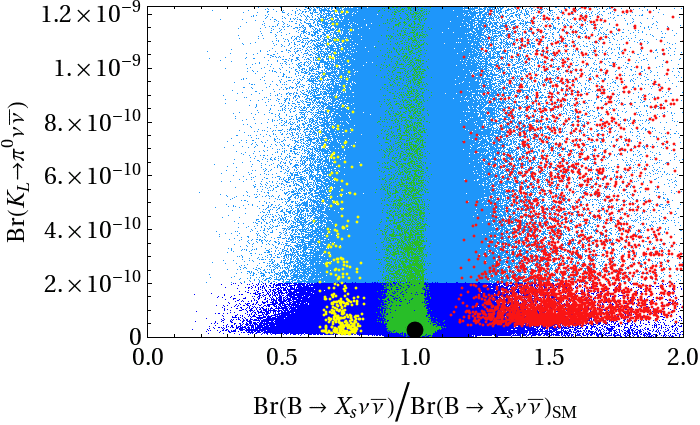}
\end{center}
\vspace{-.5cm}
\caption{${\rm Br}(K_L\to\pi^0\nu\bar\nu)$ as a function of ${\rm Br}(B\to X_s\nu\bar\nu)$.\label{fig:BXsnunu-Kpinunu}}
\end{figure}

\begin{figure}[t!!!pb] 
\begin{center}
\includegraphics[width=.48\textwidth]{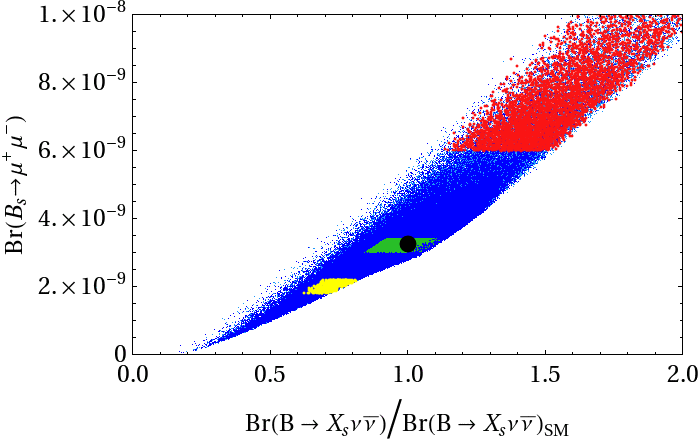}
\end{center}
\vspace{-.5cm}
\caption{${\rm Br}(B_s\to\mu^+\mu^-)$ as a function of ${\rm Br}(B\to X_s\nu\bar\nu)$.\label{fig:BXsnunu-Bsmumu}}
\end{figure}

In Fig.~\ref{fig:BXsnunu-Kpinunu}, we show ${\rm Br}(\klpn)$ as a function 
of ${\rm Br}(B\to X_s \nu\bar\nu)$. Similar to
Fig.~\ref{fig:Bsmumu-Kpinunu}, the 4G effects can be much larger in $\klpn$
than in $B \to X_s \nu \bar \nu$. In fact, 
this plot is similar to the one in Fig.~\ref{fig:Bsmumu-Kpinunu}, except that the possible 
enhancement of ${\rm Br}(B\to X_s \nu\bar\nu)$ is more modest than the one of ${\rm Br}(B_s\to\mu^+\mu^-)$. While there is no visible correlation between the two branching ratios in  Fig.~\ref{fig:BXsnunu-Kpinunu},
${\rm Br}(B\to X_s \nu\bar\nu)$ is significantly  correlated with 
${\rm Br}(B_s\to\mu^+\mu^-)$ as seen in Fig.~\ref{fig:BXsnunu-Bsmumu}.
 In particular, 
both branching ratios are most likely either simultaneously enhanced or simultaneously 
suppressed with respect to their SM3 values.

\boldmath\subsection{$K_L\to \pi^0 e^+ e^-$ vs.~$K_L\to \pi^0 \mu^+ \mu^-$}
\begin{figure}[t!!!pb] 
\begin{center}
\includegraphics[width=.48\textwidth]{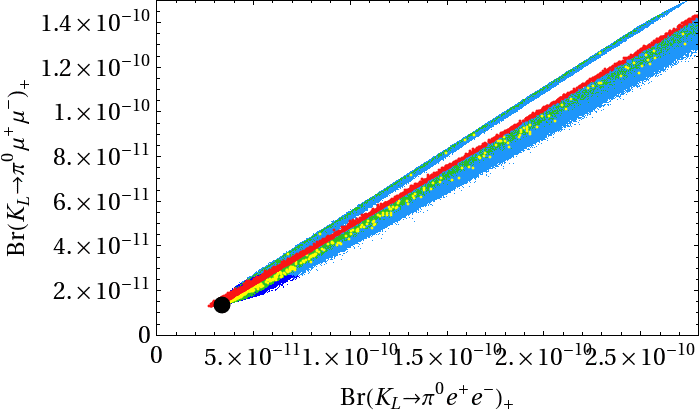}
\end{center}
\vspace{-.5cm}
\caption{${\rm Br}(K_L\to\pi^0e^+e^-)$ as a function of ${\rm Br}(K_L\to\pi^0\mu^+\mu^-)$.\label{fig:Kpimumu-Kpiee}}
\end{figure}

\unboldmath
In Fig.~\ref{fig:Kpimumu-Kpiee}, we show the correlation between 
${\rm Br}(K_L\to \pi^0 e^+ e^-)$ and ${\rm Br}(K_L\to \pi^0 \mu^+ \mu^-)$, also 
familiar from the LHT model \cite{Blanke:2006eb}.
 We show only the ``+'' solution with the SM3 values given in (\ref{eq:KLpee}) and (\ref{eq:KLpmm}). As expected, the allowed enhancements are not as pronounced as in the case of $K_L\to\pi^0\nu\bar\nu$. 
 However, they are still much larger than in the LHT model: one order of magnitude for both branching ratios with slightly larger effects for $K_L\to\pi^0\mu^+\mu^-$.

\boldmath\subsection{$K_L\to \pi^0 \ell^+ \ell^-$ vs.~$K_L\to \pi^0 \nu \bar \nu$}
\unboldmath

\begin{figure}[t!!!pb] 
\begin{center}
\includegraphics[width=.48\textwidth]{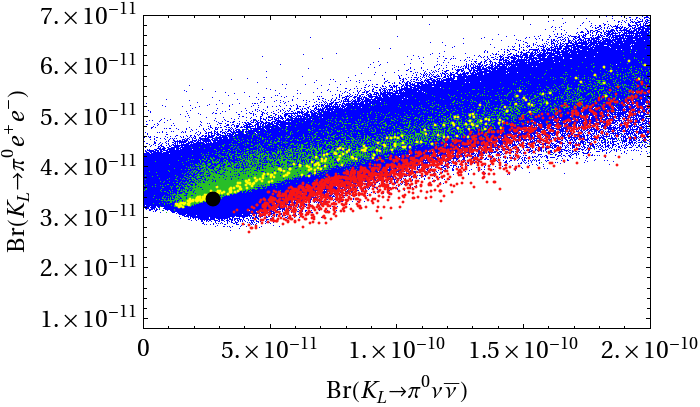}\hspace{.03\textwidth}
\includegraphics[width=.48\textwidth]{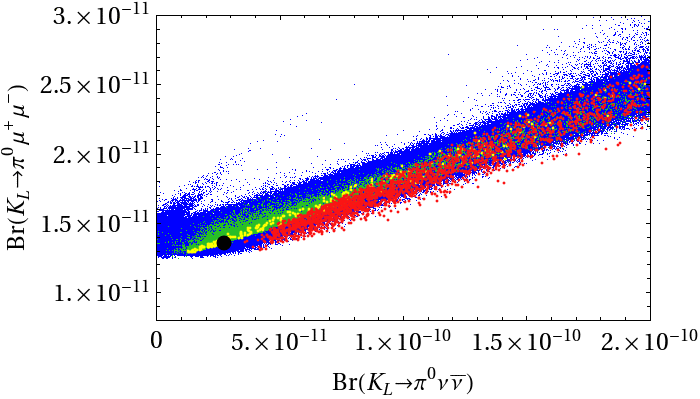}
\end{center}
\vspace{-.5cm}
\caption{${\rm Br}(K_L\to\pi^0e^+e^-)_+$ as functions of ${\rm Br}(K_L\to\pi^0\nu\bar\nu)$ (left panel),  ${\rm Br}(K_L\to\pi^0\mu^+\mu^-)_+$  as functions of ${\rm Br}(K_L\to\pi^0\nu\bar\nu)$ (right panel).\label{fig:Kpinunu-Kpill}}
\end{figure}

In Fig.~\ref{fig:Kpinunu-Kpill}, we show a correlation between
${\rm Br}(K_L\to \pi^0 \ell^+ \ell^-)$ and ${\rm Br}(K_L\to \pi^0 \nu \bar \nu)$ that
has also been found in the LHT and RSc models \cite{Blanke:2006sb,Blanke:2006eb,Blanke:2009am,Blanke:2008zb,Blanke:2008yr}.
The enhancement of one of the branching ratios implies automatically the enhancement of the 
other. We show only the results for ${\rm Br}(K_L\to\pi^0\nu\bar\nu)\leqslant2\cdot 10^{-10}$,
as the extrapolation to higher values is straightforward. 
The main message from this plot is that 
the enhancement of ${\rm Br}(\klpn)$ can be much larger than the  one of
${\rm Br}(K_L\to\pi^0\ell^+\ell^-)$, as already anticipated on the basis of analytic formulae.

\boldmath\subsection{The ratio $\epe$}\unboldmath \label{subsec:epe}

\begin{figure}[t!!!pb] 
\includegraphics[width=.48\textwidth]{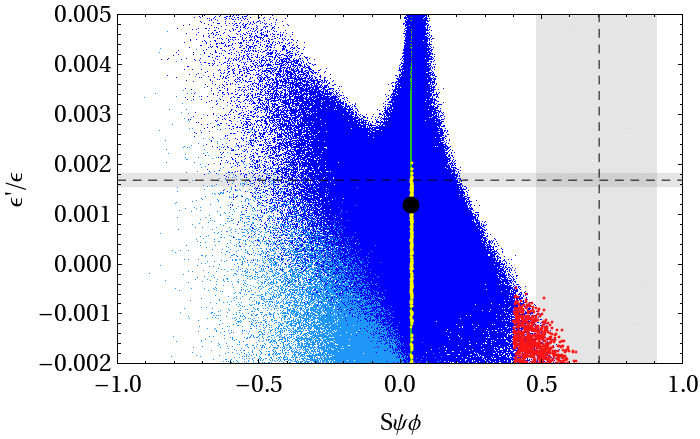}\hspace{.03\textwidth}
\includegraphics[width=.48\textwidth]{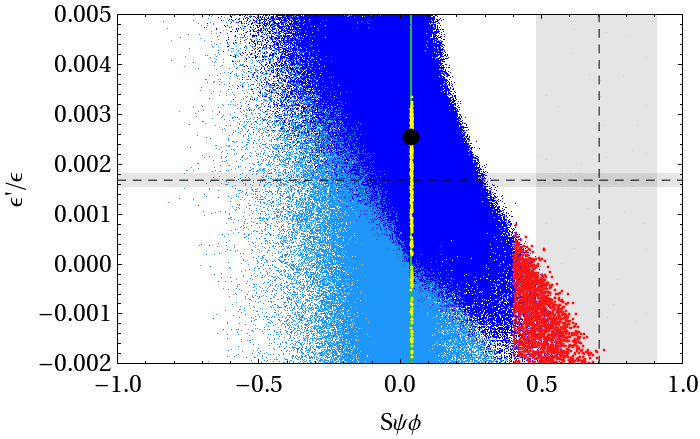}\\
\includegraphics[width=.48\textwidth]{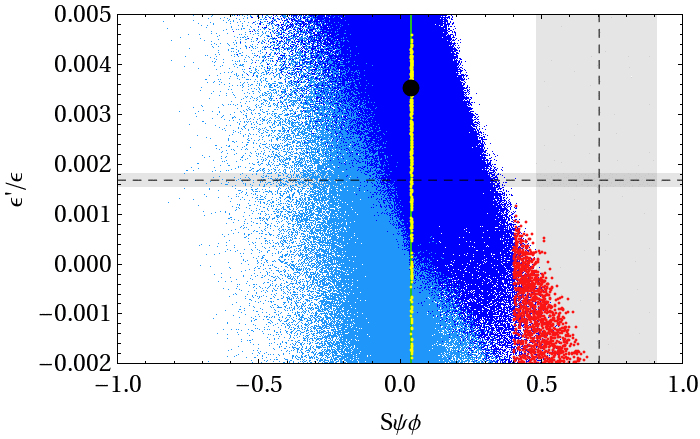}\hspace{.03\textwidth}
\includegraphics[width=.48\textwidth]{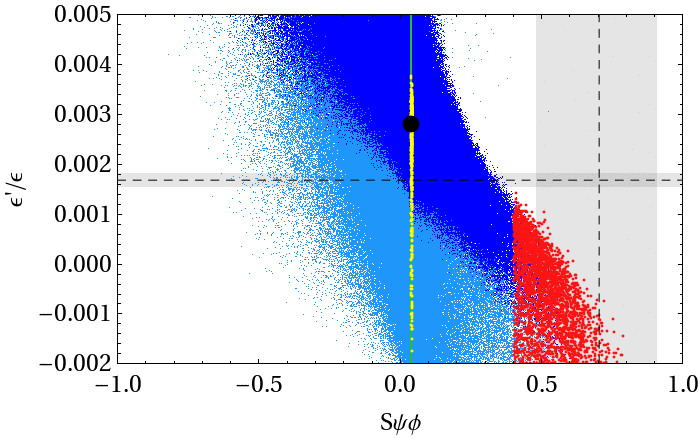}
\caption{$\epe$ as a function of the CP asymmetry $S_{\psi\phi}$ for four different scenarios of the non-perturbative parameters.
$(R_6,R_8)=(1.0,1.0)$ (upper left panel), $(1.5, 0.8)$ (upper right panel), $(2.0,1.0)$ (lower left panel) and $(1.5,0.5)$ (lower right panel).\label{fig:spsiphiepe}}
\end{figure}

In Fig.~\ref{fig:spsiphiepe},
we show $\epe$ as a function of $S_{\psi\phi}$ for four different
scenarios of the non-perturbative parameters $R_6$ and $R_8$:
$(R_6,R_8)=(1.0,1.0)$ (upper left panel), $(1.5, 0.8)$ (upper right panel), $(2.0,1.0)$ (lower left panel) and $(1.5,0.5)$ (lower right panel). Each set of points has the SM3 value
indicated by a black dot. $\Lambda_{\overline{\rm MS}}$ has been set to $340~{\rm MeV}$. The non-perturbative parameters $R_6$ and $R_8$ are defined as 
\begin{align}
 R_6&\equiv B_6^{(1/2)}\left[\frac{121~{\rm MeV}}{m_s(m_c)+m_d(m_c)}\right]^2\,, &  
 R_8&\equiv B_8^{(3/2)}\left[\frac{121~{\rm MeV}}{m_s(m_c)+m_d(m_c)}\right]^2 \,.
\end{align}

 As a general feature the SM4 can fit $\epe$ for all sets of non-perturbative parameters considered by us. However, the striking feature of these plots is the difficulty in reproducing the experimental data for $\epe$ within the SM4 when $S_{\psi\phi}$ is large and positive (red points) as suggested by the Tevatron data. Thus if the latter data will be confirmed $\epe$ can put the SM4 under pressure, unless $R_6$ is significantly larger than unity and $R_8$ sufficiently 
suppressed below one. This discussion demonstrates again that in order to use
$\epe$ to constrain NP the knowledge of the parameters $R_6$ and $R_8$ has to be improved significantly.
\update{An analysis of $\epe$ in the 4G model has been presented in \cite{Hou:2005yb}, where the hadronic uncertainties known from previous studies have been reemphasised. 
However, the direct impact of large $S_{\psi\phi}$ on $\epe$ has not been studied in details there as we do below.
}

\begin{table}[t!!!pb] 
\begin{center}
\begin{tabular}{|c|c||c|}
\hline
$R_6$	& $R_8$	&	 	\\ \hline\hline
$1.0$	& $1.0$	& dark blue	\\ \hline
$1.5$	& $0.8$	& purple	\\ \hline
$2.0$	& $1.0$	& green		\\ \hline
$1.5$	& $0.5$	& orange	\\ \hline
\end{tabular}
\caption{Four scenarios for the parameters $R_6$ and $R_8$\label{tab:Rscenarios}}
\end{center}
\end{table}

\begin{figure}[t!!!pb] 
\includegraphics[width=.48\textwidth]{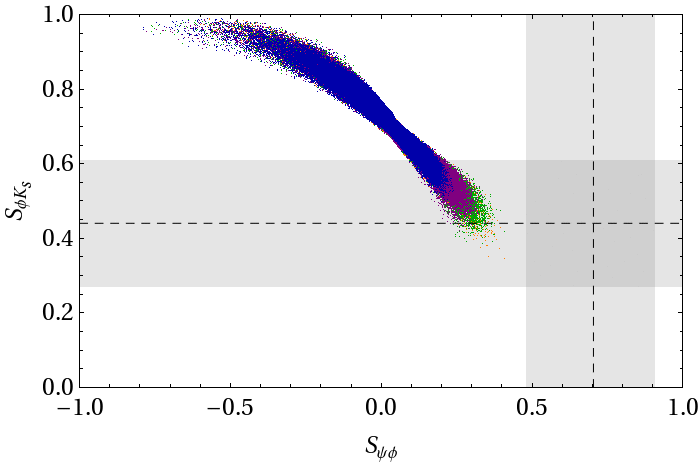}\hspace{.03\textwidth}
\includegraphics[width=.48\textwidth]{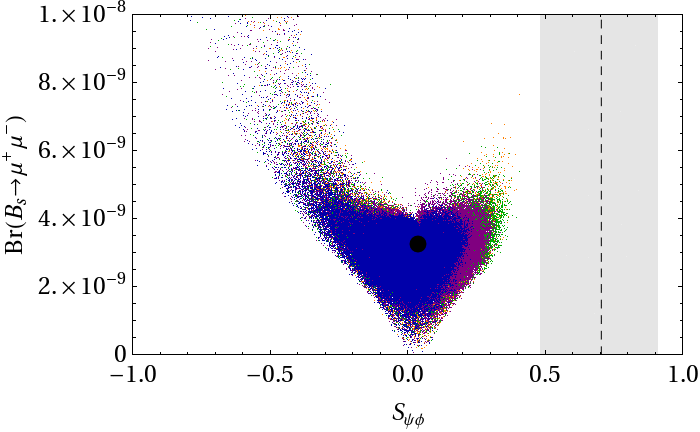}\\
\includegraphics[width=.48\textwidth]{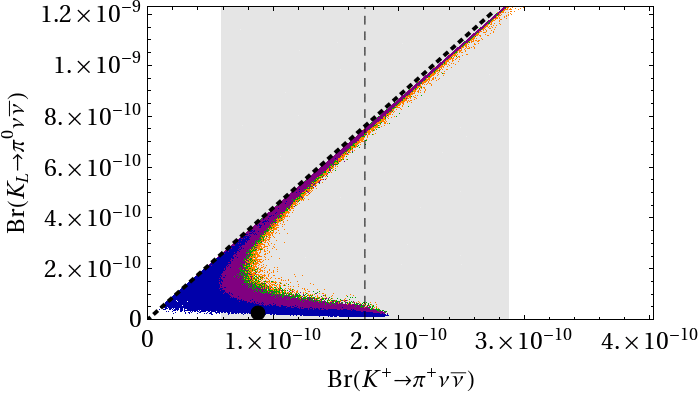}\hspace{.03\textwidth}
\includegraphics[width=.48\textwidth]{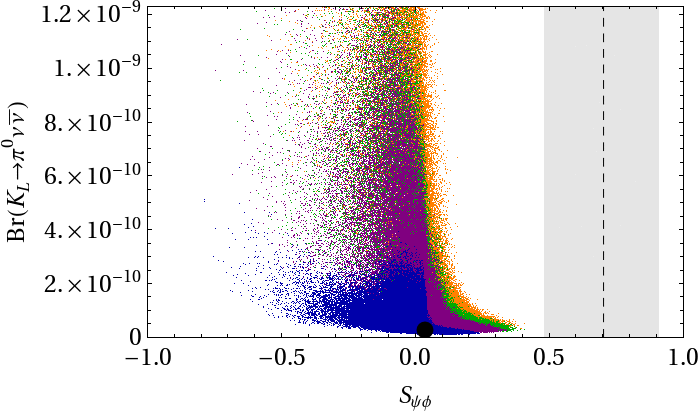}
\caption{ Correlations including the $\epe$-constraint (colour-coding according to Table~\ref{tab:Rscenarios}) .\label{fig:corr-eps}}
\end{figure}

It is then of interest to investigate what impact $\epe$ would have on our analysis when $R_6$ and $R_8$ were precisely known. To this end we introduce still another coding in Table \ref{tab:Rscenarios}, this time for different values of $R_6$ and $R_8$. In Fig.~\ref{fig:corr-eps} we show then the most interesting correlations, this time including also the $\epe$-constraint. These are $S_{\phi K_S}$ vs. $S_{\psi\phi}$, ${\rm Br}(B_s \to \mu^+\mu^-)$ vs. $S_{\psi\phi}$, ${\rm Br}(\klpn)$ vs. ${\rm Br}(\kpn)$ and  ${\rm Br}(\klpn)$ vs. $S_{\psi\phi}$. These plots should be compared to the plots in Figs.~\ref{fig:CP-asymmetries}, \ref{fig:Spsiphi-Bmumu}, \ref{fig:Kpinunu} and \ref{fig:Kpinunu-Spsiphi} respectively, where the $\epe$ constraint has not been taken into account.

We observe the following striking features already anticipated on the basis of Fig.~\ref{fig:spsiphiepe}:
\begin{itemize}
 \item $S_{\psi\phi}$ can be at most 0.4 and this upper bound is only reached for the green and orange points where the ratio $R_6/R_8 \geq 2$
 \item For the large N case $R_6=R_8=1$ represented by dark blue points, we identify the following rough absolute bounds
\begin{align*}
 S_{\psi\phi} &\lesssim 0.25 \,, &{\rm Br}(\klpn) &\lesssim 4\cdot 10^{-10}\,,\\
 {\rm Br}(\kpn) &\lesssim 2\cdot 10^{-10}\,, &{\rm Br}(B_s \to \mu^+\mu^-) &\lesssim 4.9\cdot 10^{-9}\,,
\end{align*}
where in the last case $S_{\psi\phi} > 0$ has been assumed in accordance with CDF and D0 data.
\item Weaker bounds are found for purple and in particular green and orange points where the role of electroweak penguins relative to QCD penguins in $\epe$ is suppressed and it is easier to have larger 4G effects in rare $K$ and $B$ decays, while still satisfying the $\epe$ constraint.
\end{itemize}

Finally we would like to remark that the enhancements of rare $K$ decay branching ratios found here are larger than the bounds in \cite{Buras:1998ed} would suggest. 
This is related to the fact that in the SM4 the NP effects in neutral meson mixing can be significantly larger than assumed in \cite{Buras:1998ed}, 
implying that the range for $\IM \lambda_t$ in the SM4 can be significantly larger than used in \cite{Buras:1998ed}.

\subsection{Violation of CKM Unitarity}

From our discussion of unitarity of the matrix $V_{4G}$ in Section~\ref{sec:V4G},
 we see that the sub-matrix describing the 3G mixing, for non-vanishing mixing angles $\theta_{i4}$, is necessarily non-unitary. A similar effect was observed also in the RS model with custodial protection \cite{Buras:2009ka}. In order to quantitatively describe the deviation from unitarity, we define
\begin{equation}
K^u\equiv V_\text{CKM3}V_\text{CKM3}^\dagger\,,\quad K^d\equiv V_\text{CKM3}^\dagger V_\text{CKM3}
\end{equation}
which are generally different from the $3\times 3$ unit matrix. In particular, we find
\begin{equation}
|K^u-\mathbbm{1}|_{ij}=|V_{ib^\prime}V_{jb^\prime}^\ast|\,,\quad|K^d-\mathbbm{1}|_{ij}=|V_{t^\prime i}V_{t^\prime j}^\ast|\,.
\end{equation}
In Table \ref{tab:UT_corrections}, we collect the entries of $K^{u,d}$ and give the deviations from 3G unitarity in terms of the scaling of the mixing angles $\theta_{i4}$ for the benchmark scenarios introduced in Section~\ref{sec:V4G}. 
We see that, for several unitarity relations, the violation in the SM3 can be of the 
same size as the largest individual contribution from 3G mixing angles.

\begin{table}[t!!!bp]
\renewcommand{\arraystretch}{1.1}
\begin{center}
\begin{tabular}{|rc||c|c|c|c|c|}
\hline
&&$\left|\mathbbm{1}-K\right|_{ij}$&$(a)\,431$&$(b)\,211$&$(c)\,231$&$(d)\,321$\\\hline
\hline
$\vud^2+\vcd^2+\vtd^2=$&$K^d_{11}$&$|V_{t^\prime d}|^2\!\sim\!\lambda^{2n_1}$&$\lambda^{8}$&$\lambda^{4}$&$\lambda^{4}$&$\lambda^{6}$\\
{\color{gray}\small$1\qquad\quad\lambda^2\qquad\quad\!\lambda^6\qquad\!$}&&&&&&\\
$\vus^2+\vcs^2+\vts^2=$&$K^d_{22}$&$|V_{t^\prime s}|^2\!\sim\!\lambda^{2n_2}$&$\lambda^{6}$&$\lambda^{2}$&$\lambda^{6}$&$\lambda^{4}$\\
{\color{gray}\small$\lambda^2\qquad\quad\!1\qquad\quad\ \lambda^4\qquad\!$}&&&&&&\\
$\vub^2+\vcb^2+\vtb^2=$&$K^d_{33}$&$|V_{t^\prime b}|^2\!\sim\!\lambda^{2n_3}$&$\lambda^{2}$&$\lambda^{2}$&$\lambda^{2}$&$\lambda^{2}$\\
{\color{gray}\small$\lambda^6\qquad\quad\!\lambda^4\qquad\quad\! 1\qquad\ $}&&&&&&\\\hline
$\vud^2+\vus^2+\vub^2=$&$K^u_{11}$&$|V_{ub^\prime}|^2\!\sim\!\lambda^{2n_1}$&$\lambda^{8}$&$\lambda^{4}$&$\lambda^{4}$&$\lambda^{6}$\\
{\color{gray}\small$1\qquad\quad\lambda^2\qquad\quad\!\lambda^6\qquad\!$}&&&&&&\\
$\vcd^2+\vcs^2+\vcb^2=$&$K^u_{22}$&$|V_{cb^\prime}|^2\!\sim\!\lambda^{2n_2}$&$\lambda^{6}$&$\lambda^{2}$&$\lambda^{6}$&$\lambda^{4}$\\
{\color{gray}\small$\lambda^2\qquad\quad\!1\qquad\quad\ \lambda^4\qquad\!$}&&&&&&\\
$\vtd^2+\vts^2+\vtb^2=$&$K^u_{33}$&$|V_{tb^\prime}|^2\!\sim\!\lambda^{2n_3}$&$\lambda^{2}$&$\lambda^{2}$&$\lambda^{2}$&$\lambda^{2}$\\
{\color{gray}\small$\lambda^6\qquad\quad\!\lambda^4\qquad\quad\! 1\qquad\ $}&&&&&&\\\hline
$V_{ud}V_{us}^\ast+V_{cd}V_{cs}^\ast+V_{td}V_{ts}^\ast=$&$K^d_{12}$&$|V_{t^\prime d}V_{t^\prime s}^\ast|\!\sim\!\lambda^{n_1+n_2}$&$\lambda^{7}$&$\lambda^{3}$&$\lambda^{5}$&$\lambda^{5}$\\
{\color{gray}\small$\lambda\qquad\quad\ \lambda\qquad\quad\,\lambda^5\qquad\!$}&&&&&&\\
$V_{ud}V_{ub}^\ast+V_{cd}V_{cb}^\ast+V_{td}V_{tb}^\ast=$&$K^d_{13}$&$|V_{t^\prime d}V_{t^\prime b}^\ast|\!\sim\!\lambda^{n_1+n_3}$&$\lambda^{5}$&$\lambda^{3}$&$\lambda^{3}$&$\lambda^{4}$\\
{\color{gray}\small$\lambda^3\qquad\quad\!\lambda^3\qquad\quad\!\lambda^3\qquad\!$}&&&&&&\\
$V_{us}V_{ub}^\ast+V_{cs}V_{cb}^\ast+V_{ts}V_{tb}^\ast=$&$K^d_{23}$&$|V_{t^\prime s}V_{t^\prime b}^\ast|\!\sim\!\lambda^{n_2+n_3}$&$\lambda^{4}$&$\lambda^{2}$&$\lambda^{4}$&$\lambda^{3}$\\
{\color{gray}\small$\lambda^4\qquad\quad\!\lambda^2\qquad\quad\!\lambda^2\qquad\!$}&&&&&&\\\hline
$V_{ud}V_{cd}^\ast+V_{us}V_{cs}^\ast+V_{ub}V_{cb}^\ast=$&$K^u_{12}$&$|V_{ub^\prime}V_{cb^\prime}^\ast|\!\sim\!\lambda^{n_1+n_2}$&$\lambda^{7}$&$\lambda^{3}$&$\lambda^{5}$&$\lambda^{5}$\\
{\color{gray}\small$\lambda\qquad\quad\ \lambda\qquad\quad\,\lambda^5\qquad\!$}&&&&&&\\
$V_{ud}V_{td}^\ast+V_{us}V_{ts}^\ast+V_{ub}V_{tb}^\ast=$&$K^u_{13}$&$|V_{ub^\prime}V_{tb^\prime}^\ast|\!\sim\!\lambda^{n_1+n_3}$&$\lambda^{5}$&$\lambda^{3}$&$\lambda^{3}$&$\lambda^{4}$\\
{\color{gray}\small$\lambda^3\qquad\quad\!\lambda^3\qquad\quad\!\lambda^3\qquad\!$}&&&&&&\\
$V_{cd}V_{td}^\ast+V_{cs}V_{ts}^\ast+V_{cb}V_{tb}^\ast=$&$K^u_{23}$&$|V_{cb^\prime}V_{tb^\prime}^\ast|\!\sim\!\lambda^{n_2+n_3}$&$\lambda^{4}$&$\lambda^{2}$&$\lambda^{4}$&$\lambda^{3}$\\
{\color{gray}\small$\lambda^4\qquad\quad\!\lambda^2\qquad\quad\!\lambda^2\qquad\!$}&&&&&&\\\hline
\end{tabular}
\renewcommand{\arraystretch}{1.0}
\caption{CKM unitarity relations and the amount by which they are broken in the SM4 in the four scaling scenarios introduced in Section~\ref{sec:V4G}. For comparison, in the first column we also give the scaling for the three individual terms on the l.h.s.~of the relation. \label{tab:UT_corrections}}
\end{center}
\end{table}

\boldmath\subsection{$B\to X_s \gamma$ and $B\to X_s \ell^+ \ell^-$}\unboldmath
\begin{figure}[t!!!pb] 
\begin{center}
\includegraphics[width=.48\textwidth]{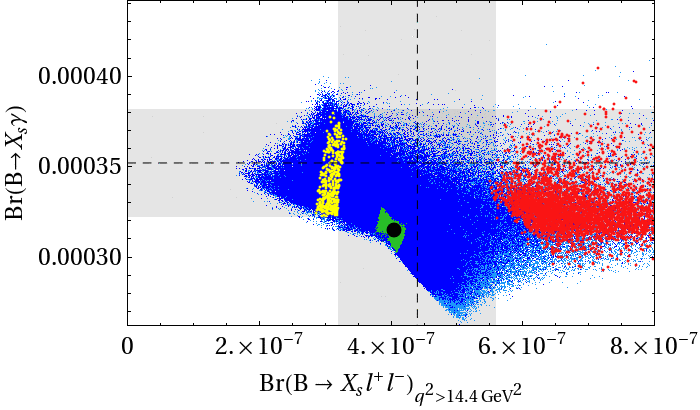}
\end{center}
\vspace{-.5cm}
\caption{${\rm Br}(B\to X_s\gamma)$ as a function of ${\rm Br}(B\to X_s\ell^+\ell^-)_{q^2>14.4{\rm GeV}^2}$.\label{fig:bsgamma-vs-bsll}}
\end{figure} 
 In Fig.~\ref{fig:bsgamma-vs-bsll} we show ${\rm Br}(B\to X_s\gamma)$ versus ${\rm Br}(B\to X_s\ell^+\ell^-)_{q^2>14.4{\rm GeV}^2}$. This plot is self-explanatory and shows that enhancements 
of $S_{\psi\phi}$ and ${\rm Br}(B_s\to \mu^+\mu^-)$ (red points) are fully consistent with the experimental data for the two branching ratios shown in the plot. 
On the other hand the reduction of the experimental error on ${\rm Br}(B\to X_s\ell^+\ell^-)$ could have a significant impact on the red points. 
\update{We confirmed that the zero $s_0$ of the forward-backward asymmetry remains SM3-like. Variations up to $10\%$ are possible, but since this is comparable with the theoretical errors (NLO calculation) we do not further discuss this issue here.}
See Section~\ref{sec:brbxsll} for details.

\newsection{Anatomy}\label{sec:Anatomy}
\subsection{Step 1}
\begin{figure}[t!!!pb]
\includegraphics[width=.48\textwidth, height=.27\textwidth]{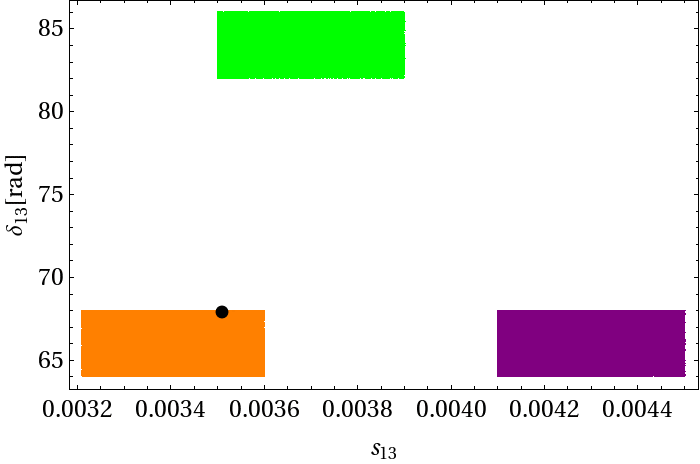}\hspace{.04\textwidth}
\includegraphics[width=.48\textwidth, height=.27\textwidth]{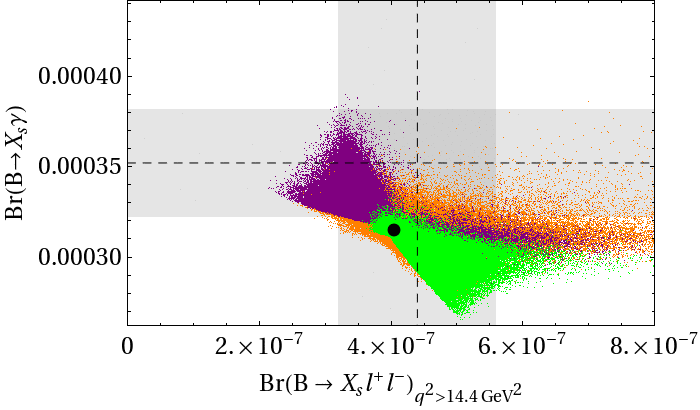}
\caption{We show the colour coding of the tension scenarios (left panel) and ${\rm Br}(B\to X_s\gamma)$ vs. ${\rm Br}(B\to X_s\ell^+\ell^-)_{q^2>14.4{\rm GeV}^2}$ (right panel) for the tension scenarios defined in Tab.~\ref{tab:scenarios}.\label{fig:tensionsColorAndBXsll}}
\end{figure}

\begin{figure}[t!!!pb]
\includegraphics[width=.48\textwidth, height=.27\textwidth]{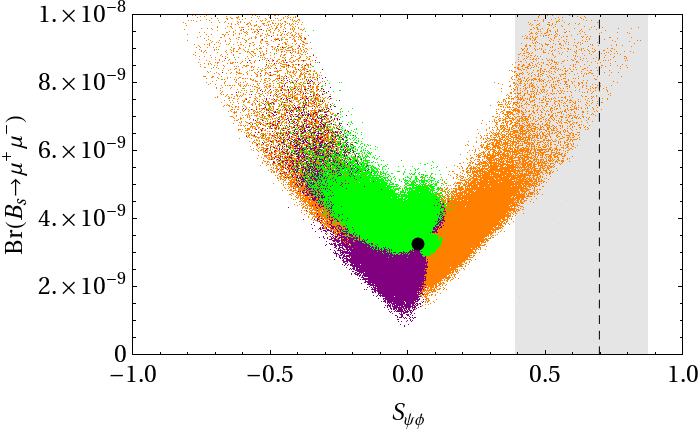}\hspace{.04\textwidth}
\includegraphics[width=.48\textwidth, height=.27\textwidth]{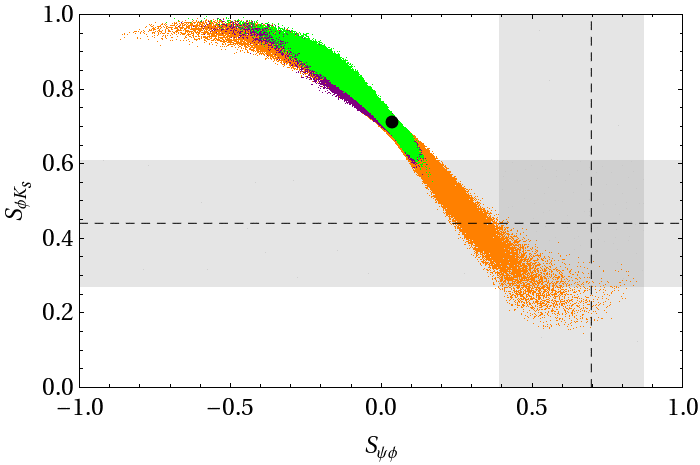}
\caption{We show ${\rm Br}(B_s\to \mu^+\mu^-)$ vs. $S_{\psi\phi}$ (left panel) and $S_{\phi K_s}$ vs. $S_{\psi\phi}$ (right panel) for the tension scenarios defined in Tab.~\ref{tab:scenarios}. The colour coding 
is defined in this table and can be read from the left panel of Fig.~\ref{fig:tensionsColorAndBXsll}.\label{fig:tensionsB}}
\end{figure}

\begin{figure}[t!!!pb]
\includegraphics[width=.48\textwidth]{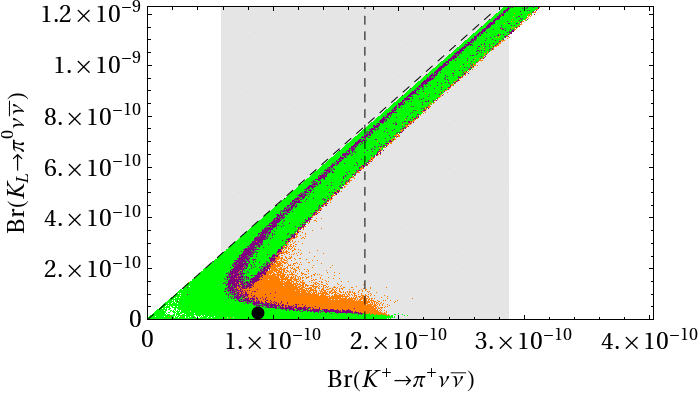}\hspace{.04\textwidth}
\includegraphics[width=.48\textwidth]{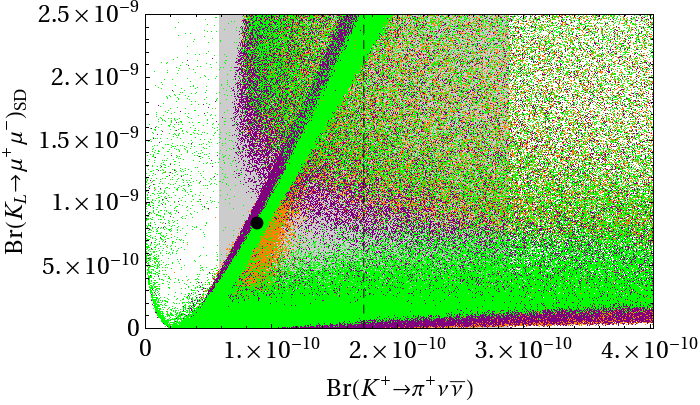}
\caption{We show ${\rm Br}(\klpn)$ vs. ${\rm Br}(\kpn)$ (left panel) and ${\rm Br}(K_L\rightarrow\mu^+\mu^-)_{\rm SD}$ vs. ${\rm Br}(\kpn)$ (right panel) for the tension scenarios defined in Tab.~\ref{tab:scenarios}. The colour coding 
is defined in the table and can be read of Fig.~\ref{fig:tensionsColorAndBXsll} left panel.\label{fig:tensionsK}}
\end{figure}

We begin the anatomy of the SM4 by analysing the three scenarios for 
$\vub$ and $\delta_{13}$ defined in 
Table~\ref{tab:scenarios}.
The three scenarios in question correspond to ones discussed in 
\cite{Altmannshofer:2009ne}
 and can be characterised as follows:
\begin{description}
 \item[S1:] $(\epsilon_K)_{\rm SM}$ is lower than the data, while $S_{\Psi K_S}$ and $\Delta M_d/\Delta M_s$ are compatible with experiment. The {\it orange} points 
 in Fig.~\ref{fig:tensionsColorAndBXsll}-\ref{fig:tensionsK} correspond to the removal of this anomaly within
the SM4.
 \item[S2:] $(S_{\psi K_S})_{\rm SM}$ is above the data, while $\epsilon_K$ and $\Delta M_d/\Delta M_s$ are compatible with experiment. The {\it purple} points 
 in Fig.~\ref{fig:tensionsColorAndBXsll}-\ref{fig:tensionsK} correspond to the removal of this anomaly within the SM4.
 \item[S3:] $(\Delta M_d/\Delta M_s)_{\rm SM}$ is much higher than the data, while $\epsilon_K$ and $S_{\Psi K_S}$ are compatible with experiment. 
The {\it green} points 
 in Fig.~\ref{fig:tensionsColorAndBXsll}-\ref{fig:tensionsK} correspond to the removal of this anomaly within the SM4.
\end{description}

\begin{table}[t!!!pb] 
\begin{center}
\begin{tabular}{|c||c|c|c|}
\hline
 		& S1 (orange)	& S2 (purple)	& S3 (green)\\ \hline\hline
$\vub$	& $0.0034\pm 0.00015$& $0.0043\pm 0.0001$ & $0.0037\pm0.0001$ 	\\ \hline
$\delta_{13}$	& $(66\pm 2)^\circ$	& $(66\pm 2)^\circ$& 
$(84\pm 2)^\circ$ 	\\ \hline
\end{tabular}
\caption{Three scenarios for the parameters $s_{13}$ and $\delta_{13}$\label{tab:scenarios}}
\end{center}
\end{table}

The clear lessons from this analysis are as follows:
\begin{itemize}
\item
 Due to the 4G contributions to $\varepsilon_K$ and $S_{\psi K_S}$ in all 
scenarios simultaneous agreement with the data for these two observables
can be achieved taking all existing constraints into account.
\item
However, only in scenario S1 values of $S_{\psi\phi}$ can be significantly 
enhanced and consequently $S_{\phi K_S}$ and $S_{\eta'K_S}$ significantly 
suppressed. As an example, we show in Fig.~\ref{fig:tensionsB} (right panel) this situation by plotting 
$S_{\phi K_S}$ as a function of $S_{\psi\phi}$.
\item
As shown in Fig.~\ref{fig:tensionsK}, in all three scenarios significant enhancements 
of ${\rm Br}(\kpn)$, ${\rm Br}(\klpn)$ and ${\rm Br}(K_L\to\mu^+\mu^-)$ are possible.
\item
On the other hand, as shown in the lower panel of Fig.~\ref{fig:tensionsB} (right panel), 
the departure of ${\rm Br}(B_s\to\mu^+\mu^-)$ can be up to a factor of 4 for all
three scenarios. However we found no points with large positive $S_{\psi\phi}$ for S2 and S3
which in turn puts a loose upper limit on ${\rm Br}(B_s\to\mu^+\mu^-)$  in these scenarios provided $S_{\psi\phi}>0$.
\end{itemize}

\subsection{Step 2}

In the next three years, LHCb should be able to provide good data on 
$S_{\psi\phi}$ and ${\rm Br}(B_s\to\mu^+\mu^-)$. As we have seen in 
 Fig.~\ref{fig:Spsiphi-Bmumu}, a measurement of $S_{\psi\phi}$ above 0.5 
accompanied by ${\rm Br}(B_s\to\mu^+\mu^-)_\text{exp}\le {\rm Br}(B_s\to\mu^+\mu^-)_{SM}$ would put 
the SM4 under pressure. Similarly for SM3-like values of $S_{\psi\phi}$, 
${\rm Br}(B_s\to\mu^+\mu^-)$ can only be slightly enhanced over its SM3 value, 
and in fact a suppression of the latter branching ratio is more likely 
in this case.

In view of this pattern, we considered three scenarios shown in 
Table~\ref{tab:Bscenarios}
and asked what they would imply for other decays. The result of this
exercise is shown in all plots of Section~\ref{sec:numerics} where
 the three scenarios of 
Table~\ref{tab:Bscenarios} are exhibited 
in three different colours indicated in this table. 
 Alternatively, the colour coding can be conveniently inferred from Fig.~\ref{fig:Spsiphi-Bmumu}.
 This results are 
self-explanatory  and have been briefly discussed already in Section~\ref{sec:numerics}. Let us  then only summarise our  observations:

\begin{itemize}
\item As seen in Fig.~\ref{fig:CP-asymmetries}, the $S_{\phi K_S}$ and $S_{\psi K_S}$ anomalies can only be explained in scenario BS3.
\item As seen in Fig.~\ref{fig:Bmumu}, in the BS1 and BS3 scenarios the branching ratio ${\rm Br}(B_d\to\mu^+\mu^-)$
remains SM3-like, while in scenario BS2 it can be enhanced by a factor of two.
\item
As seen in Figs.~\ref{fig:Kpinunu} and \ref{fig:Kpinunu-KLmumu}
 in all three scenarios large NP effects in ${\rm Br}(\kpn)$, ${\rm Br}(\klpn)$
 and ${\rm Br}(K_L\to\mu^+\mu^-)$ are possible. Moreover in scenario BS3 they are particularly 
strongly correlated with each other.
\item
 For large positive values of $S_{\psi\phi}$ the predicted value of $\epe$ is significantly below the data, unless the hadronic matrix elements of 
 the electroweak penguins are sufficiently suppressed with respect to the large $N$ result and the ones of QCD penguins enhanced.
\end{itemize}

This analysis shows that we will learn a lot  about the
SM4 when $S_{\psi\phi}$, ${\rm Br}(B_s\to\mu^+\mu^-)$, ${\rm Br}(\kpn)$ and ${\rm Br}(\klpn)$ will be precisely known.


\boldmath\newsection{Determining the $V_\text{SM4}$ matrix}\label{sec:DetV4G}\unboldmath

\subsection{Preliminaries}

 As discussed in  previous sections, the mixing between the third and fourth generation 
is bounded by the electroweak precision data and cannot be significantly larger than 
$s_{12}$. Similarly, $s_{14}$ and $s_{24}$ are bounded by (\ref{eq:sij_bounds}).
As a consequence, as can be explicitly seen by the generalised Wolfenstein
expansion for our benchmark scenarios (\ref{eqn:v4gapprox431}--\ref{eqn:v4gapprox321}),
the relation between CKM parameters and the matrix elements 
$V_{us},V_{cd},V_{ub},V_{cb}$,
\begin{align}
s_{12} & =\lambda  \simeq  |V_{us}| \simeq |V_{cd}| \,, \quad 
& 
s_{23} &= A\lambda^2 \simeq |V_{cb}| \,,\cr 
s_{13} &= A\lambda^3 |z_\rho| \simeq |V_{ub}| \,, \quad
&
\delta_{13} & = - \text{arg} (z_\rho)  \simeq - \text{arg} (V_{ub}) \,,
\label{eqn:smfundparam} 
\end{align}
are to a very good approximation unaffected by 4G contributions.
Therefore the SM3 CKM parameters can be determined from 
the corresponding tree-level decays, practically 
without  any NP pollution, with $\delta_{13}$ determined from $B\rightarrow DK$. 
In particular such a
determination does not require the $3\times 3$ CKM matrix to be unitary.\footnote{Needless
to say, $|V_{us}|,|V_{ub}|,|V_{cb}|$ and $\delta_{13}$ can be determined from tree-level 
decays even in the absence of this approximation, but the relation to
the fundamental parameters is less transparent and involves the 4G parameters.}
In the present section we will further use (\ref{eqn:relvs}) for the 
4G mixing angles, in
order to indicate how the $V_{\rm SM4}$ matrix can, in principle,
be determined from future data.

A further
comment on the determination of $\delta_{13}\approx\gamma$ is in order.

     In the approximation of neglecting the phase of $V_{cs}$ and $V_{cb}$,
     being real in the CKM convention, the $B_s\to D_s K$ complex in the SM3
     measures directly a linear combination of the phase $\gamma$ and the phase 
     of $B_s$-mixing, where the latter can be extracted from $S_{\psi\phi}$.
 Analogous comments apply to $B_d\to D\pi$
     decays. In the presence of 4G quarks, the phases in $B_s$ and $B_d$ 
     mixing may change, but again they can be determined from the $S_{\psi\phi}$
     and $S_{\psi K_S}$ asymmetries, respectively. The new feature, as seen 
     in (\ref{eqn:v4g}), are new
     phases of $V_{cb}$ and $V_{cs}$ induced by the presence of the 4G quarks.
     However, the imposition of tree-level and electroweak precision 
     constraints implies that this NP pollution amounts to significantly
     less than $1^\circ$ in the determination of $\delta_{13}$ and can be 
     safely neglected. Analogous comments apply to other tree-level methods 
     for the determination of $\delta_{13}$.

The determination of the new parameters in the $V_\text{SM4}$ matrix,
\begin{equation}
s_{14}\,, \qquad s_{24}\,, \qquad s_{34}\,, \qquad \delta_{14}\,, \qquad \delta_{24}\label{eqn:bsmfundparam} \,,
\end{equation}
is probably beyond the  scope of flavour-violating high-energy processes 
explored at the LHC and 
will have to be made through FCNC processes that, as in the SM3, appear 
first at the one-loop level due to the GIM mechanism at work.
The accuracy of this determination will 
depend on
\renewcommand{\labelenumi}{\roman{enumi})}
\begin{enumerate}
\item the precision of the relevant experimental data,
\item the theoretical cleanliness of the observables involved
 (i.e.\ observables with small
hadronic uncertainties should be favoured),
\item the  potential size of NP contributions to the considered observable.
\end{enumerate}
\renewcommand{\labelenumi}{\arabic{enumi}}
Clearly, the room for NP contributions to observables that are known already 
precisely will depend on the values of the CKM parameters in (\ref{eqn:smfundparam}).
In this context, let us recall the existing tension between the experimental values of
$\varepsilon_K$ and $S_{\psi K_S}$ within the SM3 that has been extensively discussed 
in \cite{Lunghi:2008aa,Buras:2008nn,Buras:2009pj,Lunghi:2009sm}.
 Whether the SM3 has a problem with $\varepsilon_K$, 
$S_{\psi K_S}$ or both observables depends on the values in (\ref{eqn:smfundparam}).
In turn, this will 
have an impact on the determination of the new parameters in (\ref{eqn:bsmfundparam}).
In what follows, we will first make a list of observables that could help us in the future
to determine the parameters (\ref{eqn:bsmfundparam}), subsequently illustrating such determinations on a few
examples. A more extensive general numerical analysis appears to us to be premature at present. On the other hand, 
as we will see in Section~\ref{Sec:anatomy_scenarios}, new insight can be gained by  examining the anatomy of  different scaling scenarios
for the mixing angles and their implications for various flavour observables.

\subsection{Basic Observables}

Among the FCNC observables that have already been measured, 
the values for
\begin{equation}
\varepsilon_K\,, \qquad \Delta M_{d}/\Delta M_s\,, \qquad S_{\psi K_S}\,,\qquad
{\rm Br}\left( B\rightarrow X_s\gamma \right) 
\end{equation}
have presently the dominant impact on the 
allowed structure of the $V_\text{SM4}$ matrix. 
Indeed, $S_{\psi K_S}$ is
theoretically very clean 
(see also the comment 
before Eq.~(\ref{eq:lttprimeformerfootnote})),
and the hadronic uncertainties in $\varepsilon_K$ and
$\Delta M_{d}/\Delta M_s$ are below $5\%$ already now and 
 are expected to be decreased further in the
coming years through improved lattice calculations.

Of particular interest in this decade will be the measurements of the branching 
ratios for
\begin{equation}
B_{s,d}\rightarrow \mu^+\mu^-\,,\qquad K^+\rightarrow\pi^+\nu\bar\nu\,,\qquad K_L\rightarrow\pi^0\nu\bar\nu\,,\qquad B\rightarrow X_s\nu\bar\nu\,,
\end{equation}
and, very importantly, of the CP-violating observables
\begin{equation}
S_{\psi\phi}\,,\qquad S_{\phi K_S}\,,\qquad A_\text{CP}(b\to s\gamma)\,.
\end{equation}
 In particular, various correlations between all these observables will significantly constrain the allowed range of the SM4 parameters and even have the power
to exclude this NP scenario.
 Assuming that the SM4 will survive these new tests and having at hand all these measurements,
 it will be possible to determine the matrix $V_\text{SM4}$.
Indeed, let us note on the basis of the formulae of 
Section~\ref{sec:MasterFormulae} that all these 
observables depend on six complex variables involving new parameters (see (\ref{eqn:4gunitaritydef})),
\begin{equation}
\lambda^{(K)}_{t'}\,,\qquad\lambda^{(s)}_{t'}\,,\qquad\lambda^{(d)}_{t'}\,,\qquad \lambda^{(K)}_{t}\,,\qquad\lambda^{(s)}_{t}\,,\qquad\lambda^{(d)}_{t}\,,
\end{equation}
that are not fully independent as, with (\ref{eqn:smfundparam}) being fixed, 
they depend on the five parameters in (\ref{eqn:bsmfundparam}).
It is instructive to write down the expressions for $\lambda^{(i)}_{t^{(')}}$ setting 
$c_{ij}=1$  and neglecting higher-order terms in the Wolfenstein expansion, 
leading to 
\begin{align}
\label{eqn:lambdadtpapprox}
\lambda_{t'}^{(d)}= V^*_{t'b}V_{t'd}
&\approx 
\left(s_{34}+s_{13} s_{14} e^{i \left(\delta _{13}-\delta _{14}\right)}+s_{23} s_{24} e^{-i \delta _{24}}\right)
\cr &\qquad \times
(e^{i \delta_{14}} s_{14} -
   e^{i \delta_{24}} s_{12} s_{24} - 
   e^{i \delta_{13}} s_{13} s_{34} + s_{12} s_{23} s_{34})
\cr
&\approx -e^{i \delta_{13}}s_{13} s_{34}^2 +s_{34} \left(s_{14} e^{i \delta_{14}}-s_{12} s_{24} e^{i \delta_{24}}\right)
\cr &\qquad
+s_{14} s_{23} s_{24} e^{i \left(\delta
   _{1,4}-\delta_{24}\right)} 
 -s_{12} s_{23} \left(s_{24}^2-s_{34}^2\right) \,,\\
\label{eqn:lambdastpapprox}
\lambda_{t'}^{(s)}= V^*_{t'b}V_{t's}
&\approx 
\left(s_{34}+s_{13} s_{14} e^{i \left(\delta _{13}-\delta _{14}\right)}+s_{23} s_{24} e^{-i \delta _{24}}\right)
\cr &\qquad \times
(e^{i \delta_{24}} s_{24} + e^{i \delta_{14}} s_{12} s_{14}  - s_{23} s_{34})
\cr &\approx 
s_{12} s_{14} s_{34} e^{i \delta_{14}}+s_{24} s_{34} e^{i \delta_{24}}+s_{23} \left(s_{24}^2-s_{34}^2\right)\,,\\
\label{eqn:lambdaktpapprox}
\lambda_{t'}^{(K)}= V^*_{t's}V_{t'd}
& \approx 
(e^{-i \delta_{24}} s_{24} + e^{-i \delta_{14}} s_{12} s_{14}  - s_{23} s_{34}) 
\cr & \qquad \times
(e^{i \delta_{14}} s_{14} -
   e^{i \delta_{24}} s_{12} s_{24} - 
   e^{i \delta_{13}} s_{13} s_{34} + s_{12} s_{23} s_{34})\cr
&\approx  s_{24} \left(-s_{12} s_{24}+s_{14} e^{i \left(\delta _{14}-\delta _{24}\right)}\right)-s_{24} s_{34} e^{-i \delta _{24}} \left(-s_{12} s_{23}+s_{13} e^{i \delta
   _{13}}\right) 
\cr &\qquad +\left(s_{12} s_{24} e^{i \delta _{24}}-s_{14} e^{i \delta _{14}}\right) \left(s_{23} s_{34}-s_{12} s_{14} e^{-i \delta_{14}}\right)
\cr &\approx
\frac{\lambda_{t'}^{(s)\ast}\lambda_{t'}^{(d)}}{|V_{t'b}|^2}\,,\\
\label{eqn:lambdadtapprox}
\lambda_{t}^{(d)}= V^*_{tb}V_{td}
&\approx 
-s_{13} e^{i \delta_{13}}-s_{14} s_{34} e^{i \delta_{14}}+s_{12} \left(s_{23}+s_{24} s_{34} e^{i \delta_{24}}\right)
\cr&\approx
 -s_{13} e^{i \delta_{13}}+s_{12} s_{23} - \lambda_{t'}^{(d)}\,, \\
\label{eqn:lambdastapprox}
\lambda_{t}^{(s)}= V^*_{tb}V_{ts}
&\approx 
-s_{23}-s_{24} s_{34} e^{i \delta_{24}}
\cr&\approx
-s_{23} - \lambda_{t'}^{(s)}\,, \\
\label{eqn:lambdaktapprox}
\lambda_{t}^{(K)}= V^*_{ts}V_{td}
& \approx 
\left(s_{23}+s_{24} s_{34} e^{-i \delta_{24}}\right)
\cr & \qquad \times \left(s_{13} e^{i \delta_{13}}+s_{14} s_{34} e^{i \delta_{14}}-s_{12} \left(s_{23}+s_{24} s_{34}
   e^{i \delta_{24}}\right)\right)
\cr &\approx 
\lambda_{t}^{(s)\ast} \lambda_{t}^{(d)}\,.
\end{align}

We observe that, in general, the variables $\lambda_{t^{(')}}^{(K,s,d)}$ 
involve all 5 mixing parameters associated with the 4G in a rather complicated
way. Some simplification arises if we assume a certain scaling of the mixing
angles $\theta_{i4}$, for instance with one of our 
benchmark scenarios of Section~\ref{sec:V4G}. Let us, as an example,
study the case (\ref{default2}), for which the above equations simplify
as follows:
\begin{align}
\lambda_{t'}^{(s)}
&\approx 
e^{i \delta_{24}} s_{24} s_{34}  \equiv \sigma_{23} \,,\cr
\lambda_{t'}^{(d)}
&\approx 
s_{34} \left(s_{14} e^{i \delta_{14}}-s_{12} s_{24} e^{i \delta_{24}}\right) \equiv \sigma_{13} - s_{12} \sigma_{23}
\cr
\lambda_{t'}^{(K)}
& \approx 
s_{24} \left(-s_{12} s_{24}+s_{14} e^{i \left(\delta_{14}-\delta_{24}\right)}\right)= \frac{1}{s_{34}^2} \sigma_{23}^\ast\ (\sigma_{13}-s_{12} \sigma_{23}) 
\quad [= \frac{1}{s_{34}^2} \lambda_{t'}^{(s)\ast} \lambda_{t'}^{(d)}]\,,\cr
\lambda_{t}^{(s)}
&\approx 
-s_{23}-s_{24} s_{34} e^{i \delta_{24}} = -s_{23} - \lambda_{t'}^{(s)} \,,\cr
\lambda_{t}^{(d)}
&\approx 
 -s_{13} e^{i \delta_{13}}-s_{14} s_{34} e^{i \delta_{14}}+s_{12} \left(s_{23}+s_{24} s_{34} e^{i \delta_{24}}\right)
=  -s_{13} e^{i \delta_{13}}+s_{12} s_{23} - \lambda_{t'}^{(d)} \,,\cr
\lambda_{t}^{(K)}
& \approx 
 \left(s_{23}+s_{24} s_{34} e^{-i \delta_{24}}\right) \left(s_{13} e^{i \delta_{13}}+s_{14} s_{34} e^{i \delta_{14}}-s_{12} \left(s_{23}+s_{24} s_{34}
   e^{i \delta_{24}}\right)\right)\cr
&=\left(s_{23}+\lambda_{t'}^{(s)\ast}\right) \left(s_{13} e^{i \delta_{13}}-s_{12} s_{23} + \lambda_{t'}^{(d)}\right)\,,
\end{align}
where the following two combinations of new parameters (\ref{eqn:bsmfundparam}) have been introduced:
\begin{align}
 \sigma_{13} \equiv s_{14} s_{34} e^{i \delta_{14}}\,,\qquad
 \sigma_{23} \equiv s_{24} s_{34} e^{i \delta_{24}}\,.
\end{align}
\renewcommand{\labelenumi}{\roman{enumi})}
For this particular case, we observe that
\begin{enumerate}
\item $\lambda^{(K)}_{t}$ and $\lambda^{(K)}_{t'}$ depend on $\sigma_{13}$, $\sigma_{23}$ and $s_{34}$.

\item $\lambda^{(d)}_{t}$ and $\lambda^{(d)}_{t'}$ depend on $\sigma_{13}$ and $\sigma_{23}$.

\item $\lambda^{(s)}_{t}$ and $\lambda^{(s)}_{t'}$ are sensitive to $\sigma_{23}$ only.
The determination of $\Delta M_s$, ${\rm Br}(B_s\rightarrow \mu^+\mu^-)$
and $S_{\psi\phi}$ will thus play an important role in constraining 
these parameters. Moreover for $\szv \gtrsim \szd\sdv$
and $\dzv$ not too small, $S_{\psi\phi}$ should be much larger than its SM3 value 
$\left(S_{\psi\phi}\right)_{SM}\approx 0.04$ as we have seen in Section~\ref{sec:numerics}.
\end{enumerate}
\renewcommand{\labelenumi}{\arabic{enumi}}

\begin{figure}[t!!!pb] 
\includegraphics[width=.48\textwidth]{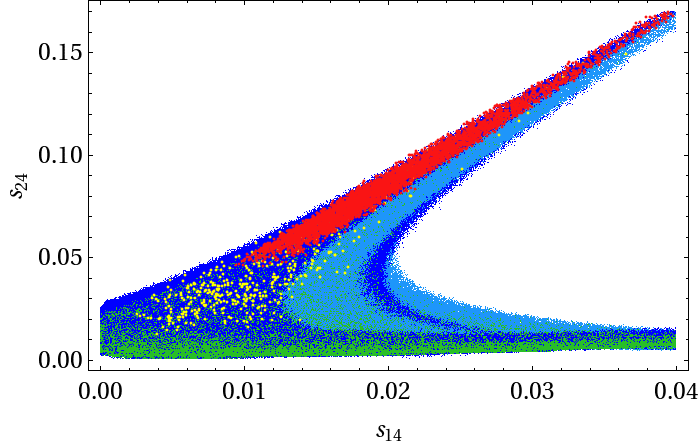}\hspace{.03\textwidth}
\includegraphics[width=.48\textwidth]{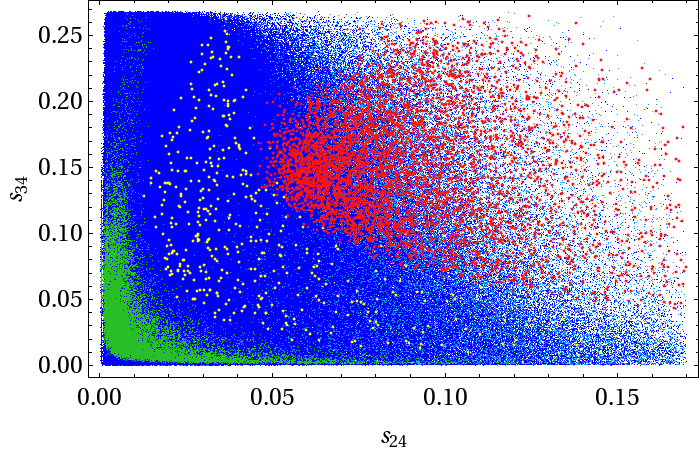}\\[1.5em]
\includegraphics[width=.48\textwidth]{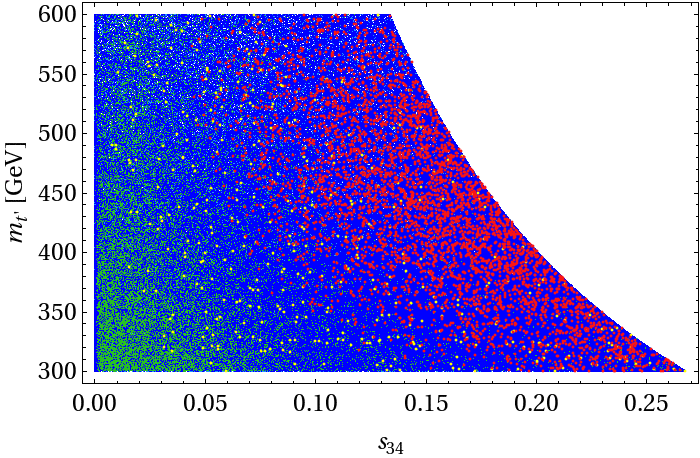}\hspace{.03\textwidth}
\includegraphics[width=.48\textwidth]{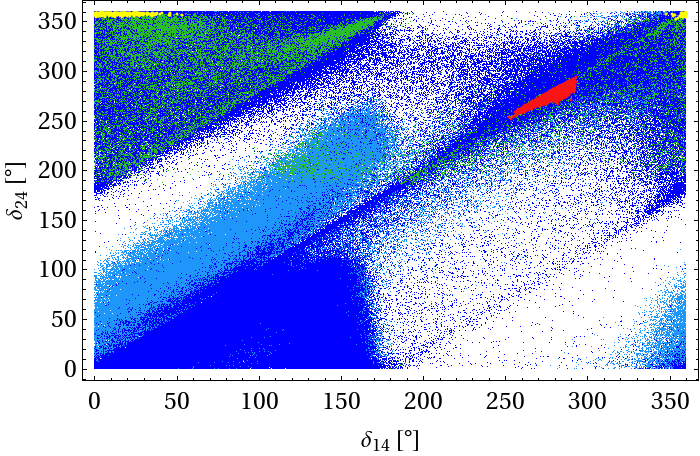}
\caption{Correlations between the new parameters for the parameter points used in our global analysis. 
The mixing angle $s_{24}$ as a function of $s_{14}$ (upper left panel), $s_{34}$ as a function of $s_{24}$ (upper right panel), $m_{t'}$ as a function of $s_{34}$ (lower left panel) and
$\delta_{24}$ as a function of $\delta_{14}$ (lower right panel). The colour coding corresponds to scenarios of  Table~\ref{tab:Bscenarios}.} \label{fig:newparam}
\end{figure}

In view of the many possibilities for theoretical parameters and
phenomenological observables, 
we will in the following develop a general procedure that allows us to
analyse future experimental data in a systematic manner.
Let us first consider the plots in Fig.~\ref{fig:newparam}, which show 
correlations between various parameters of the SM4, on the
basis of the three scenarios of Table~\ref{tab:Bscenarios}.
In particular,
\renewcommand{\labelenumi}{\roman{enumi}}
\begin{itemize}
\item in the case of the scenario BS3 of Table~\ref{tab:Bscenarios},
 there is a strong correlation between
$s_{24}$ and $s_{14}$ with $s_{24}\sim 4s_{14}$ and $s_{24}$ in the range
\begin{align}
&0.046 \leq s_{24} \leq 0.17\,.
\end{align}
\item $\delta_{14}$ and $\delta_{24}$ are also very strongly correlated in the 
BS3 scenario with $\delta_{14}\approx\delta_{24}\approx 270^\circ \pm 20^\circ$.
\item \update{in the lower left panel of Fig.~\ref{fig:newparam} one can see the strong correlation (\ref{eq:sij_bounds}) of $s_{34}$ and $m_{t^\prime}$. We checked numerically
that treating the EWPT in this approximate manner indeed gives reasonable bounds on $m_{t^\prime}$.}
\end{itemize}
\renewcommand{\labelenumi}{\arabic{enumi}}
 In the following section, we will analyse these correlations 
within specific scaling scenarios for the 4G mixing angles.


\subsection{Anatomy of 4G mixing angles and CP phases}

\label{Sec:anatomy_scenarios}

As explained in Section~\ref{sec:V4G}, the 4G mixing matrix allows for
different scalings of the 4G mixing angles $\theta_{14}$,
$\theta_{24}$ and $\theta_{34}$ with the Wolfenstein parameter
$\lambda\ll 1$. 
Among the set of parameters that fulfil the present 
constraints on flavour observables,
we may thus classify different subsets by calculating the exponents
\begin{equation}
 (n_1,n_2,n_3) = \mbox{round}\left(\log_\lambda \left[\theta_{14},\theta_{24},\theta_{34} \right]\right) \,,
\end{equation}
in (\ref{eq:wolf}) for $\lambda=s_{12} \ll 1$. 
 Furthermore (see also Fig.~\ref{fig:newparam}), for a given set $(n_1,n_2,n_3)$, the allowed values for the new CP phases $\delta_{14}$ and $\delta_{24}$
also exhibit a more or less strong correlation, which we will exploit
to further distinguish different regions in the 4G parameter space.
This enables us to identify a set of equivalence classes which
-- as we will illustrate -- share certain
characteristic features for the corresponding predictions
for various flavour observables.

\begin{table}[t!!!pb]
 
\begin{center}
 \begin{tabular}{c c c}
  $(n_1,n_2,n_3)$ & Correlation $\delta_{24}$ vs.\ $\delta_{14}$ & Assignment
\\
\hline &&
\\
\parbox[c]{0.1\textwidth}{ 
$(2,1,1)$ \\ $(2,2,1)$
}
 & 
 \parbox[c]{0.3\textwidth}{\includegraphics[width=0.3\textwidth]{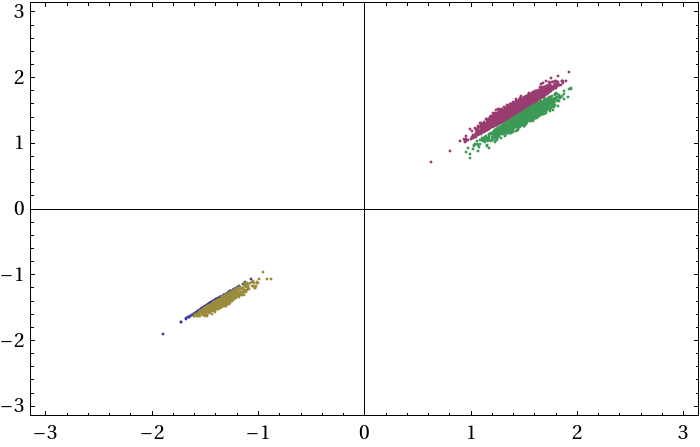}} & 
 \parbox[c]{0.5\textwidth}{
\footnotesize
$\to \left\{ 
\begin{array}{ll}
 \mbox{\color{blue} class 1a:} &
 \color{black} 
  \delta_{14} < \delta_{24} < \delta_{14} + \frac{\pi}{8},  \quad \delta_{24}<0.
\\
 \mbox{\color{brown} class 1b:} & 
 \color{black} 
  \delta_{14}-\frac{\pi}{8} < \delta_{24} < \delta_{14}, \quad \delta_{24} <0. 
\\ \mbox{\color{magenta} class 2a:} & 
 \color{black} 
 \delta_{14} < \delta_{24} < \delta_{14} + \frac{\pi}{8},  \quad \delta_{24} >0.
\\ \mbox{\color{green} class 2b:} & 
 \color{black}
 \delta_{14}-\frac{\pi}{8} < \delta_{24} < \delta_{14}, \quad \delta_{24}>0.
 \end{array} \right.$}
\\[0.1em]
\parbox[c]{0.1\textwidth}{ 
$(3,2,1)$ 
} &
 \parbox[c]{0.3\textwidth}{\includegraphics[width=0.3\textwidth]{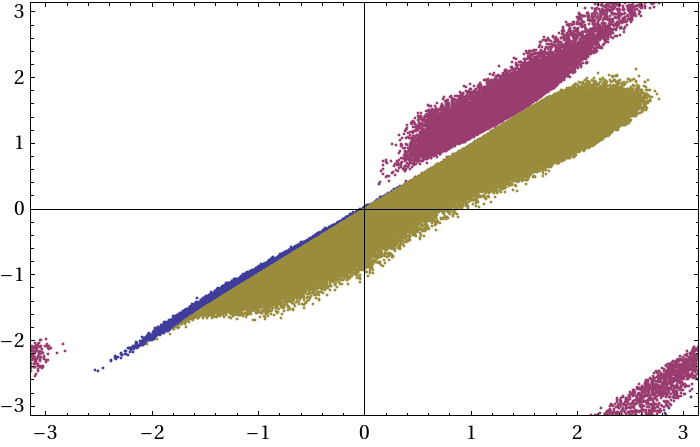}} & 
 \parbox[c]{0.5\textwidth}{
\footnotesize 
$\to \left\{ \begin{array}{ll}
\mbox{\color{blue} class 1a:} & 
\color{black} 
  \delta_{14} < \delta_{24} < \delta_{14} + \frac{\pi}{8},  \quad \delta_{24}<0.
\\ 
\mbox{\color{magenta} class 2a/3a:} & 
 \color{black} 
 \delta_{14} < \delta_{24} < \delta_{14} + \frac{3\pi}{8},  \ \ \delta_{24}>0.
\\ 
\mbox{\color{brown} class 3b:} & 
 \color{black} 
 \delta_{14}-\frac{3\pi}{8} < \delta_{24} < \delta_{14} . 
 \end{array} \right.$}
\\
\parbox[c]{0.1\textwidth}{ 
$(3,3,1)$ \\ $(3,3,2)$ 
} &
 \parbox[c]{0.3\textwidth}{\includegraphics[width=0.3\textwidth]{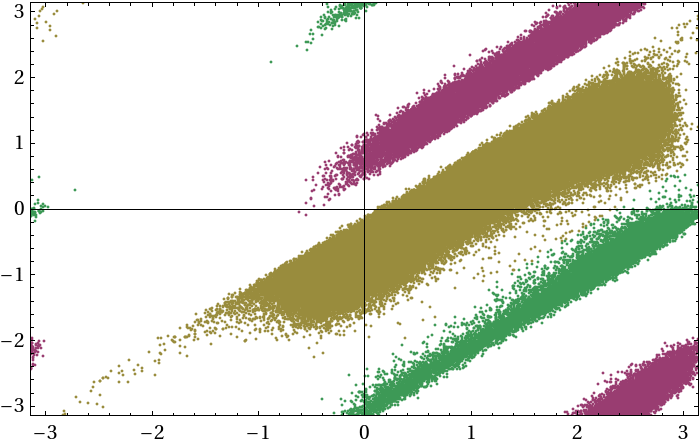}} & 
 \parbox[c]{0.5\textwidth}{
\footnotesize $\to \left\{ \begin{array}{ll}
\mbox{\color{magenta} class 3a:} & 
\color{black} 
 \delta_{14}+\frac{\pi}{8} < \delta_{24} < \delta_{14} + \frac{\pi}{2}. 
\\ \mbox{\color{brown} class 3b:} & 
 \delta_{14}-\frac{3\pi}{4} < \delta_{24} < \delta_{14} . 
\\ \mbox{\color{green} class 4:} & 
 \delta_{14}-\frac{9\pi}{8} < \delta_{24} < \delta_{14}-\frac{3\pi}{4} . 
 \end{array} \right.$}
\\
\parbox[c]{0.1\textwidth}{ 
$(4,3,1)$ \\ $(4,3,2)$
}
 & 
 \parbox[c]{0.3\textwidth}{\includegraphics[width=0.3\textwidth]{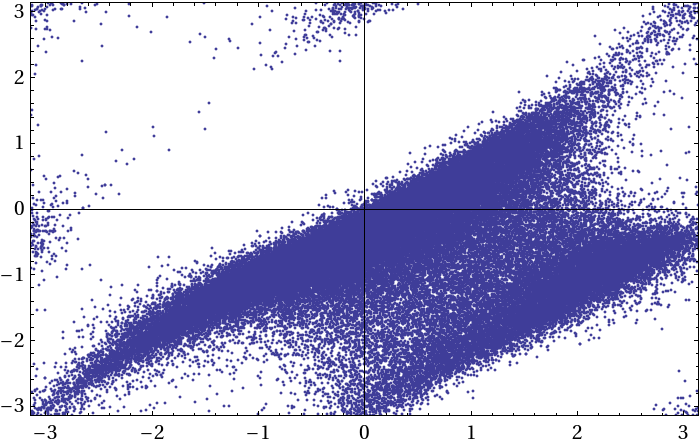}} & 
 \parbox[c]{0.45\textwidth}{
\footnotesize $\to \left\{ \begin{array}{ll}
 \mbox{\color{blue} class 5:} 
 & \delta_{14}-\frac{3\pi}{2} < \delta_{24} < \delta_{14}+\frac{\pi}{4}   
 \end{array} \right.$}
\\
\parbox[c]{0.1\textwidth}{ 
$(2,3,1)$ 
} &
 \parbox[c]{0.3\textwidth}{\includegraphics[width=0.3\textwidth]{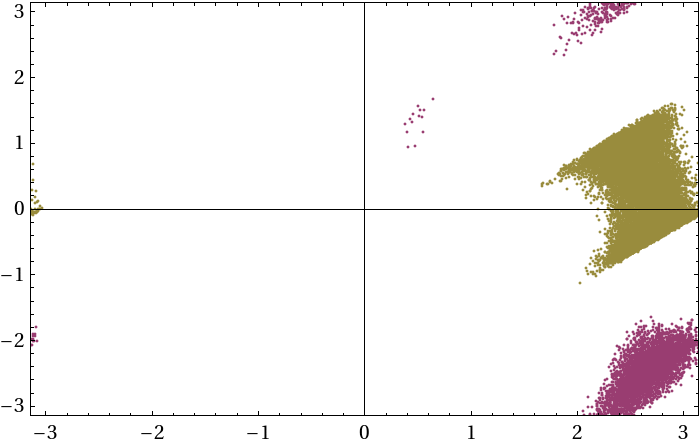}} & 
 \parbox[c]{0.45\textwidth}{
\footnotesize $\to \left\{ \begin{array}{ll}
\mbox{\color{magenta} class 6:} & 
 \delta_{14}+\frac{\pi}{8} < \delta_{24} < \delta_{14} + \frac{3\pi}{4}.
\\ \mbox{\color{brown} class 7:} & 
 \delta_{14}-\frac{9\pi}{8} < \delta_{24} < \delta_{14}-\frac{\pi}{4} . 
 \end{array} \right.$}
\\
 \end{tabular}
\caption{\label{tab:procedure} Correlations between 4G phases for different
scalings of 4G mixing angles (some selected examples). The constraints
on the phases, of course, are understood to be periodic in units of $2\pi$.}
\end{center}

\end{table}

According to our discussion above, values $(n_1,n_2,n_3)$
that do not fulfil the inequalities (\ref{consist}), from the theoretical
point of view, should be considered as
fine-tuned and will not be included in the subsequent discussion.
(We convinced ourselves that we do not miss any of the observed
 phenomenological features by doing so.)
For the remaining cases, we first identify -- for a given
scaling $(n_1,n_2,n_3)$ -- the allowed correlations between $\delta_{14}$
and $\delta_{24}$, and then determine the assignment to one or the
other equivalence class.
Our procedure is summarised for the most prominent and representative
examples in Table~\ref{tab:procedure}:
\begin{itemize}
\item
Evidently, for small values of the $n_i$ (i.e.\ large values of 4G mixing angles),
one expects the largest deviations from the SM3 predictions. In fact, the most extreme
case is characterised by example (\ref{default2}) from
Section~\ref{sec:V4G}, with values $(n_1,n_2,n_3) = (2,1,1)$. 
As we already discussed, in this scenario the new 4G phases contribute at leading
order to CP-violating observables, and therefore $\delta_{14}$ and $\delta_{24}$
are highly correlated and constrained. Separating the different relative
and absolute signs of $\delta_{i4}$ enables us to classify the subsets $1a,1b,2a,2b$
which show characteristic features in the selected set of observables shown in Tables~\ref{tab:mixanatomy1}+\ref{tab:mixanatomy2}.
Similar correlations are found for the $(2,2,1)$ scenario.

\item Decreasing the values of (some of) the mixing angles (i.e.\ increasing $n_1,n_2,n_3$),
we observe that the correlations between the 4G phases become
broader. Still, the values populate restricted areas in the $\delta_{24}$--$\delta_{14}$
plane, which again allows to identify sub-classes with definite properties. 
A typical example is the case $(3,2,1)$ from our benchmark scenario (\ref{default4}).
This scenario divides into two well-separated regions.
Among one of them, we may (or may not) identify a subset of points 
as belonging to class~1a. This kind of arbitrariness
is unavoidable (and expected), since the separation of points from scenario
$(2,2,1)$ and $(3,2,1)$ is not clear-cut.

\item For even smaller values of mixing angles, we observe 
for the cases $(3,3,1)$ and $(3,3,2)$ a separation into three sub-classes,
where two classes (3a and 3b) 
can be considered as a continuation of those from the $(3,2,1)$
scenario, and class~4 is new. 

\item Considering $(4,3,1)$ (the benchmark
scenario (\ref{default1}) discussed in Section~\ref{sec:V4G}) 
and $(4,3,2)$, the former
classes~3b and 4 merge into one class~5 which already covers  around half
of the $\delta_{24}$-$\delta_{14}$ plane. 
 Finally, for the scenario (\ref{default3}) with the scaling
$(2,3,1)$, the former class 3a continues into class~6,
while 3b and 4 merge into class~7, where the parameter space
in the $\delta_{24}$-$\delta_{14}$ plane is again somewhat more constrained.

\end{itemize}

Having identified the different sub-classes, we investigate the characteristic
features for certain phenomenological observables in Tables~\ref{tab:mixanatomy1}-\ref{tab:mixanatomy4}.
We see that each class can be distinguished from the others by at least
one of the shown correlations: 
${\rm Br}(B_s\to \mu^+\mu^-)$ vs.\ $S_{\psi\phi}$,
$S_{\phi K_S}$ vs.\ $S_{\psi\phi}$,
${\rm Br}(b\to s\gamma)$ vs.\ ${\rm Br}(B \to X_s\ell\ell)$,
${\rm Br}(K_L \to \mu^+\mu^-)_{\rm SD}$ vs.\ ${\rm Br}(K^+ \to \pi^+\nu\bar\nu)$
(The colour-coding is as defined in Table~\ref{tab:Bscenarios}).

Turning the argument around, the observation of a combination of
particular correlations in rare flavour processes can be translated
into one or the other favoured scenario for 4G mixing angles and CP~phases.
Only after a particular scaling scenario has been identified, 
the formulae for $\lambda_{t'}^{(K,d,s)}$ in (\ref{eqn:lambdaktpapprox}--\ref{eqn:lambdastapprox})
can be simplified, and we may (more or less) unambiguously determine
the 4G mixing parameters from the future experimental data.
We emphasise that, quite generally, such a procedure should be applied to
the analysis of flavour parameters in NP models without MFV. In particular,
without a specific theory of flavour at hand,
the fact that certain scaling scenarios -- like (\ref{default1}) -- are represented
by only a small number of points in the overall scan of parameter space, should not
be taken as a signal for a small probability to observe the associated correlations
in flavour observables.

\begin{table}[t!!!bp]
 \begin{center}
 \begin{tabular}{ l |ccc  }
Class
&   $ 10^8 \!\cdot\! {\rm Br}(B_s \!\to\! \mu^+\mu^-) \ 
            \mbox{vs.} \  S_{\psi\phi} $
&   $S_{\phi K_S} \ 
            \mbox{vs.} \ 
             S_{\psi\phi}$
&
\\
\hline \hline 
&&& \\[-0.7em]
1a:
& \parbox[c]{0.4\textwidth}{\includegraphics[width=0.38\textwidth]{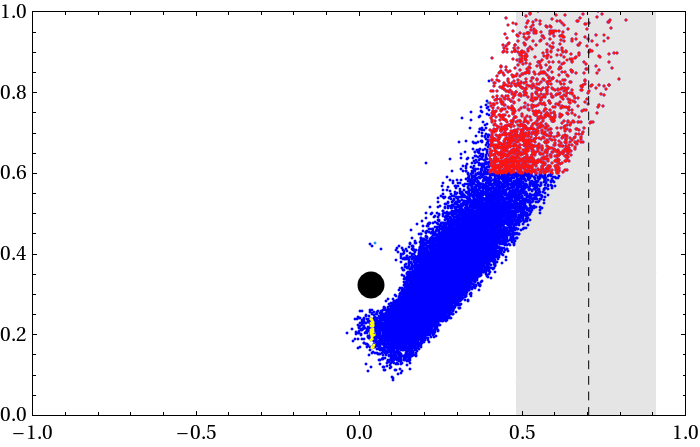}}
& \parbox[c]{0.4\textwidth}{\includegraphics[width=0.38\textwidth]{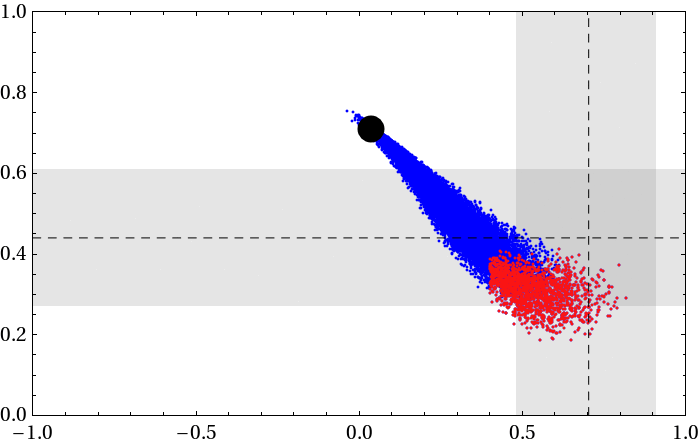}}
& $\cdots$
\\
 1b:
& \parbox[c]{0.4\textwidth}{\includegraphics[width=0.38\textwidth]{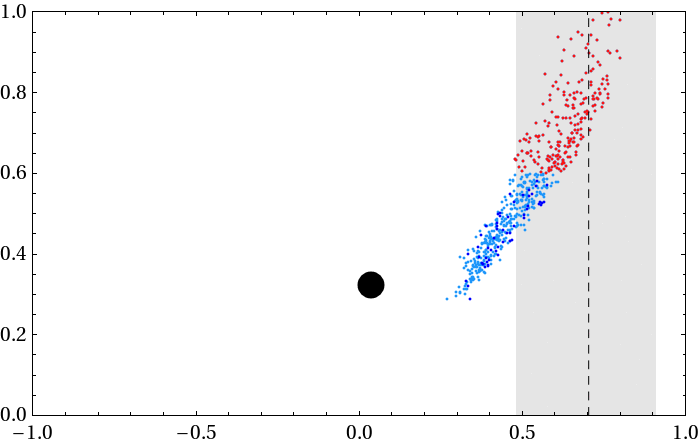}}
& \parbox[c]{0.4\textwidth}{\includegraphics[width=0.38\textwidth]{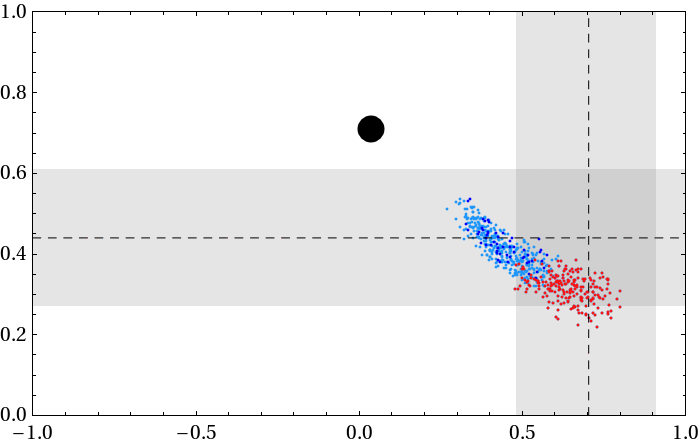}}
& $\cdots$
\\
 2a: 
& \parbox[c]{0.4\textwidth}{\includegraphics[width=0.38\textwidth]{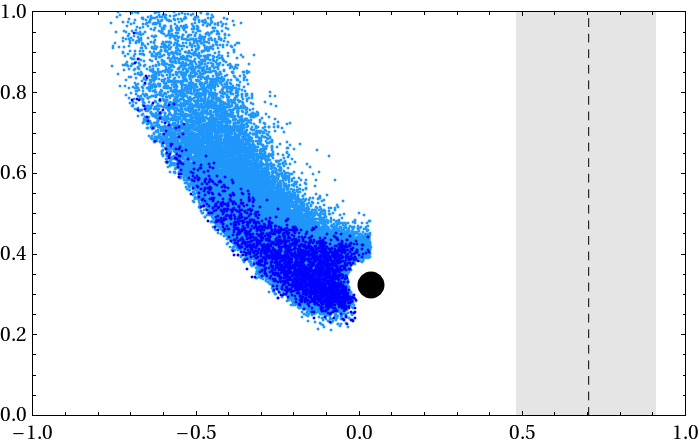}}
& \parbox[c]{0.4\textwidth}{\includegraphics[width=0.38\textwidth]{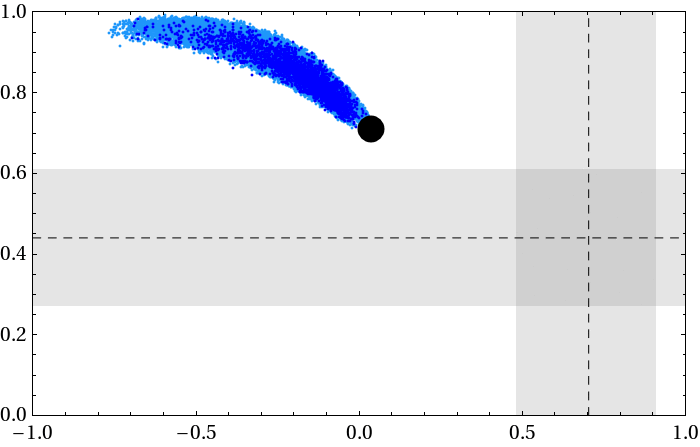}}
& $\cdots$
\\
 2b: 
& \parbox[c]{0.4\textwidth}{\includegraphics[width=0.38\textwidth]{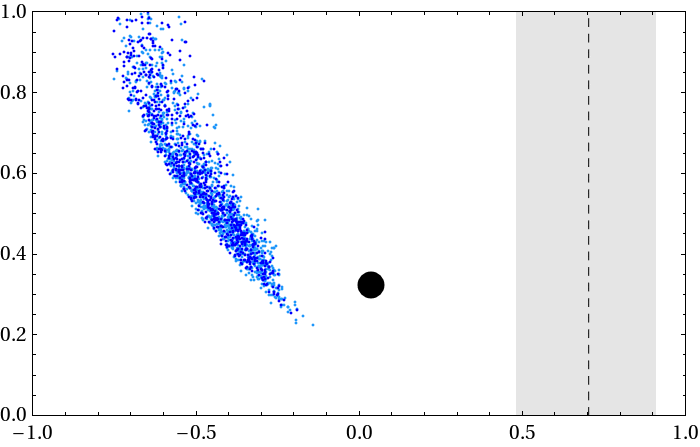}}
& \parbox[c]{0.4\textwidth}{\includegraphics[width=0.38\textwidth]{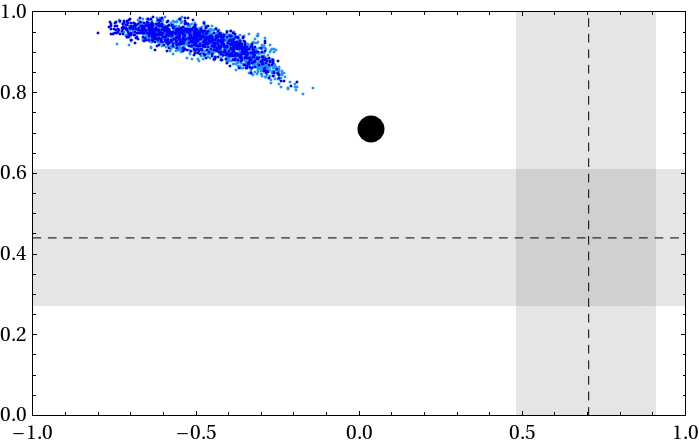}}
& $\cdots$
\\
 3a: 
& \parbox[c]{0.4\textwidth}{\includegraphics[width=0.38\textwidth]{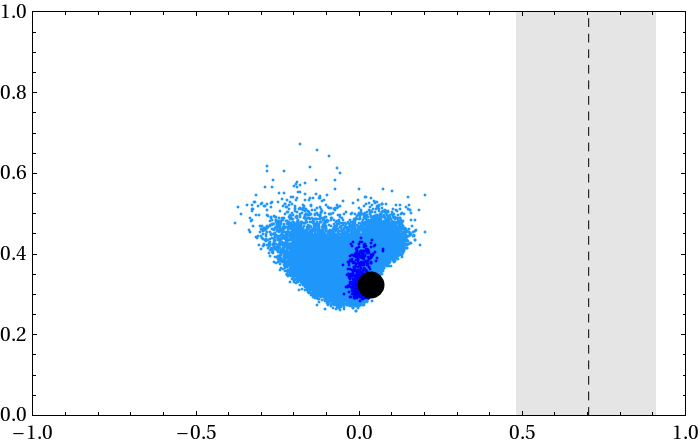}}
& \parbox[c]{0.4\textwidth}{\includegraphics[width=0.38\textwidth]{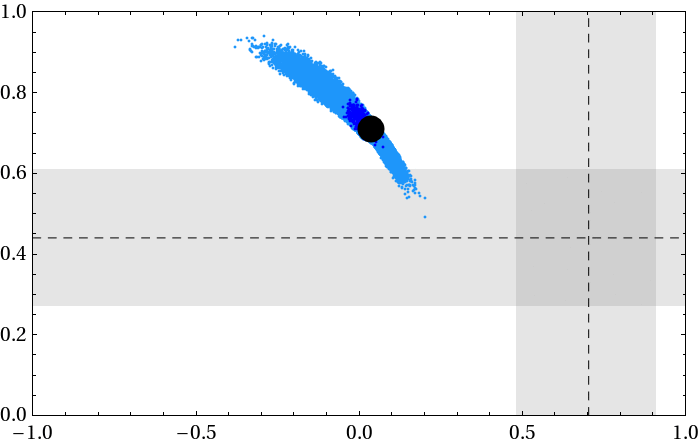}}
& $\cdots$ \\
& \ldots & \ldots 
\end{tabular}
\end{center}
\caption{\label{tab:mixanatomy1} Selected correlations for classes identified in Table~\ref{tab:procedure} (part 1 of 4).
{(The colour-coding is as defined in Table~\ref{tab:Bscenarios}).}}
\end{table}


\begin{table}[t!!!bp]
 \begin{center}
 \begin{tabular}{ ccc | r }
   & $\begin{array}{l} 10^4 \!\cdot \!{\rm Br}(b\!\to\! s\gamma) \\
    \mbox{vs.} \  
          10^6 \!\cdot \!{\rm Br}(B \!\to\! X_s\ell\ell)_{q^2>14.4\gev^2} \end{array} $
   & 
    $\begin{array}{l} 10^9 \!\cdot \!{\rm Br}(K_L \!\to\! \mu^+\mu^-)_{\rm SD} \\ \mbox{vs.} \
           10^{10}\!\cdot \!{\rm Br}(K^+ \!\to\! \pi^+\nu\bar\nu) \end{array}$
 & Class
\\
\hline \hline 
&&& \\[-0.7em]
  $\cdots$
& \parbox[c]{0.4\textwidth}{\includegraphics[width=0.38\textwidth]{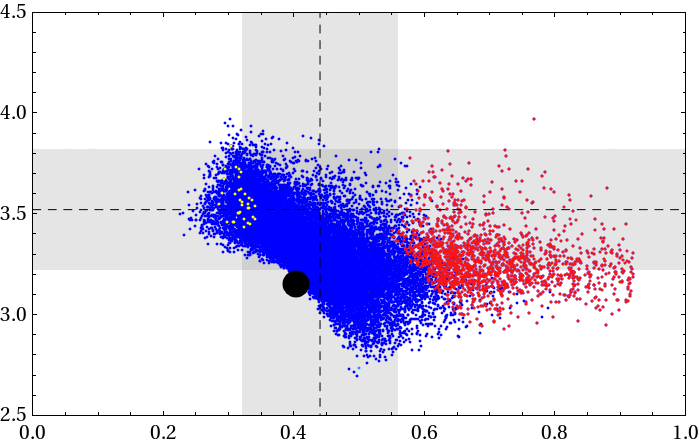}}
& \parbox[c]{0.4\textwidth}{\includegraphics[width=0.38\textwidth]{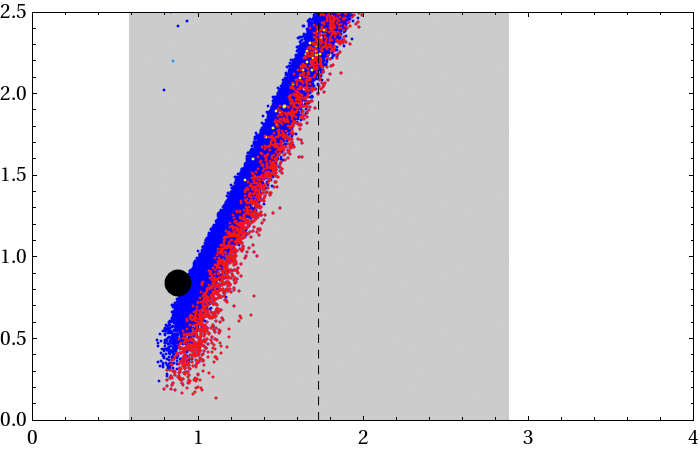}}
& 1a
\\
$\cdots$  
& \parbox[c]{0.4\textwidth}{\includegraphics[width=0.38\textwidth]{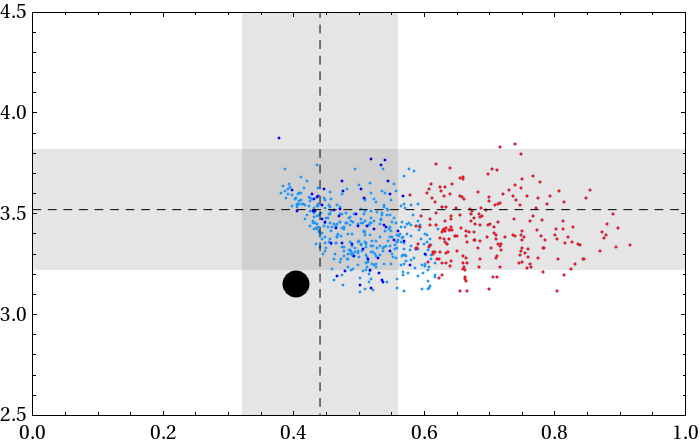}}
& \parbox[c]{0.4\textwidth}{\includegraphics[width=0.38\textwidth]{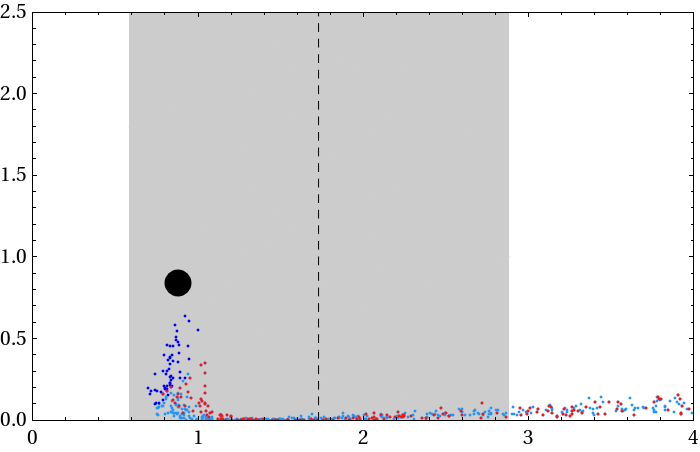}}
& 1b
\\
$\cdots$  
& \parbox[c]{0.4\textwidth}{\includegraphics[width=0.38\textwidth]{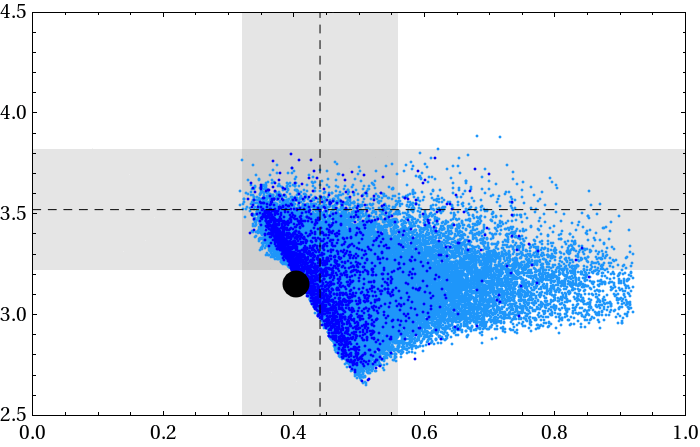}}
& \parbox[c]{0.4\textwidth}{\includegraphics[width=0.38\textwidth]{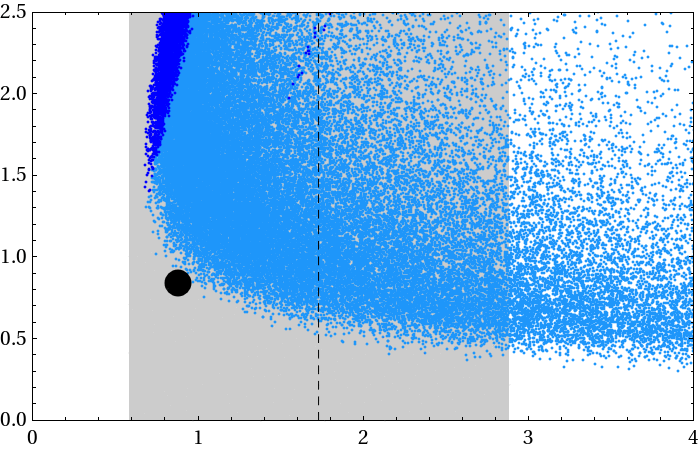}}
& 2a
\\
$\cdots$  
& \parbox[c]{0.4\textwidth}{\includegraphics[width=0.38\textwidth]{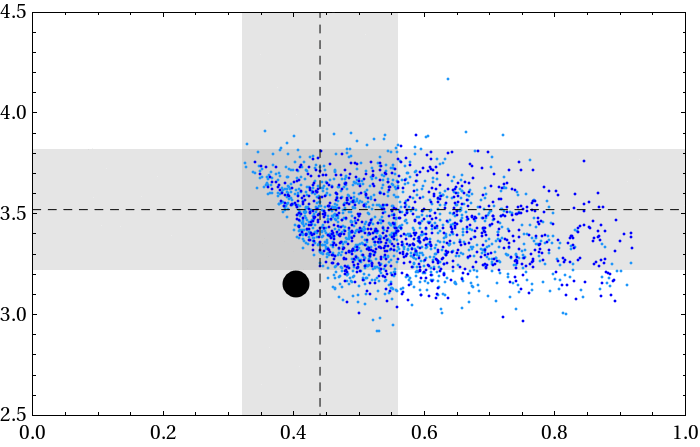}}
& \parbox[c]{0.4\textwidth}{\includegraphics[width=0.38\textwidth]{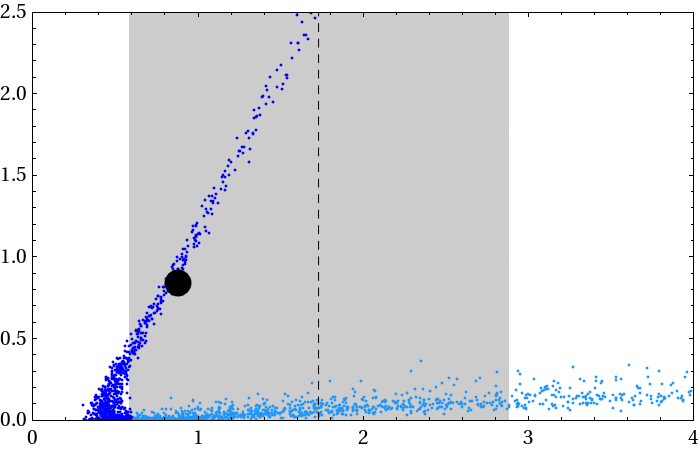}}
& 2b
\\
$\cdots$  
& \parbox[c]{0.4\textwidth}{\includegraphics[width=0.38\textwidth]{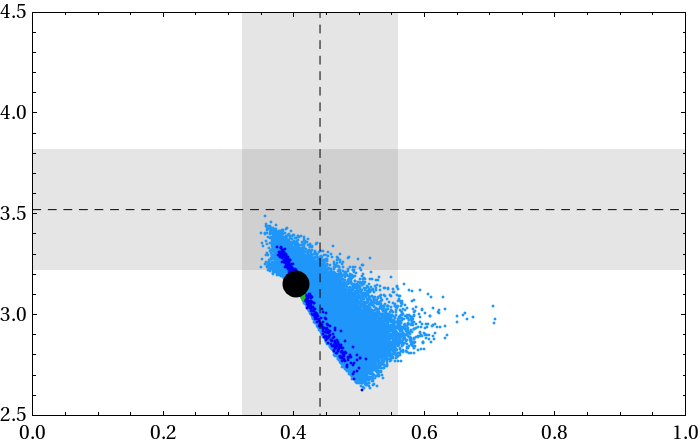}}
& \parbox[c]{0.4\textwidth}{\includegraphics[width=0.38\textwidth]{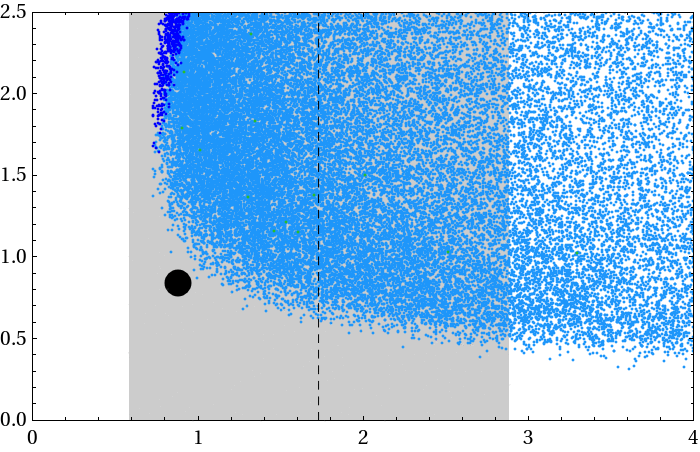}}
& 3a
\\
& \ldots & \ldots & 
  \end{tabular}
 \end{center}
\caption{\label{tab:mixanatomy2}
Selected correlations for classes identified in Table~\ref{tab:procedure} 
(part 2 of 4).}
\end{table}


\begin{table}[t!!!bp]
 \begin{center}
 \begin{tabular}{ l | ccc }
Class
   & $ 10^8 \!\cdot\! {\rm Br}(B_s \!\to\! \mu^+\mu^-) \
            \mbox{vs.} \ 
             S_{\psi\phi}$
   & $ S_{\phi K_S} \ 
            \mbox{vs.} \ 
             S_{\psi\phi}$
&
\\
\hline \hline 
&&& \\[-0.7em]
& \ldots & \ldots & 
\\
 3b: 
& \parbox[c]{0.4\textwidth}{\includegraphics[width=0.38\textwidth]{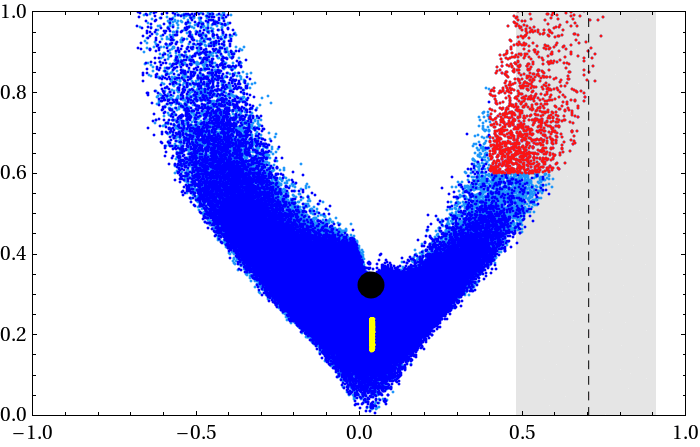}}
& \parbox[c]{0.4\textwidth}{\includegraphics[width=0.38\textwidth]{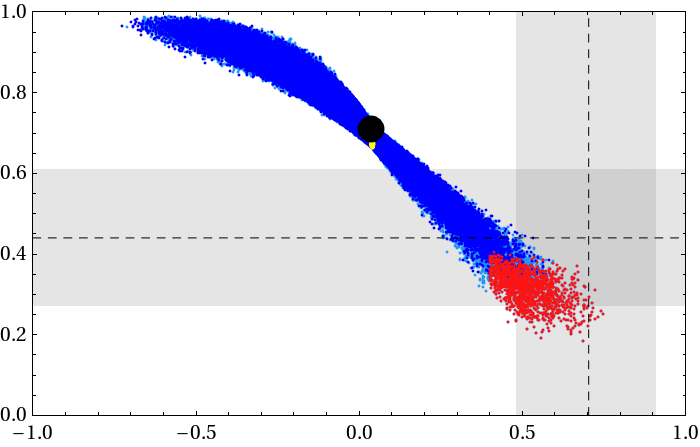}}
& $\cdots$
\\
 4: 
& \parbox[c]{0.4\textwidth}{\includegraphics[width=0.38\textwidth]{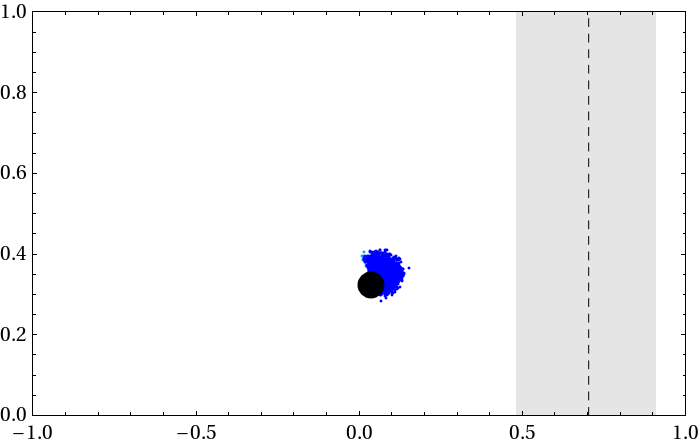}}
& \parbox[c]{0.4\textwidth}{\includegraphics[width=0.38\textwidth]{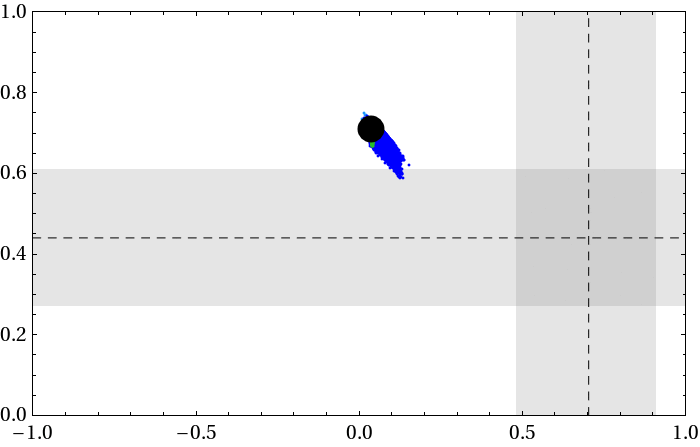}}
& $\cdots$
\\
 5: 
& \parbox[c]{0.4\textwidth}{\includegraphics[width=0.38\textwidth]{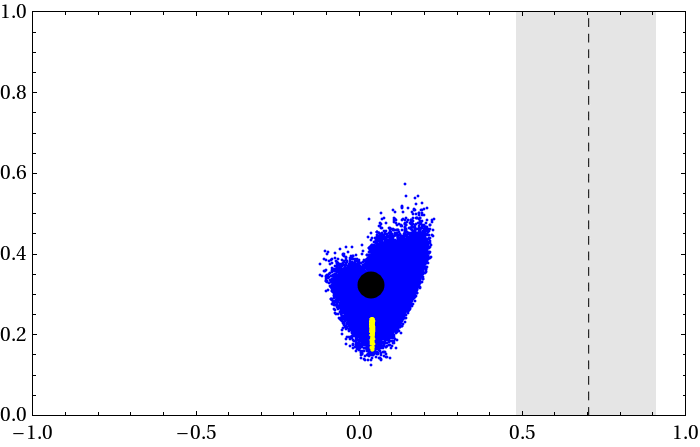}}
& \parbox[c]{0.4\textwidth}{\includegraphics[width=0.38\textwidth]{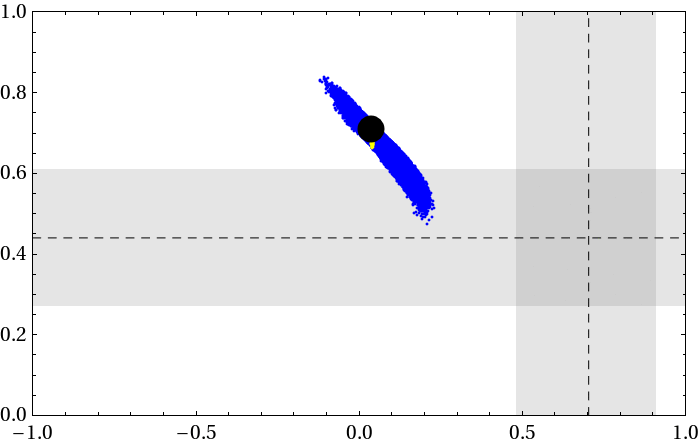}}
& $\cdots$
\\
 6: 
& \parbox[c]{0.4\textwidth}{\includegraphics[width=0.38\textwidth]{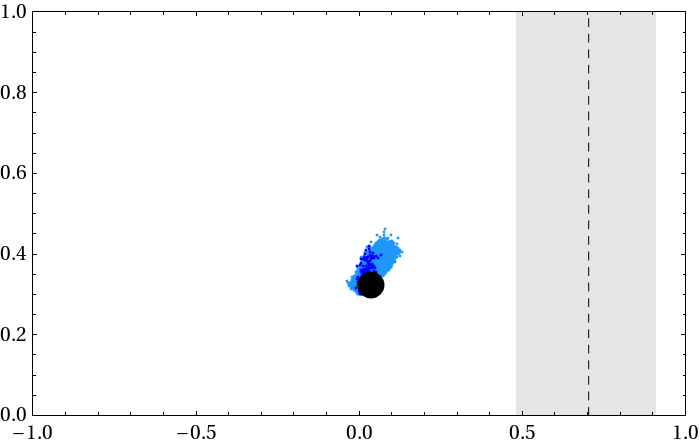}}
& \parbox[c]{0.4\textwidth}{\includegraphics[width=0.38\textwidth]{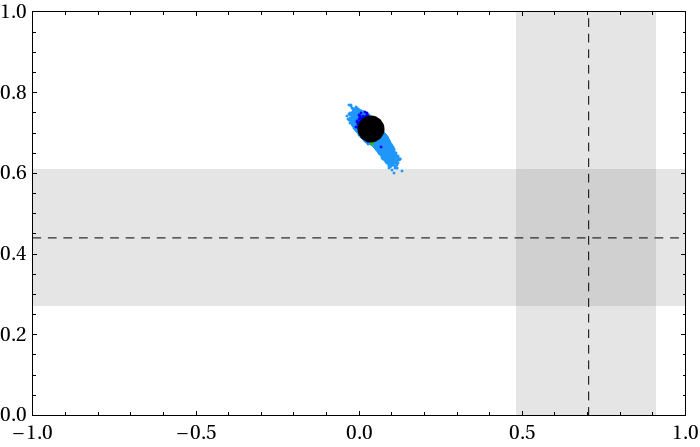}}
& $\cdots$
\\
 7: 
& \parbox[c]{0.4\textwidth}{\includegraphics[width=0.38\textwidth]{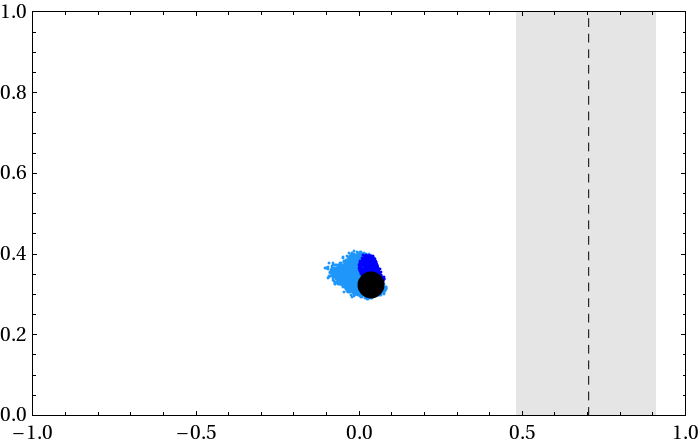}}
& \parbox[c]{0.4\textwidth}{\includegraphics[width=0.38\textwidth]{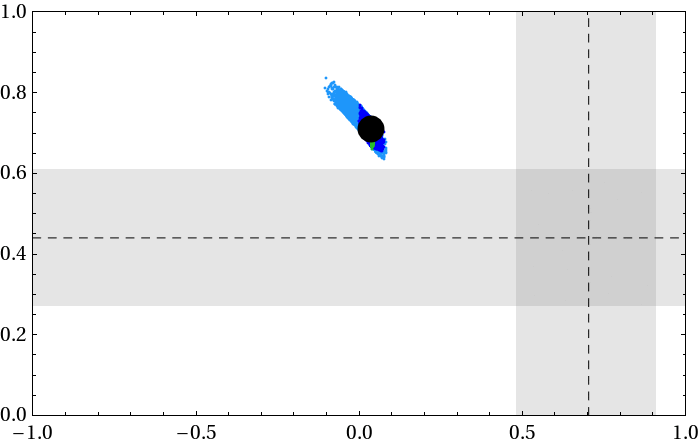}}
  \end{tabular}
 \end{center}
\caption{\label{tab:mixanatomy3}
Selected correlations for classes identified in Table~\ref{tab:procedure}
(part 3 of 4).}
\end{table}


\begin{table}[t!!!bp]
 \begin{center}
 \begin{tabular}{ ccc | r }
   & $\begin{array}{l} 10^4 \!\cdot \!{\rm Br}(b\!\to\! s\gamma) \\
    \mbox{vs.} \  
          10^6 \!\cdot \!{\rm Br}(B \!\to\! X_s\ell\ell)_{q^2>14.4\gev^2} \end{array} $
   & 
  $\begin{array}{l} 10^9 \!\cdot \!{\rm Br}(K_L \!\to\! \mu^+\mu^-)_{\rm SD} \\ \mbox{vs.} \
           10^{10}\!\cdot \!{\rm Br}(K^+ \!\to\! \pi^+\nu\bar\nu) \end{array}$
& Class
\\
\hline \hline 
&&& \\[-0.7em]
& \ldots & \ldots & 
\\
$\cdots$
& \parbox[c]{0.4\textwidth}{\includegraphics[width=0.38\textwidth]{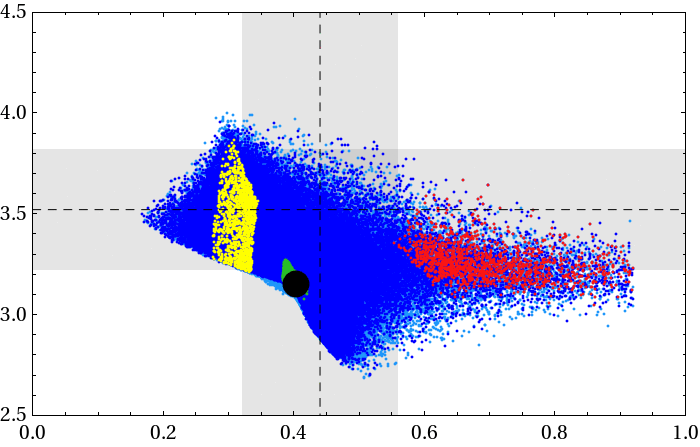}}
& \parbox[c]{0.4\textwidth}{\includegraphics[width=0.38\textwidth]{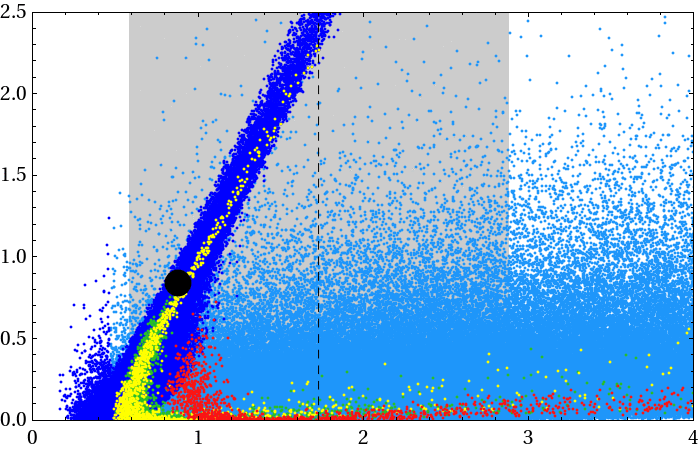}}
& 3b
\\
$\cdots$
& \parbox[c]{0.4\textwidth}{\includegraphics[width=0.38\textwidth]{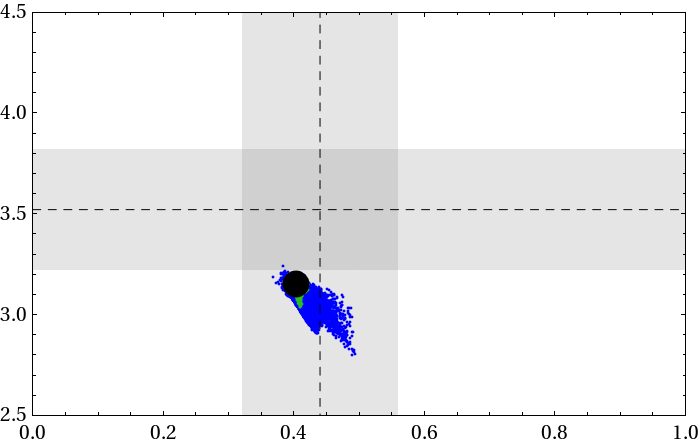}}
& \parbox[c]{0.4\textwidth}{\includegraphics[width=0.38\textwidth]{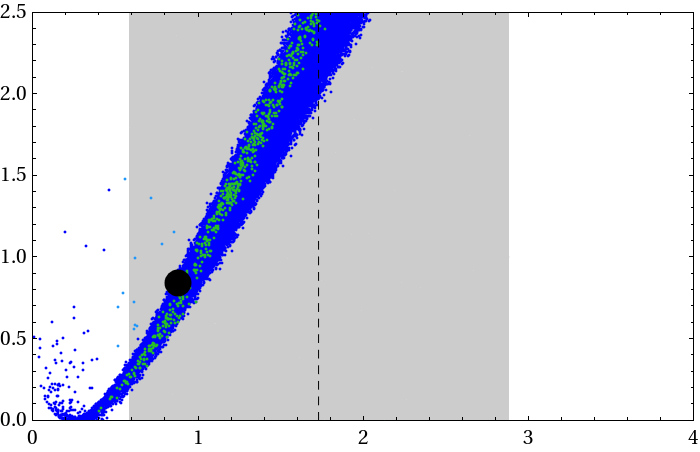}}
& 4
\\
$\cdots$
& \parbox[c]{0.4\textwidth}{\includegraphics[width=0.38\textwidth]{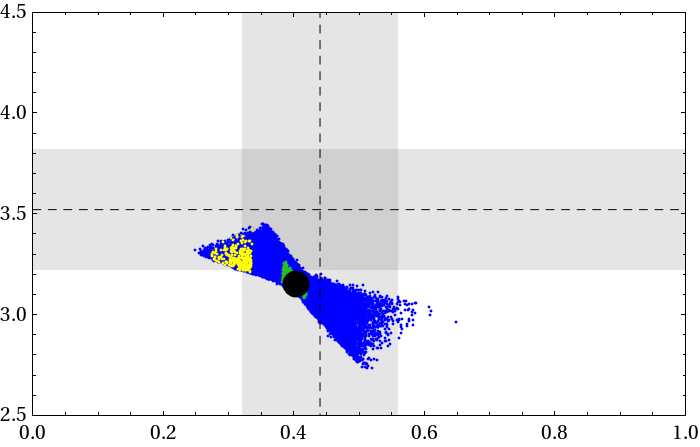}}
& \parbox[c]{0.4\textwidth}{\includegraphics[width=0.38\textwidth]{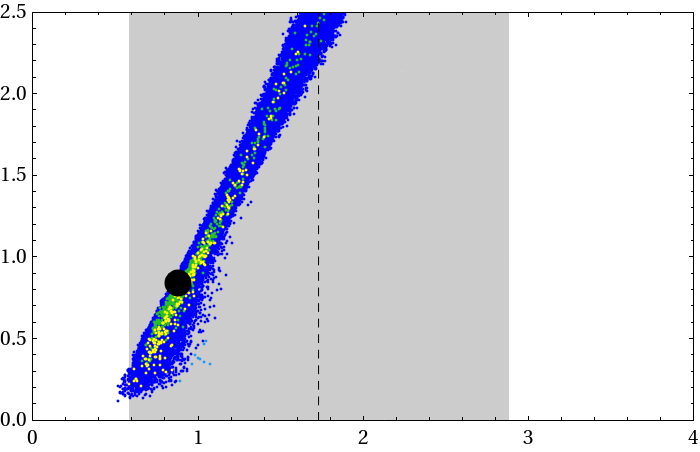}}
& 5
\\
$\cdots$
& \parbox[c]{0.4\textwidth}{\includegraphics[width=0.38\textwidth]{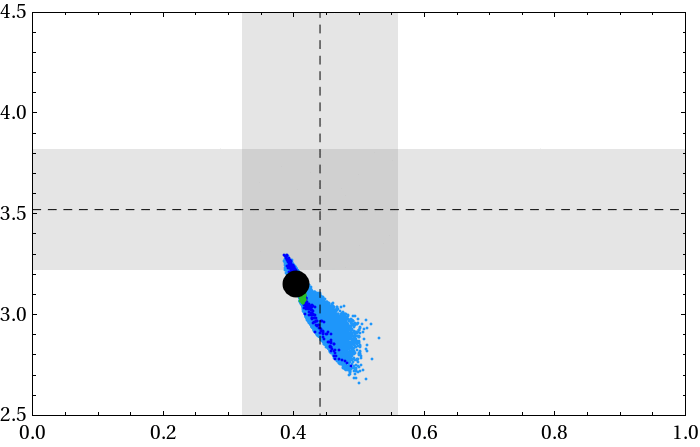}}
& \parbox[c]{0.4\textwidth}{\includegraphics[width=0.38\textwidth]{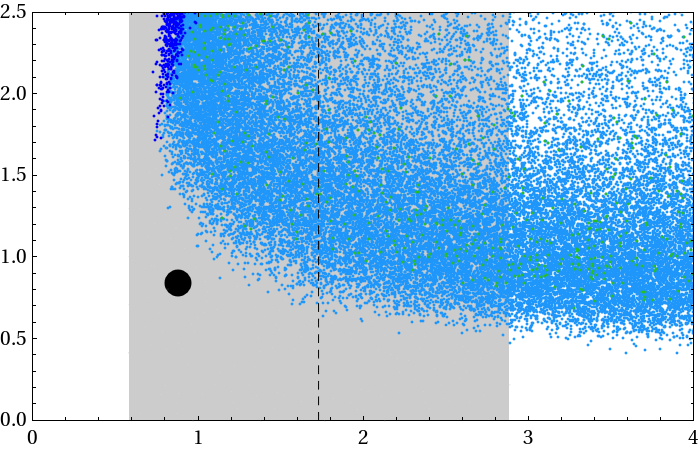}}
& 6
\\
$\cdots$
& \parbox[c]{0.4\textwidth}{\includegraphics[width=0.38\textwidth]{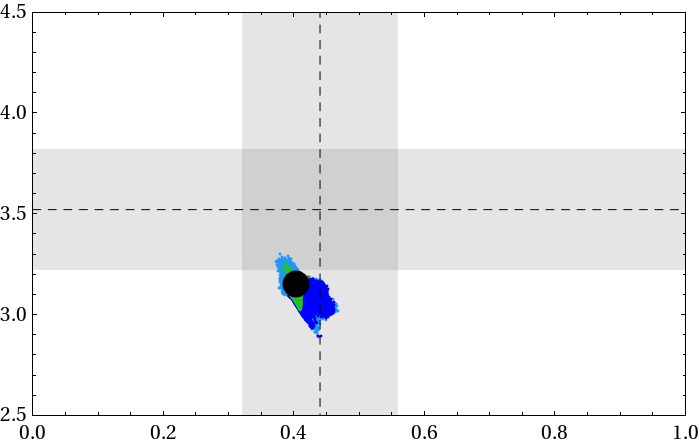}}
& \parbox[c]{0.4\textwidth}{\includegraphics[width=0.38\textwidth]{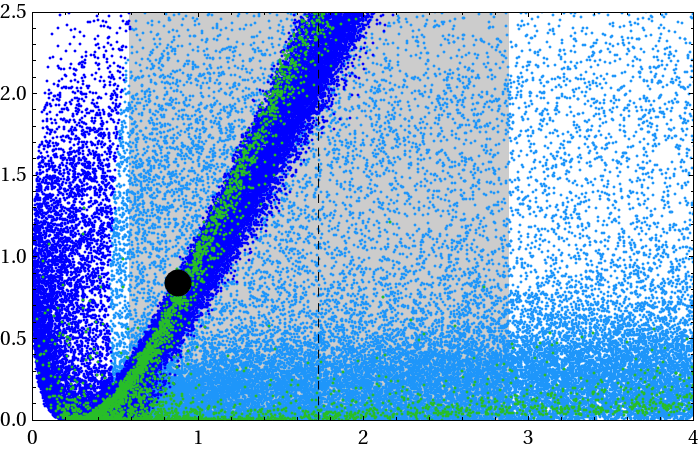}}
& 7
  \end{tabular}
 \end{center}
\caption{\label{tab:mixanatomy4}
Selected correlations for classes identified in Table~\ref{tab:procedure}
(part 4 of 4).}
\end{table}

\FloatBarrier


\newsection{Summary and Conclusions}
\label{sec:concl}
In the present paper we have performed a detailed analysis of non-MFV effects
in the $K$, $B_d$ and $B_s$ systems in the SM4,  paying particular
 attention to correlations between various flavour observables and addressing within 
this framework a number of anomalies present in the experimental data. 

Similarly  as in the LHT model, the RSc model and
supersymmetric flavour models that have been analysed in \cite{Blanke:2006sb,Blanke:2006eb,Blanke:2009am,Blanke:2008zb,Blanke:2008yr,Altmannshofer:2009ne}, the correlations between various observables that are characteristic for models
with MFV and in particular with CMFV can be strongly violated in the SM4 
while still satisfying all existing data on flavour violating processes and
electroweak precision observables.

Probably, the most striking signature of the SM4, compared to the LHT, RSc and
SUSY flavour models, is the possibility of having simultaneously sizable 
NP effects in the $K$, $B_d$ and $B_s$ systems, even if truly spectacular 
effects are only possible in rare $K$ decays with the prominent exception
of $S_{\psi\phi}$ which can also be enhanced by an order of magnitude.
This different pattern can be traced back to the fact that the mass scales involved in the SM4 are generally significantly lower than in the LHT and in particular in the RSc. 
\update{Also the non-decoupling of the new heavy fermions $t'$ and $b'$ plays a role here, whereas an analogous effect does not occur in the LHT, RSc and SUSY.
}

We recall that in SUSY flavour models, NP effects in the $K$ system are
small but can be large in $B$ physics observables to which dipole operators 
and scalar operators contribute. In models with right-handed currents 
also $S_{\psi\phi}$ can be large. In the LHT and RSc models, large 
effects are found predominantly in rare $K$ decays, although $S_{\psi\phi}$
can also be sizable. These three different global patterns of flavour 
violation in the models in question could help one day to distinguish 
between these NP scenarios.

The most interesting patterns of flavour violation in the SM4 are the 
following ones.
\begin{itemize}
\item
All existing tensions in the UT fits can be removed in this NP scenario.
\item
In particular the desire to explain the $S_{\psi\phi}$ anomaly implies,
 as seen in Fig.~\ref{fig:CP-asymmetries}, uniquely the suppressions of 
 the CP asymmetries $S_{\phi K_S}$ and $S_{\eta' K_S}$ in agreement with the data.
 This correlation has been pointed out in \cite{Soni:2008bc,Hou:2006jy,Hou:2006zza}, and we confirm 
 it here. However we observe that for $S_{\psi\phi}$ significantly larger
 than 0.6 the values of $S_{\phi K_s}$ and $S_{\eta' K_s}$ are below their central values indicated by the data.
\item
The same anomaly implies a sizable enhancement of ${\rm Br}(B_s\to \mu^+\mu^-)$
over the SM3 prediction although this effect is much more modest than 
in SUSY models where the Higgs penguin  with large $\tan\beta$ is at work. Yet,
values as high as $8\cdot 10^{-9}$ are certainly possible in the SM4, 
which is well beyond those attainable in the LHT model and  the RSc model.
\item
Possible enhancements of ${\rm Br}(\kpn)$ and ${\rm Br}(\klpn)$ over the SM3 values
are much larger than found in the LHT and RSc models and in particular in
SUSY flavour models where they are SM3 like. Both branching ratios as high
as several $10^{-10}$ are still possible in the SM4.
 Moreover, in this case, the two branching ratios are strongly correlated
as seen in Fig.~\ref{fig:Kpinunu} and close to the GN bound. 
\update{For an earlier analysis see \cite{Hou:2005yb}.}
\item
Interestingly, in contrast to the LHT and RSc models, a high value of 
$S_{\psi\phi}$ does not preclude a sizable enhancements of ${\rm Br}(\kpn)$, 
and ${\rm Br}(\klpn)$.
\item
NP effects in $K_L\to\pi^0\ell^+\ell^-$ and $K_L\to\mu^+\mu^-$ can be visibly
larger than in the LHT and RSc models. In particular 
${\rm Br}(K_L\to\mu^+\mu^-)_{\rm SD}$ can easily violate the existing bound of
$2.5\cdot 10^{-9}$. Imposition of this bound on top of other constraints 
results in a characteristic shape of the correlation between 
${\rm Br}(\kpn)$ and ${\rm Br}(\klpn)$ shown in Fig.~\ref{fig:Kpinunu}.
\item 
The magnitude of the CP asymmetry $A_{\rm CP}^{bs\gamma}$ remains 
  small, but the desire to explain large values of $S_{\psi\phi}$ reverses its sign.
\item 
 Even in the presence of SM-like values for $S_{\psi\phi}$ and ${\rm Br}(B_s\to \mu^+\mu^-)$, large effects in the $K$-system are possible as illustrated by green points in numerous plots.
\item
For large positive values of $S_{\psi\phi}$ the predicted value of $\epe$ is significantly below the data, unless the
 hadronic matrix elements of the electroweak penguins are sufficiently suppressed with respect to the large $N$ result and the ones of QCD penguins enhanced.
\item 
{
We have also reemphasised \cite{Buras:1998ed,Buras:1999da} the important role $\epe$ will play in bounding rare $K$ decay branching ratios once the relevant hadronic matrix elements in $\epe$ will be precisely known (see Sec.~\ref{subsec:epe} for more details).
}
\end{itemize}

 Other 4G effects in various observables and correlations between them
can be found in numerous plots in Sections~\ref{sec:numerics}--\ref{sec:DetV4G}.
In particular, in Section~\ref{sec:DetV4G}, we have addressed the question of the determination of the parameters of the SM4 with the help of future measurements
employing various scenarios for the scalings of the 4G mixing matrix angles $\theta_{14}, \theta_{24}$ and $\theta_{34}$ with the Wolfenstein parameter $\lambda\ll 1$.

In summary, the SM4 offers a very rich pattern of flavour violation which can be tested already in the coming years, in particular through precise measurements
of $S_{\psi\phi}$, ${\rm Br}(B_q\to \mu^+\mu^-)$, ${\rm Br}(K^+\to \pi^+\nu\bar\nu)$
and, later, $S_{\phi K_s}$, $S_{\eta' K_s}$ and ${\rm Br}(K_L\to \pi^0\nu\bar\nu)$. 
Also, precise measurements of the phase $\gamma\approx \delta_{13}$ 
will be important for this investigation.
We close our detailed analysis with the following important question:
Can the SM4 be excluded by precise measurements of FCNC processes? 
The answer is positive, provided
large departures from the SM3 are observed with a different pattern of deviation from the SM3 predictions than found in our analysis. Let us just list three examples:
\begin{itemize}
\item Large values of $S_{\psi\phi} > 0.6$ accompanied with SM-like values of ${\rm Br}(B_s\to \mu^+\mu^-)$ will clearly disfavour the SM4 as the explanation of the $S_{\psi\phi}$
anomaly.
\item Similarly, such large values accompanied by the observation of $S_{\phi K_s}\approx S_{\eta' K_s}\approx S_{\psi K_s}$ would also put the SM4 into difficulties.
\item Finally, $S_{\psi\phi} > 0.3$ accompanied by the $\epe$ relevant $R_6 \approx R_8 \approx 1$ from future lattice calculations will disfavour SM4 by means of the measured $\epe$. 
\end{itemize}
It is evident from our analysis and other analyses mentioned in the beginning of our paper that FCNC processes will contribute in a profound manner to the search for the
4G quarks and leptons and, in the case of direct discovery,
 to the exploration of the structure of their weak mixings.
We hope that our numerous plots will help in monitoring the future data
from LHC and high-intensity experiments, with the hope to firmly establish 
or firmly exclude the presence of the fourth sequential generation of quarks in
nature.

\subsection*{Note added}
Very recently, during the final stages of our work, a detailed study of rare $K$ and $B$ decays has been presented in \cite{Soni:2010xh}, see also the discussion in Section~\ref{sec73}.

\subsection*{Acknowledgements}
We would like to thank Wolfgang Altmannshofer, Monika Blanke, Amarjit Soni and Soumitra Nandi for illuminating discussions.
This research was partially supported by the Cluster of Excellence `Origin and Structure of the Universe', the Graduiertenkolleg 
GRK 1054 of DFG and by the German `Bundesministerium f{\"u}r Bildung und
Forschung' under contract 05H09WOE.

\clearpage

\begin{appendix}

\newsection{Relevant Functions\label{app:functions}}
\begin{eqnarray}
\label{X0}X_0(x_i)&=&\frac{x_i}{8}\;\left[\frac{x_i+2}{x_i-1}
+\frac{3 x_i-6}{(x_i -1)^2}\; \log x_i\right],\\
\label{Y0}Y_0(x_i)&=&\frac{x_i}{8}\; \left[\frac{x_i-4}{x_i-1}
+ \frac{3 x_i}{(x_i -1)^2} \log x_i\right],\\
\label{Z0}Z_0(x_i)&=&-\frac{1}{9}\log x_i+\frac{18 x_i^4-163x_i^3+259
  x_i^2-108x_i}{144(x_i-1)^3}\nn\\
&&+\frac{32x_i^4-38x_i^3-15x_i^2+18x_i}{72(x_i-1)^4}\log x_i\,.\\
\label{C0}C_0(x_i)&=&\frac{x_i}{8}\left[\frac{x_i-6}{x_i-1}+\frac{3x_i+2}{(x_i-1)^2}\log x_i\right],\\
\label{D0}D_0(x_i)&=&-\frac{4}{9}\log
x_i+\frac{-19x_i^3+25x_i^2}{36(x_i-1)^3}+\frac{x_i^2(5x_i^2-2x_i-6)}{18(x_i-1)^4}\log x_i\,,\\
\label{E0} E_0(x_i)&=&-\frac{2}{3}\log x_i + \frac{x_i^2(15-16x_i+4x_i^2)}{6(x_i-1)^4}\log x_i
+ \frac{x_i(18-11x_i-x_i^2)}{12(1-x_i)^3}\,,\\
\label{Dp0}D'_0(x_i)&=&-\frac{(3x_i^3-2x_i^2)}{2(x_i-1)^4}\log x_i +
\frac{(8x_i^3+5x_i^2-7x_i)}{12(x_i-1)^3}\,,\\
\label{Ep0}E'_0(x_i)&=&\frac{3x_i^2}{2(x_i-1)^4}\log x_i +
\frac{(x_i^3-5x_i^2-2x_i)}{4(x_i-1)^3}\,.
\end{eqnarray}

\begin{eqnarray}
B^{\mu\bar\mu}(x_i,y_j) &=& \frac{1}{4}\left[U(x_i,y_j)+\frac{x_i y_j}{4}U(x_i,y_j) - 2 x_i y_j \tilde U(x_i,y_j) \right]\,, \label{Bmumu}\\
B^{\nu\bar\nu}(x_i,y_j) &=& \frac{1}{4}\left[U(x_i,y_j)+\frac{x_i y_j}{16}U(x_i,y_j) +\frac{x_i y_j}{2} \tilde U(x_i,y_j) \right]\,, \label{Bnunu}
\end{eqnarray}
with
\begin{eqnarray}
 U(x_1,x_2) 	&=& \frac{x_1^2 \log x_1}{(x_1-x_2) (1-x_1)^2} + \frac{x_2^2 \log x_2}{(x_2-x_1) (1-x_2)^2} + \frac{1}{(1-x_1)(1-x_2)}\,,\\
\tilde U(x_1,x_2) &=& \frac{x_1 \log x_1}{(x_1-x_2) (1-x_1)^2} + \frac{x_2 \log x_2}{(x_2-x_1) (1-x_2)^2} + \frac{1}{(1-x_1)(1-x_2)}\,.
\end{eqnarray}
$S_0(x_i,x_j)$ for arbitrary $x_i,x_j$ is given by \cite{Buras:1983ap}
\begin{eqnarray}
	S_0 (x_i,x_j) &=& x_i x_j\left(\frac{(4 - 8 x_j + x_j^2)\log x_j}{4 (x_j-1)^2 (x_j-x_i)} +(i\leftrightarrow j) - \frac{3}{4 (x_i-1) (x_j-1)}\right)\,.
\end{eqnarray}
In the limit of $\varepsilon \rightarrow 0$ 
in $S_0(x_i+\varepsilon,x_i-\varepsilon)$ one recovers the SM3 version of $S_0(x_i)$,
\begin{eqnarray}
	S_0 (x_i) &=& \frac{x_i}{4}\;\frac{-4 + 15x_i - (12 - 6 \log x_i )x_i^2 + x_i^3}{(x_i-1)^3}\,.
\end{eqnarray}

\end{appendix}

\clearpage

\providecommand{\href}[2]{#2}\begingroup\raggedright\endgroup
\end{document}